*UNIVERSITE DE TUNIS EL MANAR*
FACULTÉ DES SCIENCES DE TUNIS

ÉCOLE DOCTORALE EN INFORMATIQUE

# MÉMOIRE DE MASTÈRE

présenté en vue de l'obtention du
Diplôme de Mastère en Informatique

par

Leila BEN OTHMAN
(Maîtrise en Informatique Appliquée à la Gestion, ISG de Tunis)

# Une nouvelle approche de complétion des valeurs manquantes dans les bases de données

Soutenu le 14 Novembre 2006, devant le jury d'examen

| | | |
|---|---|---|
| MM. | Yahya SLIMANI | Président |
| | Kamel BEN SALEM | Membre |
| | Khaled BSAIES | Directeur |
| | Sadok BEN YAHIA | Invité |

# Remerciements

Je tiens tout d'abord à remercier Mr. Yahya Slimani, Maître de Conférence à la Faculté des Sciences de Tunis qui me fait le grand honneur de présider le jury. Je le remercie en particulier pour l'évaluation qu'il a portée à mon travail.

Toute ma gratitude est également adressée à Mr. Kamel Ben Salem, Professeur à la Faculté des Sciences de Tunis, de bien vouloir accepter de rapporter ce travail. Je le remercie pour son attention accordée à ce mémoire.

Mes sincères remerciements sont de même adressés à Mr. Khaled Bsaïes, Professeur à la Faculté des Sciences de Tunis, pour la confiance qu'il m'a accordée pour mener à bien ce mémoire et pour son soutien tout au long de l'élaboration de ce travail.

Je remercie spécialement Mr. Sadok Ben Yahia, Maître assistant à la Faculté des Sciences de Tunis, de m'avoir permis de découvrir ce passionnant domaine de recherche, qui est la fouille de données ainsi que d'évoluer tant scientifiquement que personnellement. Je le remercie en particulier pour l'intérêt permanent qu'il a porté à mon travail, pour sa disponibilité ainsi que pour la richesse de ses réflexions scientifiques.

Je souhaite par la même occasion exprimer toute ma reconnaissance envers Mr. François Rioult, Maître assistant à l'Université de Caen Basse-Normandie, qui s'est montré plus que disponible à répondre à toutes mes questions et d'avoir mis à ma disposition son prototype MVminer utilisé dans le cadre de mon travail.

Je tiens également à remercier Mlle Yosr Slama, Assistante à la Faculté des Sciences de Tunis, pour son aide à l'élaboration de l'étude théorique de la complexité de l'algorithme proposé.

Je remercie tout spécialement Tarek Hamrouni pour la relation amicale et scientifique que nous entretenons. J'ai tant profité de son esprit d'analyse et de ses conseils constructifs.

Je souhaite de même exprimer toute ma gratitude envers ma famille. Je remercie affectueusement mes parents. Je les remercie pour le cadre agréable qu'ils m'ont toujours offert pour mener à bien ce travail. Toute ma reconnaissance va à mes frères et leur femme respective, pour leur présence quotidienne. Une pensée particulière est de même adressée à mon fiancé, pour ses encouragements et sa patience tout au long de l'aboutissement de ce mémoire. Enfin, une dernière pensée à mon cher neveu Youssef, pour sa belle présence.

# Table des matières

# Table des figures

# Liste des tableaux

# Introduction générale

Avec une évolution rapide des outils informatiques, de nombreux organismes construisent des bases de données de plus en plus volumineuses concernant leurs activités. En effet, nous sommes aujourd'hui entrés dans l'ère de l'information. Dans tous les secteurs économiques, l'information devient un capital en tant que tel et il s'avère de plus en plus simple de collecter des données à des coûts qui se rapprochent de la gratuité [31]. Il en résulte que la prise de décision devient de plus en plus complexe. Ainsi, nous sommes face à un enjeu décisionnel, matérialisé par "les risques décisionnels à l'ère de l'hyper choix" [31]. L'attention des entreprises s'est alors tournée vers des systèmes décisionnels qui contribuent véritablement à la différenciation stratégique [22]. Cependant, cette orientation ne résout pas le problème. En effet, il ne suffit pas de mettre en place un système décisionnel pour garantir une bonne prise de décision. Il faudra plutôt tirer le meilleur parti avec des outils d'analyse. C'est le but de la fouille de données (ou Data Mining). L'idée sous-jacente de la fouille de données est l'extraction de connaissances à partir d'un grand volume de données. Dans le domaine de l'analyse de risques, les établissements financiers peuvent par exemple vouloir déterminer si un crédit peut ou non être accordé à une personne. La fouille de données est apparue avec la promesse de fournir des outils et des techniques d'analyse et d'aide à la décision. Cependant, le terme fouille de données est fréquemment utilisé comme synonyme de l'ECD (Extraction de Connaissances à partir de Données). Il constitue en réalité, la phase d'extraction de connaissances à partir de données prépa- rées dans le processus de l'ECD [35]. L'ECD est un processus semi-automatique, itératif, constitué de plusieurs étapes allant de la sélection et le prétraitement des données jus- qu'à l'extraction (fouille), l'interprétation et la visualisation des résultats [35]. Cependant, même si les techniques dédiées à l'extraction de ces connaissances arrivent maintenant à maturité, elles n'en restent pas sans faille dans le cas où les données sont incomplètes. Ce problème classique, est inévitable dans le domaine des bases de données. Par conséquent, des techniques de prétraitement doivent être appliquées. Cette phase est primordiale, puisque la qualité des connaissances ainsi extraites est étroitement liée à la qualité des données [41]. Ceci explique le fait que cette phase constitue 60% de l'effort [23]. C'est dans le cadre de prétraitement des données incomplètes que se situe ce travail. La complétion[1] des valeurs manquantes en constitue la principale problématique.

# Motivations

Les données collectées dans le but d'analyse sont souvent susceptibles d'être bruitées, incomplètes ou inconsistantes [23]. Certainement, parce qu'elles proviennent de différentes sources et parce qu'elles constituent des données réelles [46]. Dans le cadre de ce mémoire, nous nous intéresserons particulièrement au prétraitement des données incomplètes. En effet, nous verrons au chapitre 1, que dans la vie courante, les phénomènes conduisant à ce genre de perturbations sont inévitables et imprévisibles. Ainsi, et avec l'émergence des techniques d'extraction de connaissances à partir de données, ce problème a commencé à intéresser la communauté de la fouille de données. En effet, dans un des livres de référence sur l'ECD, Fayyad, Shapiro et Smyth classent cette problématique comme la 5ème tâche prioritaire en ECD [41].

Nous verrons dans le cadre de ce mémoire que ce problème a suscité l'intérêt de plusieurs communautés scientifiques, de nombreuses solutions ont été ainsi proposées. Parmi ces travaux, des approches proposent d'utiliser une des techniques de la fouille de données, à savoir la découverte des règles associatives. Ces approches se basent sur le fait que les règles associatives décrivent des corrélations entre les données. La complétion des valeurs manquantes devient alors possible par une prise en considération des corrélations ainsi induites. Une étude bibliographique a montré que les approches classiques de complétion se sont focalisées sur l'extraction des règles associatives, procédant par une suppression préalable de toute donnée incomplète. Ces solutions sont loin d'assurer un traitement optimal puisque ignorer des données introduit un biais[2] dans l'analyse [48]. Ceci conduit à l'obtention de règles peu fiables affectant l'efficacité du processus de complétion. De plus, le défi majeur face aux valeurs manquantes ne réside-t-il pas dans l'exploitation même des données incomplètes? Conscients de cet enjeu, d'autres travaux, se sont focalisés sur l'extraction des règles à partir de la base totale, incluant les données incomplètes. Cependant, ces approches pèchent par leur inefficacité dans la résolution du problème de conflit. En effet, ce problème présente un impact capital lors de l'étape de complétion, i.e., pouvant conduire à la détérioration de l’efficacité du processus.

[1] Trouver une valeur de remplacement à une valeur manquante.

[2] Tout fait susceptible de rendre une base non représentative.

Pour remédier aux insuffisances susmentionnées, notre contribution consiste en la proposition d'une nouvelle approche de complétion des valeurs manquantes, appelée $GBAR_{MVC}$. La particularité de cette approche est qu'elle ne procède pas à la complétion d'une manière brutale. Le processus de complétion est préalablement associé à une prise en considération de l'aspect incomplet des données. De plus, cette approche se démarque lors de l'étape de complétion par l'utilisation d'une base générique de règles associatives. Ces règles montrent plusieurs intérêts quant à la complétion des valeurs manquantes. En effet, cette base générique renvoie un nombre minimal de règles, démunies de redondance et pré- sentant le minimum de contraintes à satisfaire lors de la complétion. De plus, ces règles assurent une réduction considérable des conflits. Ces conflits sont inévitables à toute approche de complétion. Pour cette même raison, nous proposons également une nouvelle mesure de résolution de conflit.

## Structure du mémoire

Les travaux réalisés sont résumés dans ce mémoire qui est composé de quatre chapitres.

Le premier chapitre introduit le problème des valeurs manquantes, les différentes typologies associées ainsi que quelques solutions dédiées au traitement des valeurs manquantes. Ensuite, il présente le formalisme de dérivation des règles associatives, ainsi que celui d'extraction des bases génériques de règles associatives. Finalement, des adaptations proposées dans le cadre de l'extraction des règles associatives, en présence de valeurs manquantes sont présentées.

Le deuxième chapitre compare sur la base de critères définies, les différents travaux proposés dans la littérature, dédiés à la complétion des valeurs manquantes et utilisant la technique des règles associatives. Ce chapitre constitue un état de l'art en la matière.

Dans le troisième chapitre, nous commençons par montrer l'intérêt de l'utilisation d'une base générique de règles associatives quant à la complétion des valeurs manquantes. Ensuite, nous proposons une nouvelle approche de complétion, appelée $GBAR_{MVC}$, basée sur une base générique de règles associatives et nous introduisons une nouvelle mesure de résolution de conflit, intitulée Robustesse. Ce chapitre propose également une étude de la complexité théorique de l'approche $GBAR_{MVC}$.

Par le quatrième chapitre, nous présentons une évaluation expérimentale de l'approche $GBAR_{MVC}$, réalisée sur des bases benchmark. Cette évaluation porte sur plusieurs axes définis sur la base de critères d'évaluation. Ces expérimentations permettent d'une part d'analyser les caractéristiques de l'approche proposée et d'autre part de la comparer à une approche existante dans la littérature.

La conclusion du mémoire dresse un état récapitulatif de notre travail et propose de nouvelles perspectives et axes de recherche ultérieurs.

# Chapitre 1

# Valeurs manquantes et règles associatives

## 1.1 Introduction

Les données collectées dans le but d'analyse sont souvent bruitées, incomplètes ou inconsistantes [23]. Ce problème est classique dans le domaine des bases de données. Puisque celles-ci proviennent de processus réels d'acquisition et de techniques d'intégration [46]. En effet, dans la pratique, ce genre de perturbation est inévitable et imprévisible. Une erreur de saisie, un oubli de la part de l'utilisateur ou un dysfonctionnement d'un équipement de mesure sont des exemples classiques de tels phénomènes. L'application d'un processus d'extraction de connaissances à partir de telles données va avoir comme conséquence, l'obtention d'un résultat aberrant et peu able. Ainsi, l'efficacité du processus est considérablement affectée par la qualité des données [23]. Par conséquent, des techniques de prétraitement doivent être appliquées afin d'améliorer la qualité des données et des connaissances extraites. Pour cela, il existe diverses techniques de prétraitement. Le nettoyage des données est appliqué dans le cas où les données sont inconsistantes, incomplètes ou bruitées. La technique d'intégration doit être mise en œuvre afin de fusionner des don- nées provenant de différentes sources. La transformation a pour but la modification du format des données afin de les rendre appropriées à l'analyse. La réduction des données permet de réduire la taille de celles-ci par application de techniques diverses, telles que l'agrégation, l'échantillonnage ou le regroupement[23].

## 1.2 Le nettoyage des données

Les données recueillies par les systèmes décisionnels sont loin d'être parfaites. Cet aspect d'imperfection a été désigné par l'aspect dangereux des données [10]. En effet, dans certaines applications, l'erreur est fatale. Il s'agit généralement de systèmes d'assistance médicale où la vie humaine est en jeu. Ou encore de systèmes bancaires et financiers telle que la détection de fraudes. Le nettoyage des données, qui est une technique de pré- traitement, est apparu avec la promesse d'améliorer et de corriger cet aspect d'imperfection. En effet, dans la vie courante, les données sont souvent bruitées, i.e., contiennent des erreurs et des valeurs aberrantes, inconsistantes, i.e., manquent de cohérence, ou incomplètes, i.e., présentent des valeurs manquantes [23].
Dans ce qui suit, nous présentons ces différents phénomènes, susceptibles de perturber la qualité des données. Cependant, dans le cadre de ce mémoire, nous nous intéressons essentiellement au prétraitement des données incomplètes.

1. Les données bruitées : une donnée bruitée est définie, comme étant une valeur qui semble être correcte, mais qui en réalité, ne l'est pas [10]. Ces valeurs peuvent être identifiées, car elles constituent des valeurs aberrantes. Un exemple de ce phénomène a été observé dans une entreprise, où les dates de naissance de 5% des clients étaient identiques [10]. Non seulement identiques, mais exactement le 11 Novembre 1911. Il s'est avéré, que le fait de saisir une suite de "11" était le moyen le plus simple et le plus rapide de remplir le champ date de naissance.
2. Les données inconsistantes : les données inconsistantes sont généralement des données présentant de la discordance [23]. L'inconsistance des données peut être une conséquence directe du processus d'intégration. C'est le cas par exemple d'une donnée présentant la même information et possédant différentes désignations dans les différents systèmes d'une entreprise [23].
3. Les données incomplètes : une donnée incomplète, i.e., présentant des valeurs manquante a été définie comme l'absence de réponse à un ou plusieurs descripteurs[1] dans une base de données [41].

Dans ce qui suit, nous présentons les phénomènes classiques conduisant à la présence de valeurs manquantes et les différentes typologies associées.

[1]Ou attributs.

### 1.2.1 Phénomènes conduisant à la présence de valeurs manquantes

Une donnée manquante peut se produire dans l'un des cas suivants [21] :

Une donnée perdue : une donnée peut être saisie, ensuite effacée suite à une mauvaise manipulation.

Une donnée jugée non pertinente : la valeur de cette donnée n'a aucune influence sur la prise de décision. Par exemple, un médecin peut être en mesure de faire le diagnostic d'un malade bien que certains tests sont non effectués. Les résultats de ces tests manquants constituent des données non pertinentes pour le diagnostic de la maladie [21].

Une donnée non spécifiée : pour spécifier une donnée, on est parfois amené à accéder à des bases situées au niveau du serveur. Une panne du serveur ou un dysfonctionnement lors du transfert des données peut introduire la non spécification d'une donnée. En outre, en ce qui concerne les sondages d'opinions, il est générale- ment rare qu'une personne remplisse complètement le questionnaire par manque de temps, ou parce qu'elle refuse tout simplement de répondre [48].

En n, un dernier exemple évoqué dans [41] concerne la non adéquation des choix de réponses avec la réponse réelle. L'utilisateur sera obligé de laisser le champ vide, puisqu'il n'existe pas de réponse adéquate.

### 1.2.2 Typologies des valeurs manquantes

L'absence de réponse à un descripteur cache trois types de valeurs manquantes. Vu l'absence de terminologie Française en littérature de ces différentes typologies, nous utiliserons la terminologie telle que présentée par Little et Rubin [32]. Nous distinguons les valeurs manquantes de type MCAR [2], les valeurs manquantes de type MAR[3] et les valeurs manquantes de type NMAR[4] :

– Valeur Manquante MCAR : une donnée ou une observation est dite affectée par une valeur manquante de type MCAR, lorsque la probabilité d'avoir un descripteur manquant est indépendante de la valeur même du descripteur. D'ailleurs, la majeure partie des travaux dédiés aux traitements des valeurs manquantes le suppose [50].

[2]Missing Completely At Random.
[3]Missing At Random.
[4]Not Missing At Random.

C'est le cas où la valeur manquante n'est associée à aucun phénomène particulier, une erreur de saisie, une impossibilité d'effectuer une mesure sont des exemples de valeurs manquantes MCAR.

- Valeur Manquante MAR : une observation est dite affectée par une valeur manquante de type MAR, lorsque la probabilité d'avoir un descripteur manquant ne dépend pas de la valeur même du descripteur, mais d'autres valeurs observées. Un exemple illustratif de ce type de valeur manquante est quand une pièce n'est pas défectueuse, la taille du défaut serait nécessairement absente [46].
- Valeur Manquante NMAR : une observation est dite affectée par une valeur manquante de type NMAR, lorsqu'elle cache une certaine valeur qu'on peut à priori déterminer. Dans ce cas, une valeur manquante est dite informative. Un exemple de ce type de valeur manquante est lors des sondages sur les salaires, les refus de réponses, se traduisant par des valeurs manquantes, indiquent généralement des personnes ayant un salaire élevé [41]. Un autre exemple est lorsque les valeurs sont absentes à cause du fait qu'elles présentent des valeurs impossibles à mesurer, soit parce qu'elles sont très élevées, soit parce qu'elles sont trop basses [33].

## 1.3 Traitement classique des valeurs manquantes

Le problème des valeurs manquantes est un problème largement étudié par différentes communautés scientifiques. Dans ce qui suit, nous présentons des solutions qui ont été proposées dans le domaine de la statistique, étant donné que le problème des valeurs manquantes a été étudié en premier lieu par la communauté des statisticiens depuis les années 80 [32]. Ensuite, nous présentons des solutions proposées dans le cadre des travaux issus du domaine de la classification supervisée[5], vu l'abondance de ces travaux dans ce domaine. Cependant, nous nous intéresserons plus particulièrement au chapitre 2 aux solutions proposées dans le contexte des règles associatives. Cette technique d'extraction a été utilisée pour la première fois à des fins de complétion [42, 43] et semble depuis intéresser la communauté de la fouille de données pour résoudre le problème des données manquantes [47]. Cependant, il s'avère utile de présenter en premier lieu quelques solutions adoptées dans les domaines que nous avons déjà mentionnés.

[5] Une des techniques de la fouille de données.

### 1.3.1 Solutions simples

Les solutions les plus simples de traitement des valeurs manquantes consistent à :

Supprimer toute observation présentant au moins une valeur manquante. Cette solution est connue sous le nom LD[6]. Cependant, dans certains cas, cette solution peut conduire à la suppression de 98% des observations totales de la base. Par exemple, dans le cas de la base de la maladie de Hodgkin[7], 808 observations présentent des valeurs manquantes, parmi un total de 824 observations [41]. Ainsi, cette solution s'avère inadaptée dans le cas où la base contient plusieurs observations manquantes. Par conséquent, cette technique rend la base de données non représentative et introduit un biais dans l'analyse.

Supprimer tout descripteur affecté par des valeurs manquantes. Cette solution est connue sous le nom VD[8]. De la même façon, une telle solution conduit à rendre la base non représentative. Cependant, à l'inverse de la première solution, cette méthode introduit un biais non pas au niveau de la population mais au niveau de la description de la base. Ces descripteurs supprimés peuvent s'avérer intéressants lors de l'analyse.

Considérer la valeur manquante comme une nouvelle valeur spéciale [38]. Cette solution se propose d'affecter à toutes les valeurs manquantes une même valeur. C'est une solution valable seulement dans le cas où les valeurs manquantes sont informatives et cachent une même valeur [33, 41]. Cependant, généralement, les valeurs manquantes sont de type MCAR [50]. Dans ce cas, la solution indiquée devient in- appropriée puisque cela revient à affecter toutes les valeurs manquantes une même valeur.

### 1.3.2 Analyse statistique des valeurs manquantes

Le problème des valeurs manquantes a suscité l'intérêt des statisticiens depuis les années 80 [32]. Pour pallier la faiblesse des méthodes dites simples, des solutions qui au lieu de supprimer toute observation présentant une valeur manquante, proposent de déterminer la valeur cachée derrière celle qui manque. Ces solutions viennent renforcer le constat

[6]List-wise deletion.

[7]Disponible à l'adresse suivante http://ics.uci.edu/mlearn/MLRepository.html.

[8]Variable deletion.

"Améliorer la qualité des données manquantes est meilleur que de les ignorer"[9]. En effet, ceci permettra, d'une part, de travailler sur des bases représentatives. D'autre part, d'obtenir des données complètes pouvant être utilisées lors de toute analyse des connaissances en aval.

Les principales solutions proposées dans le cadre de la statistique sont : l'imputation[10] par la moyenne, la médiane, ou la valeur la plus commune [32]. Toutes ces solutions consistent à affecter la moyenne, la médiane ou la valeur la plus commune à une valeur manquante. Cependant, ces solutions sont jugées non efficaces par les statisticiens puisqu'elles introduisent une sous-estimation/surestimation dans les données [34]. Une autre solution proposée dans le domaine de la statistique est l'algorithme EM[11], qui permet d'estimer la distribution d'un échantillon de données manquantes se basant sur la technique des voisins les plus proches [17]. Une méthode basée sur la même technique a été de même proposée dans la littérature [7]. Le principe de cette solution réside sur le fait qu'elle procède par l'imputation des valeurs manquantes considérées comme des trous dans un tableau de données par la recherche des k-voisins les plus proches itérativement. L'avantage de cette solution, est que cette imputation est faite en un nombre réduit d'itérations [7]. L'imputation par la méthode de régression est une autre solution qui consiste à déterminer la corrélation qui existe entre les variables pour imputer une valeur manquante. La méthode des moindres carrés peut être utilisée dans ce cas. L'inconvénient de cette méthode, est qu'elle se base sur des modèles mathématiques non justifiés [34]. Cependant, il est important de signaler l'existence de plusieurs autres travaux dédiés au traitement des valeurs manquantes dans le domaine de la statistique tels que [20, 44].

### 1.3.3 Techniques de traitement des valeurs manquantes dans la classification supervisée

La classification supervisée est parmi les techniques utilisées dans l'ECD. Elle est constituée de deux étapes [35] :

---

[9] Making up missing data is better than throwing out it away.

[10] Le terme imputation est utilisé par les statisticiens pour désigner le terme complétion.

[11] L'acronyme EM désigne Expectation Maximization.

1. Construction d'un modèle de classification, en choisissant un échantillon significatif d'objets[12] de la base que l'on appelle jeu d'essai. Dans cet échantillon, chaque objet va être associé à une classe. Ce modèle peut se présenter sous différentes formes : arbres de décisions [38], règles de classifications [3], réseaux neuronaux [49], etc.
2. Utilisation de ce modèle afin de classifier de nouveaux objets, dont la classe d'appartenance est inconnue.

Le problème des valeurs manquantes peut alors se poser lors de la première étape, i.e., lors de la construction du modèle de classification, ainsi que lors de la deuxième étape, i.e., lors de la classification [33]. Nous présentons brièvement dans ce qui suit les principaux travaux effectués dans le cadre des arbres de décision. Ces travaux et bien d'autres sont présentés dans [33, 39, 41] et constituent un état de l'art en la matière.

- L'approche C4.5 : l'approche C4.5 est une extension de l'approche ID3, qui est un algorithme de classification basé sur les arbres de décision [40]. Le traitement des valeurs manquantes a été introduit selon une approche probabiliste, qui consiste à envoyer dans différentes branches de l'arbre un objet présentant une valeur manquante. Cela est fait par affectation d'une probabilité à chaque valeur d'un descripteur manquant. Cette probabilité est calculée à partir de toutes les observations appartenant à une même classe et ayant des valeurs du descripteur connues. Ainsi, une donnée peut se retrouver fragmentée dans plusieurs sous-branches de l'arbre avec différentes probabilités. Dans ce cas, l'algorithme C4.5 renvoie la valeur qui présente la probabilité la plus élevée. Cependant, Il a été démontré dans [41] que cette approche se base sur le principe de la valeur la plus commune, qui est loin d'assurer une complétion correcte.
- Une autre méthode proposée par Shapiro et décrite par Quinlan dans [38] consiste à utiliser un arbre de décision pour compléter une valeur manquante. Cette méthode se base sur le constat suivant si un arbre est capable de déterminer la classe d'un objet à partir de l'information disponible, alors il peut être utilisé, pour déterminer les valeurs manquantes d'un descripteur $X$, par permutation de la classe et du descripteur $X$ [41]. Ainsi, la classe devient un descripteur alors que le descripteur présentant la valeur manquante devient la classe à déterminer. C'est une technique qui s'avère adéquate seulement dans le cas où les valeurs manquantes affectent un seul

[12] C'est le terme utilisé dans la classification pour désigner une observation.

descripteur [33, 41].

Bien que, les solutions procédant par la suppression des données manquantes s'avèrent simples, elles ne constituent pas en réalité une solution au problème. Puisqu'elles se contentent simplement de supprimer toute observation (ou descripteur) présentant une donnée manquante. Ceci aura comme conséquence, l'analyse d'une base dite biaisée, i.e., non représentative [48, 53]. De même, le fait de considérer la valeur manquante comme étant une nouvelle valeur est limité aux cas où la valeur manquante est informative [33, 41]. Quant aux approches statistiques, même si elles se basent sur un formalisme mathématique puissant, elles présentent l'inconvénient qu'elles sont assez difficiles à appliquer [26]. De plus, il a été montré que les approches de traitement des valeurs manquantes dans le cadre de la classification se basent essentiellement sur la valeur la plus commune [41]. Ainsi, ces approches sont loin d'assurer une complétion fiable.

Partant du fait que les règles associatives décrivent des relations entre les données, d'autres approches de complétion des valeurs manquantes ont été proposées [8, 24, 41, 53]. Ces approches se basent sur le fait qu'il devient facile de déterminer les valeurs qui se cachent derrière les valeurs manquantes par l'exploitation des règles associatives. Dans toute la suite de ce mémoire, nous nous placerons dans le cadre de complétion des valeurs manquantes par l'exploitation des règles associatives.

La technique d'extraction des règles associatives a été introduite par Agrawal et al. [1]. A l'origine, cette technique a été développée pour l'analyse de bases de données de transactions de vente, connue sous le nom de l'analyse du panier de la ménagère. Une règle associative est une relation d'implication $X \Rightarrow Y$ entre deux ensembles d'articles $X$ et $Y$. Cette règle indique que les transactions qui contiennent les articles de l'ensemble $X$ ont tendance à contenir les articles de l'ensemble $Y$ [35]. L'analyse d'une base de transactions d'un supermarché permettra par exemple, d’affirmer que 95% des clients qui achètent du café achètent du sucre. L'extraction des règles associatives a donc pour intérêt l'identification de corrélations significatives cachées entre les données. Ces corrélations peuvent s'avérer utiles pour les utilisateurs finaux [22]. La règle énoncée précédemment permettra par exemple, de suggérer à un gestionnaire de rapprocher la disposition du café et du sucre, ou au contraire de les éloigner s'il souhaite que le client soit tenté par d'autres produits au cours de son passage entre les rayons.

Dans ce qui suit, nous présentons le formalise de base de dérivation des règles associatives, tel que introduit dans [1].

## 1.4 Les règles associatives

Dans cette section, nous allons nous intéresser à la problématique de dérivation des règles associatives. Cette problématique repose sur le concept d'itemset fréquent. Nous nous intéresserons de même à la problématique de dérivation des bases génériques de règles associatives. Cependant, cette dérivation repose sur la notion d'itemset fermé fréquent [35] et des générateurs minimaux associés [4]. En outre, nous passerons en revue les différentes redéfinitions de ces concepts dans le cadre de données incomplètes. Ces redéfinitions ont été proposées afin de s'adapter à la présence des valeurs manquantes.

### 1.4.1 Formalisme de base

Notation 1 Tout au long de ce mémoire, nous utiliserons les notations suivantes :
$T(A)$ désigne une table relationnelle où :

- $T$ est un ensemble ni et non vide de tuples, $T = \{t_1, t_2,..., t_n\}$.
- $A$ est un ensemble ni et non vide d'attributs, $A = \{X_1, X_2,..., X_n\}$.
- À chaque attribut $A_i$, tel que $(i = 1, 2, ..., n)$, est associé un domaine de valeurs, noté $dom(A_i)$ qui peut être soit un ensemble discret de valeurs soit un ensemble continu.
- Un attribut $X_i$ peut présenter une valeur manquante, notée " ?".
- Une table présentant des valeurs manquantes est dite contexte incomplet, sinon elle est dite contexte complet.

Cependant dans le cadre des règles associatives, les données doivent être transformées sous format transactionnel. Cette transformation constitue le point de départ de toute méthode d'analyse, par la technique des règles associatives. Cependant, même si la binarisation[13] des attributs discrets ne pose pas un problème. La discrétisation des attributs continus constitue un axe de recherche à part entière qui sort du cadre de ce mémoire. Nous désignerons par contexte formel, une table sous son format transactionnel. Dans ce qui suit, nous présentons la définition d'un contexte formel.

[13] La transformation du format relationnel au format transactionnel.

Définition1 (Contexte formel)
Un contexte formel est un triplet $K = (O, I, R)$, décrivant deux ensembles finis $O$ et $I$ et une relation (d'incidence) binaire R, entre $O$ et $I$ tel que $R \subseteq O \times I$. L'ensemble $O$ est appelé ensemble d'objets (ou transactions) et $I$ est appelé ensemble d'items. Chaque couple $(o, i) \subseteq R$ désigne que l'objet $o \subseteq O$ possède l'item $i \subseteq I$.

Un item dans un contexte formel est donc associé à un couple (attribut, valeur) dans une table relationnelle. Cependant, même si le problème de valeurs manquantes ne se pose pas dans les bases transactionnelles [29, 41], une valeur manquante dans un contexte formel, cache la présence de l'un des items associés à l'attribut en question. Dans la suite du mémoire, nous désignerons par $K_{VM}$ tout contexte incomplet et par $K$ tout contexte complet.

Définition2 (Support)
Soit un itemset $X$ contenant $k$ items. Un objet $o \subseteq O$ contient l'itemset $X$ si $X \subseteq o$. Le support relatif de $X$, noté $Supp(X)$ est défini par, $\text{Supp}(X) = \frac{|\{o \in O \mid X \subseteq o\}|}{|O|}$.
Un itemset $X$ est dit fréquent si son support relatif est supérieur ou égal à un seuil noté minsup spécifié par l'utilisateur.

Définition3 (Règle d'association)
Une règle d'association basée sur Z est une implication entre itemsets de la forme R : $X \Rightarrow Y$ tels que $X \subseteq Z$ et $Y = Z \backslash X$. les itemsets X et Y sont respectivement appelés prémisse et conclusion de la règle R. Le support de la règle R est le support de l'itemset Z. La confiance de la règle définit une probabilité conditionnelle, qui correspond au nombre de fois où Z est présent, rapporté au nombre de présence de X. La confiance d'une règle R : $X \Rightarrow Y$ est définie comme suit : $Conf(R) = \frac{Supp(Z)}{Supp(X)}$. Une règle associative est dite exacte si $Conf(R) = 1$, sinon elle est dite approximative.

### 1.4.2 Extraction des itemsets fréquents et dérivation des règles associatives

Le problème d'extraction des règles associatives consiste à déterminer l'ensemble des règles associatives valides, i.e., dont le support et la confiance sont au moins égaux respectivement à minsup et minconf, fixés par l'utilisateur. Ce problème repose sur l'extraction des itemsets fréquents et a été résolu par l'un des algorithmes pionniers dans l'extraction des règles associatives, à savoir Apriori [2]. Ce problème peut être décomposé en deux sous-problèmes comme suit [1] :

1. Déterminer l'ensemble des itemsets fréquents.
2. Pour chaque itemset fréquent $I_1$, générer toutes les règles associatives de la forme $R$ : $I_2 \Rightarrow I_1$ tel que $I_2 \subset I_1$ et dont la confiance est supérieure ou égale à minconf.

Cependant, la technique d'extraction des règles associatives à partir des itemsets fréquents s'avère insuffisante lorsque l'espace de recherche est important ou la base est très dense [46]. De plus, cette technique d'extraction engendre des milliers, voire des millions de règles dont la plupart sont redondantes, i.e., convoyant la même information [9, 35]. Les représentations condensées se proposent de fournir une solution au problème d'extraction des itemsets fréquents. Ces représentations fournissent un résumé des motifs fréquents tout en assurant la reconstruction de ces derniers au besoin [46]. La représentation condensée la plus connue est celle des motifs[14] fermés [36, 54]. D'autres représentations ont été introduites telles que les motifs $\delta-$libres [11], les motifs libres disjonctifs [14] et les motifs *k*-libres [15]. Toutes ces représentations offrent les avantages suivants :

1. La concision de l'ensemble des motifs obtenus: une représentation $R$ est dite plus concise qu'une représentation $R'$ si le nombre de motifs extraits à partir de $R$ est inférieur ou égal au nombre de motifs extrais à partir de $R'$ pour une même valeur de minsup [15].

2. L'efficacité d'extraction lorsque les techniques classiques d'extraction des motifs fréquents deviennent impraticables [12].

3. La possibilité de reconstruire à partir de ces représentations, l'ensemble de motifs fréquents sans effectuer un accès supplémentaire à la base [25].

Dans ce qui suit, nous allons nous intéresser à la représentation condensée des motifs fermés [36] ainsi que leurs générateurs minimaux associés [4]. L'approche d'extraction des motifs fermés repose sur une théorie mathématique, connue sous le nom de l'analyse formelle de concepts [52]. Cette théorie traite des concepts formels, considérés en tant qu'objets auxquels s'appliquent un ensemble d'items [22]. L'analyse formelle de concepts se base essentiellement sur la notion de Correspondance de Galois [18]. La section suivante

[14] Un motif désigne un itemset.

est consacrée à la présentation de cette notion.

### 1.4.3 Extraction des itemsets fermés fréquents et dérivation des bases génériques de règles associatives

Définition4 (Correspondance de Galois)
Soit un contexte d'extraction $K = (O, I, R)$. Soient $f$ et $g$ deux opérateurs définis comme suit [18] :

$$f(O) = \{i \in I \mid \forall o \in O, (o, i) \in R\}$$
$$g(I) = \{o \in O \mid \forall i \in I, (o, i) \in R\}$$

$f$ représente l'ensemble de tous les items communs à un ensemble d'objets $O$ (intention) et $g$ l'ensemble des objets qui contiennent tous les items de $I$ (extension). Le couple $(f, g)$ définit une Correspondance de Galois entre $O$ et $I$. $h = f \circ g$ et $h' = g \circ f$ sont les opérateurs de fermeture de la connexion de Galois.

Les opérateurs $h$ et $h'$ sont des opérateurs de fermeture, puisqu'ils vérifient les propriétés suivantes [16] : $\forall I_1, I_2 \subseteq I$ (respectivement $\forall o_1, o_2 \subseteq O$).

1. Extensivité: $I_1 \subseteq h(I_1)$.
2. Isotonie: $I_1 \subseteq I_2 \Rightarrow h(I_1) \subseteq h(I_2)$.
3. Idempotence: $h(h(I)) = h(I)$.

Définition 5 [35] (Itemset fermé) Étant donné l'opérateur de fermeture $h$ de la connexion de Galois. Un itemset $X \subseteq I$ est dit itemset fermé ssi $h(X) = X$.

Ainsi, un itemset fermé est l'ensemble maximal d'items communs à un ensemble d'objets [35]. L'itemset X est dit fermé fréquent s'il est fermé tel que $Supp(X) \geq minsup$.

Définition6 [4] (Générateur minimal) Un itemset $g \subseteq I$ est dit générateur minimal d'un itemset fermé $f$, si et seulement si $h(g) = f$ et il n'existe pas $g_1 \subseteq I$ tel que $g_1 \subset g$ et $h(g_1) = f$. L'ensemble $GM_f$ des générateurs minimaux d'un itemset fermé $f$ est défini comme suit :

$$GM_f = \{ g \subseteq I \mid h(g)=f \wedge \nexists\, g_1 \subset g \text{ tel que } h(g_1) = f \}.$$

Ainsi, ces nouvelles définitions ont donné lieu à une sélection d'un sous-ensemble de règles (les plus pertinentes), appelé base générique, à partir duquel toutes les autres règles pourraient être dérivées [35]. Plusieurs définitions de bases génériques ont été proposées dans la littérature [4, 19, 30]. Ces bases permettent de présenter à l'utilisateur le minimum de règles, tout en véhiculant le maximum d'informations [19]. Bastide et al. [4, 5] ont proposé deux bases génériques qui sont définies comme suit :

1. La Base générique de règles associatives exactes, est définie comme suit :

Définition7 Soit $IFF_K$ l'ensemble des itemsets fermés fréquents extrait d'un contexte d'extraction $K$. Pour chaque itemset fermé fréquent f $\in IFF_K$, nous désignons par $GM_f$ l'ensemble de ses générateurs minimaux. La base générique de règles associatives exactes est donnée par :

$$BG = \{R : g \Rightarrow (f - g) \mid f \in IFF_K \text{ et } g \in GM_f \text{ et } g = f\}.$$[15]

2. La base générique de règles associatives approximatives appelée Base informative de règles associatives approximatives, est définie comme suit [4, 5] :

Définition8 Soit $GMF_K$ l'ensemble des générateurs minimaux fréquents extrait d'un contexte d'extraction $K$. La base informative de règles associatives approximatives BI est donnée par : $BI = \{R : g \Rightarrow (f - g) \mid f \in IFF_K$ et $g \in GMF_K$ et $h(g) \subset f$ et $\mathrm{Conf}(R) \geq$ minconf$\}$.

## 1.5 Extraction de connaissances en présence de valeurs manquantes

Dans le cadre d'un contexte incomplet, la présence des valeurs manquantes cause un désagrément lors de l'extraction de connaissances. Par exemple, les valeurs manquantes posent un problème lors de l'évaluation du support d'un itemset. Par conséquent, cette évaluation doit être aménagée en fonction de ces valeurs manquantes. Ainsi, plusieurs solutions ont été proposées dans ce cadre [29]. Kryszkiewicz propose de considérer deux stratégies lors de l'évaluation du support. Une stratégie dite optimiste, pour laquelle l'itemset est supposé présent. Ceci donne la valeur optimiste du support. En outre, une stratégie dite pessimiste, pour laquelle l'itemset est supposé absent, permet de donner la valeur pessimiste du support. Ces solutions s'avèrent inadaptées dans le cas d'une base contenant plusieurs valeurs manquantes [29]. Une autre solution proposée par le même auteur, consiste en la redéfinition de la notion de support pour l'adapter à la présence de valeurs manquantes [27]. Dans ce qui suit, nous présentons brièvement cette solution.

[15] La condition $g = f$ permet de ne pas retenir les règles de la forme $g \Rightarrow \emptyset$.

### 1.5.1 Approche probabiliste de calcul de support dans une base incomplète

Toutes les définitions que nous allons introduire à la présentation de cette approche sont relatives à un contexte incomplet $K_{VM}$.

**Définition 9** [27] L'ensemble de tuples qui contiennent nécessairement la valeur $v_i$ pour un attribut $X_i$, noté $n(X_i, v_i)$, est défini par :

$$n(X_i, v_i) = \{t \in T \,|\, t(X_i) = v_i\}.$$

L'ensemble maximal de tuples qui contiennent nécessairement et qui peuvent contenir la valeur $v_i$ pour l'attribut $X_i$, noté $m(X_i, v_i)$, est défini par :

$$m(X_i, v_i) = \{t \in T \,|\, t(X_i) = v_i \vee t(X_i) = ?\}.$$

L'ensemble de tuples où la valeur de l'attribut $X_i$ est manquante, noté $d(X_i, v_i)$, est défini par :

$$d(X_i, v_i) = m(X_i, v_i) \setminus n(X_i, v_i).$$

Pour déterminer le support d'un itemset $X$, tel que $X = \{(X_1, v_1), (X_2, v_i), \ldots, (X_n, v_n)\}$, l'auteur propose de calculer une quantité appelée $probSup_t$ relativement à chaque tuple $t$. La définition de $probSup_t$ est donnée dans ce qui suit :

**Définition 10** [27]
La probabilité qu'un tuple $t$ contienne l'item $(X_i, v_i)$, notée $probSup_t(X_i, v_i)$, est défini comme suit :

$$probSup_t(X_i, v_i) = \begin{cases} 1 & \text{si } t(X_i) = v_i \\ 0 & \text{si } t(X_i) = v_i \\ \mu(X_i, v_i) & \text{si } t(X_i) = ?. \end{cases}$$

où $\mu(X_i, v_i)$ désigne la probabilité d'apparition de la valeur $v_i$ pour l'attribut $X_i$ parmi l'ensemble des valeurs connues de l'attribut $X_i$. La probabilité $\mu(X_i, v_i)$ est donnée par :

$$\mu(X_i, v_i) = \frac{|n(X_i, v_i)|}{|T \setminus d(X_i, v_i)|} .$$

La probabilité qu'un tuple t contienne l'itemset $X$, est égale à :

$$probSup_t(X) = probSup_t(X_1, v_1) \times probSup_t(X_2, v_2) \times \ldots \times probSup_t(X_n, v_n).$$

Le support probable d'un itemset $X$, noté par $probSup(X)$, est défini comme suit : $probSup(X) = \sum_{t \in T} probSup_t(X)$. Un itemset $X$ est considéré fréquent si $probSup(X) > minsup$.

De la même façon que la redéfinition du support, les concepts mêmes sur lesquels se base la dérivation des bases génériques de règles associatives, i.e., itemsets fermés et générateurs minimaux ont été redéfinis dans le cadre d'un contexte incomplet. Dans ce qui suit, nous présentons une approche proposée par Rioult permettant l'extraction des motifs $\delta$-libres en présence de valeurs manquantes [48].

### 1.5.2 Représentation condensée en présence de valeurs manquantes

Les motifs $\delta$-libres ont été introduits dans [11]. Les motifs $0-$ libres sont aussi appelés les générateurs minimaux ou motifs clés. Ces motifs sont associés au concept de fermeture lorsque ($\delta = 0$) et au concept de presque-fermeture dans le cas où ($\delta > 0$). Par la suite, nous utiliserons le terme presque-fermeture, pour désigner à la fois le concept de fermeture et de presque-fermeture, puisque la notion de presque-fermeture correspond au cas particulier de fermeture lorsque ($\delta = 0$).
Cette approche de traitement des valeurs manquantes se propose de redéfinir le concept de presque-fermeture, pour l'adapter à la présence des valeurs manquantes.

Pour présenter cette approche, nous commençons d'abord par introduire quelques concepts présentés dans [11].

Définition11 (Règle $\delta$-forte) Une règle $\delta$-forte sur un motif Z=XY est une règle d'association de la forme $X \Rightarrow Y$ où $Y = \emptyset$, et $X \cap Y = \emptyset$, , qui admet au plus $\delta$ exceptions.

Il est important de signaler que lorsque ($\delta = 0$), ces règles correspondent aux règles exactes (0 exceptions). Lorsque ($\delta > 0$), la confiance de telles règles est au moins égale à $1 - (\delta/support(X))$ [11].

Définition12 (Motif $\delta$-libre) Un motif $Z$ est $\delta$-libre s'il n'existe sur $Z$ aucune règle $\delta$-forte $X \Rightarrow Y$ (avec $X \subset Z$ et $Y = Z \backslash X$).

D'après la Définition 12, un motif 0-libre, ou générateur minimal est un motif à partir duquel aucune règle exacte ne peut être construite. Nous remarquons que cette définition est équivalente à celle introduite dans [35]. En effet, l'auteur considère qu'un motif de taille $k$ n'est pas un générateur minimal si son support est égal au support de l'un de ses sous-ensembles de taille $k - 1$. En effet, si un motif $AX$ possède un support égal à l'un de ses sous-ensembles $X$, alors la règle $X \Rightarrow A$ est une règle exacte. Cependant, la particularité des motifs $\delta$-libres, est qu'ils introduisent une part d'incertitude liée à la notion d'exception (lorsque $\delta > 0$), d'où la notion de presque-fermeture [12]. De plus, ces motifs vérifient la propriété d'anti-monotonie. Ceci rend l'extraction possible lorsque les algorithmes d'extraction des motifs fréquents deviennent impraticables [11].

Définition13 (Presque-fermeture) Soit $\delta$ un entier positif, la presque-fermeture de $X$ dans un contexte complet $K$, notée par $AC(X, K)$ rassemble les items $A$ tels que :

$$supp(X, K) - supp(XA, K) \leq \delta.$$

où $supp(X, K)$ désigne le support absolu de $X$ dans le contexte $K$ et est défini par $supp(X, K) = |\{o \in O | X \subseteq o\}|$.

Le travail de Rioult [45, 48], consiste à redéfinir la notion de presque-fermeture en prenant en considération les transactions désactivées. La notion de transaction désactivée, a été introduite pour la première fois par Ragel dans [41]. Cependant, dans son travail, Rioult utilise la définition de transaction désactivée introduite par Kryszkiewicz dans [28].

Définition 14 (Transaction désactivée) Une transaction $t$ dans un contexte incomplet $K_{VM}$ est désactivée pour $X$ si tous les items de X sont présents dans $t$, sauf au moins l'un d'entre eux est déclaré manquant. Dans ce qui suit, nous allons noter par $Des(X, K_{VM})$, les transactions désactivées pour $X$ dans le contexte $K_{VM}$.

En présence de valeurs manquantes, Rioult propose donc de redéfinir la notion de presque-fermeture. La redéfinition de la presque-fermeture est donnée par la Définition 15.

Définition 15 La presque-fermeture de $X$ dans un contexte incomplet $K_{VM}$, notée par $AC(X, K_{VM})$ rassemble les items $A$ tels que :

$$supp(X, K_{VM}) - supp(XA, K_{VM}) \leq \delta + |Des(A, K_{VM}(X)|$$

où $|Des(A, K_{VM}(X)|$ désigne le nombre de transactions dans $K_{VM}$, contenant l'itemset $X$ et désactivées pour $A$.

La redéfinition de la presque-fermeture adopte une stratégie optimiste. Ceci est matérialisé par une prise en considération des données désactivées relativement à un item $A$ lors du calcul de $AC(X, K_{VM})$. Ces données désactivées sont considérées comme étant des données contenant l'item $A$.

## 1.6 Conclusion

Dans ce chapitre, nous avons présenté le problème des valeurs manquantes dans les données et les différentes typologies associées. Nous avons également présenté les solutions classiques dédiées à cette problématique. En particulier, nous avons montré que les solutions qui consistent à ignorer les valeurs manquantes sont inefficaces en pratique [41, 53]. De même, les solutions basées sur la valeur la plus commune sont loin d'assurer un traitement optimal [41]. Partant du fait que les règles associatives décrivent des corrélations entre les données, des approches proposent de déterminer les valeurs qui se cachent derrière celles qui manquent par l'exploitation des règles associatives [8, 24, 41, 53]. Ainsi, après avoir présenté le formalisme de base de dérivation des règles associatives et des bases génériques, nous avons passé en revue les différentes adaptations de la notion du support, des itemsets fermés et des générateurs minimaux dans le cadre d'un contexte présentant des valeurs manquantes. Le chapitre suivant se propose de présenter un état de l'art des différents travaux dédiés à la complétion des valeurs manquantes, utilisant la technique des règles associatives.

# Chapitre 2

# Règles associatives pour la complétion des valeurs manquantes : État de l'art

## 2.1 Introduction

Le problème des valeurs manquantes a longtemps intéressé une multitude de communautés scientifiques. Ceci s'est matérialisé par la richesse de contributions faites par ces différentes communautés. Cet intérêt montre bien l'importance de cette problématique. En effet, nous avons montré dans le chapitre 1 qu'en pratique, ce problème classique, est inévitable et imprévisible. De même, nous avons passé en revue des solutions proposées dans le domaine de la statistique et de la classification supervisée. Nous avons montré également que la technique des règles associatives est de plus en plus utilisée pour proposer des solutions à ce problème. Ainsi, après avoir introduit le formalisme de base de dérivation des règles associatives, nous avons présenté les diverses adaptations faites dans le cadre de cette dérivation, en présence de valeurs manquantes. Dans ce chapitre, nous présentons les approches dédiées à la complétion des valeurs manquantes se basant sur la technique des règles associatives. Ces approches, nous les avons regroupées en deux grandes stratégies. La première stratégie consiste à compléter les valeurs manquantes par l'exploitation seule des données complètes. En effet, cette stratégie consiste en premier lieu, à ignorer déf i nitivement toute donnée manquante pour ensuite appliquer un processus de complétion, i.e., trouver une valeur de remplacement pour chaque valeur manquante. Se basant sur le fait qu'il n'est pas raisonnable d'ignorer des données au prétexte qu'elles sont incomplètes [46], des approches faisant partie d'une deuxième stratégie viennent pallier la faiblesse de la première stratégie. Ces dernières se proposent d'extraire des connaissances directement à partir d'une base incomplète, i.e., sans suppression préalable des données manquantes. Ainsi, les connaissances extraites seront qualifiées relativement à la base totale [46].

La section suivante est consacrée à la présentation de ces différentes approches.

## 2.2 Complétion des valeurs manquantes par l'utilisation des règles associatives

Partant du fait que les règles associatives décrivent des corrélations entre les données, des approches se proposent d'exploiter les corrélations ainsi induites pour compléter les valeurs manquantes. Ainsi, toutes ces approches préconisent le fait que la connaissance d'une valeur manquante devient de plus en plus aisée par une prise en considération des autres valeurs. Un exemple illustratif donné par l'un des premiers auteurs à avoir introduit ce principe, qu'il a désigné par "Des relations pour compléter ", est lorsque la capacité à voler d'un animal est manquante, alors il est possible de s'appuyer sur le fait que l'animal possède quatre pattes [41]. Ainsi, la capacité à voler est facilement déduite.

Dans ce qui suit, nous commençons par la présentation des approches faisant partie de la première stratégie.

### 2.2.1 Stratégie 1 : Complétion des valeurs manquantes par l'exploitation des données complètes

Le principe de fonctionnement de cette stratégie se déroule suivant les trois étapes suivantes :

1. Supprimer d'un contexte incomplet $K_{VM}$, tous les tuples présentant au moins une valeur manquante.
2. À partir du sous-ensemble du contexte ainsi obtenu (qu'on désignera par $\overline{K}_{VM}$), extraire les connaissances sous forme de règles associatives.
3. Compléter les valeurs manquantes par l'exploitation des règles ainsi extraites.

Exemple 1 Considérons l'exemple du contexte incomplet $K_{VM}$ donné par la Figure 2.1 (Gauche). Chaque tuple est représenté par quatre attributs $X_1$, $X_2$, $X_3$ et $X_4$. À chaque attribut est associé un domaine de valeurs : $dom(X_1) = \{a, b\}$, $dom(X_2) = \{b, c\}$, $dom(X_3) = \{c, d, g, h, i\}$ et $dom(X_4) = \{c, d, e, f\}$.

Le contexte $\overline{K}_{VM}$, présentant le sous-ensemble de données complet, donné par la Figure 2.1 (Droite), est utilisé pour extraire les règles associatives à n de compléter les valeurs manquantes. Dans ce qui suit, nous présentons une première approche proposée dans le cadre de cette stratégie.

| | $X_1$ | $X_2$ | $X_3$ | $X_4$ |
|---|---|---|---|---|
| 1 | A | c | c | e |
| 2 | A | b | h | c |
| 3 | A | b | h | d |
| 4 | B | b | d | e |
| 5 | B | c | c | c |
| 6 | B | c | d | d |
| 7 | B | c | g | f |
| 8 | ? | b | c | c |
| 9 | A | b | h | ? |
| 10 | ? | c | h | e |
| 11 | B | c | i | ? |
| 12 | A | b | g | ? |

| | $X_1$ | $X_2$ | $X_3$ | $X_4$ |
|---|---|---|---|---|
| 1 | a | c | c | e |
| 2 | a | b | h | c |
| 3 | a | b | h | d |
| 4 | b | b | d | e |
| 5 | b | c | c | c |
| 6 | b | c | d | d |
| 7 | b | c | g | f |

Fig. 2.1 Gauche : Contexte incomplet($K_{VM}$). Droite : Sous-ensemble de données complètes associé ($\overline{K}_{VM}$).

Approche 1 : Complétion des valeurs manquantes par l'exploitation des règles associatives

Cette première approche de complétion proposée dans le cadre de la première stratégie se base sur l'extraction des règles associatives exactes, dont la partie conclusion est constituée d'un seul attribut [8]. Il faut noter que l'attribut sur lequel est basée la partie conclusion est supposé non affecté par des valeurs manquantes et il est fixé d'avance. Les attributs présentant des valeurs manquantes apparaissent dans la partie prémisse d'une règle. Aucune notion de support ni de confiance n'ont été utilisés lors de l'extraction des règles. Cette extraction se déroule en trois étapes d'extraction, suivies par une étape de complétion des valeurs manquantes :

1. Étape d'extraction

   a- Étape de décomposition : à cette étape, un attribut supposé non affecté par des valeurs manquantes est fixé. Cet attribut est appelé attribut cible, noté $AC$, qui figurera exclusivement dans la partie conclusion des règles extraites. Pour chaque valeur $v_i$ de l'attribut $AC$, le contexte est décomposé en deux sous-ensembles $B_0$ et $B_1$. $B_1$ constitue l'ensemble des tuples, tels que l'attribut $AC$ présente la valeur $v_i$ et $B_0$ constitue les tuples, tels que $AC$ présente une valeur différente de $v_i$.

   b- Étape de génération : durant cette étape, toutes les règles ayant dans la partie prémisse $AC = v_i$ sont générées à partir de $B_1$.

   c- Étape d'élagage : il s'agit de vérifier, pour chaque règle extraite, s'il existe au moins un tuple dans la base $B_0$ qui vérifie la partie prémisse de la règle. Si c'est le cas, la règle est éliminée. La présence d'un tuple vérifiant la prémisse d'une règle dans la base $B_0$ indique un contre-exemple de la règle en question, puisque la base $B_0$ présente des valeurs différentes de $v_i$ pour l'attribut $AC$.

2. Étape de complétion

   Quant à la complétion, elle se fait par la recherche de correspondance entre une règle et un tuple présentant une valeur manquante. Une règle peut être utilisée pour compléter une observation si elle présente la même valeur de l'attribut cible (qui constitue la partie conclusion de la règle). De plus, le tuple doit vérifier la partie prémisse de la règle pour les attributs sans valeurs manquantes.

Exemple 2 Considérons l'exemple du contexte incomplet $K_{VM}$ représenté par la Figure 2.1 (Gauche). Nous pourrons distinguer les trois étapes déjà citées.

- Étape de décomposition : L'attribut $X_2$ est supposé fixé, il constitue l'attribut cible $AC$. Si nous considérons la valeur $b$ de $X_2$, le contexte $\overline{K_{VM}}$ associé, donné par la Figure 2.1 (Droite) peut-être décomposé en deux sous-ensembles $B_0$ et $B_1$ représentés par la Figure 2.2.
- Étape de génération : les règles, ayant comme conclusion $(X_2 = b)$, sont ensuite générées à partir de $B_1$. Ces règles sont présentées par la Table 2.1.

| | $X_1$ | $X_2$ | $X_3$ | $X_4$ |
|---|---|---|---|---|
| 2 | a | b | H | c |
| 3 | a | b | H | d |
| 4 | b | b | D | e |

| | $X_1$ | $X_2$ | $X_3$ | $X_4$ |
|---|---|---|---|---|
| 1 | a | c | c | e |
| 5 | b | c | c | c |
| 6 | b | c | d | d |
| 7 | b | c | g | f |

Fig. 2.2 Gauche : Ensemble $B_1$. Droite : Ensemble $B_0$.

| | | | |
|---|---|---|---|
| $R_1$ | $(X_1 = a) \Rightarrow (X_2 = b)$ | $R_{10}$ | $(X_1 = a) \wedge (X_4 = d) \Rightarrow (X_2 = b)$ |
| $R_2$ | $(X_1 = b) \Rightarrow (X_2 = b)$ | $R_{11}$ | $(X_1 = b) \wedge (X_3 = d) \Rightarrow (X_2 = b)$ |
| $R_3$ | $(X_3 = d) \Rightarrow (X_2 = b)$ | $R_{12}$ | $(X_1 = b) \wedge (X_3 = e) \Rightarrow (X_2 = b)$ |
| $R_4$ | $(X_3 = h) \Rightarrow (X_2 = b)$ | $R_{13}$ | $(X_3 = h) \wedge (X_4 = c) \Rightarrow (X_2 = b)$ |
| $R_5$ | $(X_4 = c) \Rightarrow (X_2 = b)$ | $R_{14}$ | $(X_3 = h) \wedge (X_4 = d) \Rightarrow (X2 = b)$ |
| $R_6$ | $(X_4 = d) \Rightarrow (X_2 = b)$ | $R_{15}$ | $(X_3 = d) \wedge (X_4 = e) \Rightarrow (X_2 = b)$ |
| $R_7$ | $(X_4 = e) \Rightarrow (X_2 = b)$ | $R_{16}$ | $(X_1 = a) \wedge (X_3 = h) \wedge (X_4 = c) \Rightarrow (X_2 = b)$ |
| $R_8$ | $(X_1 = a) \wedge (X_3 = h) \Rightarrow (X_2 = b)$ | $R_{17}$ | $(X_1 = a) \wedge (X_3 = h) \wedge (X_4 = d) \Rightarrow (X2 = b)$ |
| $R_9$ | $(X_1 = a) \wedge (X_4 = c) \Rightarrow (X_2 = b)$ | $R_{18}$ | $(X_1 = b) \wedge (X_3 = d) \wedge (X_4 = e) \Rightarrow (X2 = b)$ |

Tab. 2.1 Règles générées à partir de $B_1$ concluant sur $(X_2 = b)$.

- **Étape d'élagage** : à ce niveau, il s'agit d'élaguer toute règle pour laquelle il existe au moins un tuple dans la base $B_0$, vérifiant la partie prémisse de la règle en question. Ainsi, les règles $R_1$, $R_2$, $R_3$, $R_5$, $R_6$, $R_7$ et $R_{11}$ sont élaguées.
- **Étape de complétion** : étant donné les règles qui ont été retenues durant la phase d'élagage. Nous constatons que, pour compléter le tuple $t_9$ : $(X_1 = a, X_2 = b, X_3 = h, X_4 = ?)$, les règles $R_{12}$, $R_{15}$ et $R_{18}$ n'en vérifient pas. En outre, les règles $R_4$ et $R_8$ ne permettent pas de compléter la valeur manquante sur l'attribut $X_4$. Seules les règles $R_9$, $R_{10}$, $R_{13}$, $R_{14}$, $R_{16}$ et $R_{17}$ permettent de le faire. Par exemple, la règle $R_9$ permet de compléter la valeur de l'attribut $X_4$ par la valeur "c".

Cependant, les inconvénients qui peuvent être dégagés de cette approche sont comme suit :

1. Cette approche procède par l'extraction des règles sans aucune mesure de support ni de confiance (ce qui les réduit à des implications). Ainsi, les données sur lesquelles se base l'extraction peuvent être infréquentes. Ceci va engendrer la génération d'un nombre exorbitant de règles, et impliquera un problème de conflit lors de l'étape de complétion. Ce problème se traduit par la présence de plusieurs règles susceptibles de compléter une valeur manquante, chacune avec une valeur différente. Le problème qui va donc se poser c'est évidemment, quelle valeur choisir ? Ce problème de conflit n'a pas été évoqué ni traité par les auteurs. Dans le cadre de notre exemple, les règles qui permettent de compléter la valeur de l'attribut $X_4$ engendrent le problème de conflit. En effet, les règles $R_9$, $R_{13}$ et $R_{16}$ permettent de compléter la valeur de l'attribut $X_4$ par la valeur "c", alors que les règles $R_{10}$, $R_{14}$ et $R_{17}$ se proposent de compléter l'attribut $X_4$ par la valeur "d".

2. L'approche suppose une hypothèse très contraignante, i.e., l'existence d'un attribut cible non affecté par des valeurs manquantes sur lequel va se baser l'extraction des règles.
3. D'autre part, cette solution souffre d'un principe de fonctionnement répétitif. En effet, pour compléter un attribut manquant, il faut passer par les quatre étapes sur-citées et cela pour chacune des valeurs de l'attribut cible. Dans notre exemple, nous avons traité qu'une seule valeur de l'attribut en question, i.e., $(X_2 = b)$.

Dans ce qui suit, nous présentons une deuxième approche située dans le cadre de la première stratégie.

Approche 2 : Extraction des règles associatives pour la prédiction des valeurs manquantes

Cette deuxième approche proposée dans le cadre de complétion des valeurs manquantes, se base sur l'extraction des règles exactes. La particularité de ces règles est que leurs parties conclusions se présentent sous la forme d'intervalles ou d'ensemble de valeurs, selon que le domaine de l'attribut est continu ou discret [24]. La complétion se fait sur la base d'intersection de la partie conclusion. Cette solution présente l'avantage qu'elle utilise la notion de support et de confiance lors de l'extraction des règles, en plus d'une autre mesure qui est le gain de précision [24]. Cette nouvelle mesure a été introduite afin d'améliorer l'étape de complétion, puisqu'elle apporte une réduction de l'intervalle ou de l'ensemble de valeurs prédites pour permettre ainsi une prédiction plus précise. De plus, le gain de précision vérifie la propriété d'anti-monotonie permettant ainsi une extraction nivelée des règles (de type Apriori) [24].

Ainsi, étant donné les notations suivantes :

- $A_{i0}$ de $A$ est un attribut particulier supposé fixé, appelé attribut de prédiction, sur lequel les prédictions sont effectuées. Contrairement à l'approche précédente, cet attribut constituant la partie conclusion des règles, est l'attribut présentant les valeurs manquantes.
- Nous appelons le domaine actif de l'attribut $A_i$, noté $adom(A_i)$, l'ensemble des valeurs de $dom(A_i)$ présentes dans $K_{VM}$ si $dom(A_i)$ est de type discret. Autrement, ça sera l'intervalle $[\mu_i, \nu_i]$, où $\mu_i$ et $\nu_i$ sont, respectivement, la plus petite et la plus grande valeur de $dom(A_i)$ présentes dans $K_{VM}$, si $dom(A_i)$ est de type continu.

Les règles de prédiction sont définies comme suit :

**Définition 16 (Règle de prédiction)** On appelle règle de prédiction, toute règle associative de la forme $T \Rightarrow A_{i0} \in E_T$ où :

- T est la conjonction de conditions élémentaires de la forme $(A_i = v_i)$, où $A_i$ est un attribut de $A$ différent de $A_{i0}$ et $v_i \in dom(A_i)$.

- $E_T$ est défini par $E_T = \{v \in dom(A_i) \mid \exists t \in \overline{K}_{VM}(t \models T \; et \; t.A_{i0} = v)\}$[1] si $dom(A_i)$ est discret. Si $dom(A_i)$ est continu, alors $E_T$ est défini par $E_T = [\mu_i, \nu_i]$, où $\mu_i = min\{v \in dom(A_{i0}) \mid (\exists t \in \overline{K}_{VM})(t \models T \; et \; t.A_{i0} = v)\}$ et $\nu_i = max\{v \in dom(A_{i0}) \mid (\exists t \in \overline{K}_{VM})(t \models T \; et \; t.A_{i0} = v)\}$.

**Exemple 3** Considérons l'exemple du contexte $\overline{K}_{VM}$ donné par la Figure 2.1 (Droite) (page 24). En utilisant la Définition 16 et en considérant l'attribut $X_4$ comme attribut de prédiction, la règle $R$ : $(X_1 = a) \Rightarrow X_4 \in \{c, d, e\}$ est une règle de prédiction. Les notions de support et de confiance d'une règle sont définies de la manière classique, comme cela a été introduit dans [1]. Ainsi, il est facile de constater que :

- Le support de $(X_1 = a)$ dans $\overline{K}_{VM}$ est égal à 3/7.
- Le support de la règle $R$ est égal à 3/7.
- La confiance de la règle $R$ est égale à 1.

Il est important de noter que cette approche se base seulement sur l'extraction des règles exactes. Cela est vérifié grâce à la définition même d'une règle de prédiction et des définitions des mesures de support et de confiance.

Dans ce qui suit, nous introduisons la mesure de gain de précision qui permet d'améliorer le processus de complétion par la réduction de l'ensemble de valeurs ou de l'intervalle prédit [24].

[1] $t \models T$ signifie que le tuple $t$ satisfait les conditions élémentaires de $T$.

**Définition 17 (Gain de précision)** Soient $R$ et $R'$ deux règles de formes respectives $(T \Rightarrow A_{i0} \in E_T)$ et $(T' \Rightarrow A_{i0} \in E_{T'})$, telle que $T \subseteq T'$.

Le gain de précision de $R$ par rapport à $R'$, noté $gain(R, R')$, est défini comme suit :

- Si $T$ n'est pas vide alors $gain(R, R') = (|E_T| - |E_{T'}|)/|E_T|$
- Si $T$ est vide alors $gain(\emptyset, R') = (|adom(A_{i0})| - |E_{T'}|)/|adom(A_{i0})|$

Dans la Définition 17, $|E|$ désigne la cardinalité de $E$ si $E$ est un ensemble discret. Par contre, si $E$ est l'intervalle $[\mu, \nu]$, alors $|E|$ désigne la différence $\nu - \mu$.
Le gain de précision, mesure la réduction relative de l'intervalle ou de l'ensemble des valeurs lorsque la condition d'une règle est raffinée par l'ajout d'une ou plusieurs conditions de la forme $(A_i = v_i)$ [24].

**Exemple 4** Reprenons le cas de la règle $R : (X_1 = a) \Rightarrow X_4 \in \{c, d, e\}$. Si nous raffinons la condition de la règle par l'ajout de la condition $(X_2 = b)$.
La règle devient $R'$: $(X_1 = a) \wedge (X_2 = b) \Rightarrow X_4 \in \{c, d\}$. Nous remarquons une réduction de l'ensemble de valeurs prédites qui est passé de $\{c, d, e\}$ à {c, d}. Ainsi, $gain(R, R') = (3 - 2)/3 = 1/3$. Ceci indique une réduction de 33% de l'ensemble de valeurs prédites.

Dans ce qui suit, nous présentons le critère utilisant la mesure gain de précision qui a été introduite par les auteurs afin de sélectionner une règle. En effet, ce critère va permettre de ne retenir une règle que si elle apporte effectivement une réduction de la taille de l'ensemble de valeurs ou de l'intervalle prédit.

**Définition 18 (Règle retenue)** Soient $G$ un seuil de gain de précision et une règle R : $T \Rightarrow A_{i0} \in E_T$ telle que $T$ est la conjonction de conditions élémentaires de la forme $(A_i = v_i)$ où i=1,2,...,k. $R$ est retenue par rapport à $G$ :
- Si k=1 alors, $gain(\emptyset, R) \geq G$
- Si k> 1 alors, $\forall (A_i = v_i)$ de $T$, $R \backslash (A_i = v_i)$ est retenue par rapport à $G$ et $gain(R \backslash (A_i = v_i), R) \geq G$.

**Exemple 5** Si nous considérions les règles suivantes :
- $R_1 : (X_1 = a) \Rightarrow X_4 \in \{c, d, e\}$
- $R_2 : (X_2 = b) \Rightarrow X_4 \in \{c, d, e\}$

- $R_3$ : $(X_1 = a) \wedge (X_2 = b) \Rightarrow X_4 \in \{c, d\}$

Pour un seuil de gain de précision de 15% et sachant que $admo(X_4) = \{c, d, e, f\}$, l'application de la définition 18 à la règle $R_3$ nécessite de connaître si $R_1$ et $R_2$ sont retenues. Puisque c'est le cas, i.e., $(gain(\emptyset, R_1) = gain(\emptyset, R_2) = 1/4)$, nous calculons ensuite $gain(R_1, R_3)$ et $gain(R_2, R_3)$. Comme leur valeurs respectives sont égales à 1/3, alors la règle $R_3$est retenue.

L'extraction des règles dans cette approche se base sur l'algorithme Apriori [1]. Cependant, il faut noter les différences suivantes :

- Les règles sont générées dans la même phase que celle d'extraction des itemsets fréquents.
- En plus du support, cette approche se base sur la propriété d'anti-monotonie de la mesure de gain de précision, pour retenir une règle.

Exemple 6 Si nous considérons le même exemple du contexte $\overline{K}_{VM}$ présenté par la Figure 2.1 (Droite) (page 24). Une application de l'approche avec une valeur de minsup égale à 15% et une valeur de gain de précision égale à 15% est donnée comme suit : Pour k=1, nous avons les règles candidates notées par $C_1$, données par la Table 2.2. Les valeurs entre parenthèses désignent respectivement le support et le gain de précision des règles. Nous remarquons que la règle $R_2$, ainsi que la règle $R_4$ présentent un gain de précision nul. Ces règles ne seront pas retenues. Par conséquent, les règles fréquentes de l'itération 1 notées par $L1$, sont $L_1 = C_1 \setminus \{R_2, R_4\}$. Les règles candidates de taille 2 sont ensuite générées et leurs intervalles calculés. Nous constatons que les conditions $[(X_1 = a) \wedge (X_3 = d)]$, $[(X_1 = a) \wedge (X_3 = g)]$, $[(X_2 = b) \wedge (X_3 = c)]$ et $[(X_2 = b) \wedge (X_3 = g)]$ ne sont pas fréquentes et que les règles $(X_1 = a) \wedge (X_3 = h) \Rightarrow X_4 \in \{c, d\}$ et $(X_2 = b) \wedge (X_3 = h) \Rightarrow X_4 \in \{c, d\}$ ne sont pas retenues (n'apportant pas de réduction par rapport à la règle $R_8$) .

Pour la règle $R$ : $(X_1 = a) \wedge (X_3 = h) \Rightarrow X_4 \in \{c, d\}$, le premier calcul e ectué est $gain(R_1, R)$. Ce gain est égal à 1/3. Le deuxième calcul efectué est $gain(R_8, R)$. Ce gain étant nul, la règle < n'est pas retenue. Finalement, nous avons $L2 = \{(X_1 = a) \wedge (X_2 = b) \Rightarrow X_4 \in \{c, d\}, (X_1 = a) \wedge (X_3 = c) \Rightarrow X_4 \in \{e\}, (X_2 = b) \wedge (X_3 = d) \Rightarrow X_4 \in \{e\}\}$.

Pour k=3, nous n'avons aucune règle candidate. D'où, $C3 = L3 = \emptyset$.

Ainsi, si nous considérons le contexte incomplet $K_{VM}$ donné par la Figure 2.1 (Gauche) (page 24), et étant donné un tuple, présentant une valeur manquante sur l'attribut $X_4$. Une complétion d'une valeur manquante sur l'attribut $X_4$ peut être faite par l'intersection de tous les ensembles (ou intervalles) qui représentent la partie conclusion des règles extraites vérifiant le tuple en question. Dans le cas où cette intersection est vide, la complétion n'est pas possible. Lorsque cette intersection est non vide, seules les règles présentant une condition maximale vont être utilisées pour la complétion.

| | | | |
|---|---|---|---|
| $R_1$ | $(X_1 = a) \Rightarrow X_4 \in \{c, d, e\}$ (3⁄7, 1⁄4) | $R_5$ | $(X_3 = c) \Rightarrow X_4 \in \{c, e\}$ (2⁄7, 1⁄2) |
| $R_2$ | $(X_1 = b) \Rightarrow X_4 \in \{c, d, e, f\}$ (4⁄7, 0) | $R_6$ | $(X_3 = d) \Rightarrow X_4 \in \{d, e\}$ (2⁄7, 1⁄2) |
| $R_3$ | $(X_2 = b) \Rightarrow X_4 \in \{c, d, e\}$ (3⁄7, 1⁄4) | $R_7$ | $(X_3 = g) \Rightarrow X_4 \in \{f\}$ (1⁄7, 3⁄4) |
| $R_4$ | $(X_2 = c) \Rightarrow X_4 \in \{c, d, e, f\}$ (4⁄7, 0) | $R_8$ | $(X_3 = h) \Rightarrow X_4 \in \{c, d\}$ (2⁄7, 1⁄2) |

Tab. 2.2 Règles candidates $C_1$.

Exemple 7 Dans le cadre de notre exemple, les règles permettant de compléter le tuple $t_9$ sont : $(X_1 = a) \Rightarrow X_4 \in \{c, d, e\}$(3⁄7, 1⁄4),
$(X_2 = b) \Rightarrow X_4 \in \{c, d, e\}$(3⁄7, 1⁄4),
$(X_3 = h) \Rightarrow X_4 \in \{c, d\}$(2⁄7, 1⁄2),
$(X_1 = a) \wedge (X_2 = b) \Rightarrow X_4 \in \{c, d\}$(2⁄7, 1⁄3).
C'est la dernière règle qui va être utilisée pour la complétion, puisqu'elle présente une partie prémisse maximale. L'ensemble de valeurs prédites pour l'attribut $X_4$ est alors $\{c, d\}$.

Cependant, nous constatons que cette approche souffre des limites suivantes :

1. Cette approche se propose de résoudre le problème de conflit en attribuant à chaque valeur manquante, non pas une seule valeur de remplacement, mais un intervalle ou un ensemble discret de valeurs (le plus réduit possible). Cependant, nous constatons que ce problème existe toujours. En effet, le fait d'attribuer un ensemble de valeurs ou un intervalle de valeurs à une valeur manquante ne semble pas être une solution au problème, puisqu'il s'agit de compléter une valeur qui manque par plusieurs autres.
2. De plus, comme nous l'avons déjà mentionné, cette solution peut parfois être incapable de compléter une valeur manquante dans le cas où les conclusions des règles sont disjointes. C'est le cas par exemple du tuple $t_{12}$. En effet, nous avons seulement les règles suivantes vérifiant $t_{12}$ : $(X1 = a) \Rightarrow X4 \in \{c, d, e\}$(3⁄7, 1⁄4) et $(X3 = g) \Rightarrow X4 \in \{f\}(1/7, 3/4)$. Par conséquent, il devient impossible de compléter la valeur de $X_4$ étant donné que l'intersection des conclusions des deux règles est vide.

3. Un autre inconvénient majeur de cette solution réside dans le fait d'élaguer des règles qui à un niveau donné ne permettent pas d'apporter une réduction de l'intervalle ou de l'ensemble prédit. En revanche, combinées à d'autres conditions, les conditions de ces règles peuvent apporter une meilleure précision. Cette constatation peut être observée par l'exemple du tuple $t_{11}$. Nous remarquons qu'aucune règle ne permet la complétion de $X_4$ à cause du fait que la règle $R_2$ : $(X_1 = b) \Rightarrow X_4 \in \{c, d, e, f\}(4/7, 0)$, ainsi que la règle $R_4$ : $(X_2 = c) \Rightarrow X_4 \in \{c, d, e, f\}(4/7, 0)$ n'ont pas été retenues au niveau $k = 1$ puisqu'elles présentent un gain de précision nul. Cependant, si nous considérons la conjonction de ces deux conditions, alors la règle $(X_1 = b) \wedge (X_2 = c) \Rightarrow X_4 \in \{c, d, f\}$ apportera effectivement une réduction de l'ensemble de valeurs prédites et permettra ainsi la complétion du tuple $t_{11}$.

Cependant, le défi face à la problématique des valeurs manquantes réside dans le fait que l'exploitation de toutes les données manquantes est nécessaire. En revanche, les deux approches sus-décrites ont ignoré cet enjeu. En effet, elles ont procédé par une élimination, que nous jugeons brutale, de toute observation manquante. Ceci implique un résultat peu fiable, limitant l'efficacité du processus de complétion [41]. Pour pallier cet inconvénient, d'autres travaux ont été proposés dans le cadre d'une deuxième stratégie. La section suivante est consacrée à la présentation de ces travaux.

### 2.2.2 Stratégie 2 : Complétion des valeurs manquantes par l'exploitation de l'ensemble total des données

D'autres travaux dans le cadre de complétion des valeurs manquantes par la technique des règles associatives ont été menés. Ces travaux se basent sur le fait qu'il n'est pas raisonnable d'ignorer des données au prétexte qu'elles sont incomplètes [46]. L'objectif devient alors l'exploitation de l'ensemble total des observations de la base, même celles présentant des valeurs manquantes. L'enjeu de la problématique des valeurs manquantes est ainsi respecté. Dans ce qui suit, nous présentons deux approches proposées dans le cadre de cette deuxième stratégie.

Approche 1 : La méthode RAR_MVC

La méthode RAR_MVC est une méthode proposée par Ragel [41, 42, 43]. L'extraction des connaissances est effectuée à partir de la base totale, données manquantes comprises, grâce à la méthode RAR[2]. Cette méthode est couplée avec un processus de complétion, MVC[3], permettant la complétion des valeurs manquantes.

### La méthode RAR

L'idée principale de la méthode RAR consiste à extraire les règles à partir de bases valides i.e., ne comportant pas de valeurs manquantes [42]. Contrairement, aux techniques classiques qui procèdent par la suppression de toutes les données incomplètes, cette méthode ne fait qu'ignorer ces dernières que temporairement. Il est à signaler que cette nouvelle approche nécessite une redéfinition de la notion du support et de la confiance [41, 42].
Dans ce qui suit, nous présentons les concepts de base de la méthode RAR [41, 42].

**Définition 19 (Transaction désactivée)** une transaction t est désactivée pour un item- set $X$ dans un contexte incomplet $K_{VM}$, si t contient un item i de $X$ qui est déclaré manquant.

**Exemple 8** Soit la transaction $t$ : $(X_1 = a, X_2 =?, X_3 = a)$. $t$ est désactivée pour les itemsets $(X_1 = a \wedge X_2 = c)$ et $(X_1 = b \wedge X_2 = c)$ puisqu'il existe une valeur manquante sur l'attribut $X_2$. Par contre, $t$ n'est pas désactivée pour l'itemset $(X_1 = a \wedge X_3 = b)$. Les attributs $X_1$ et $X_3$ ne présentent pas de valeurs manquantes dans $t$.

**Notation 2** Soit $Des(X, K_{VM})$ l'ensemble de transactions désactivées pour un itemset $X$. L'ensemble de transactions restantes de l'itemset X, après désactivation, va constituer sa base valide, noté $vdb(X)$[4].

**Définition 20 (Base valide)** La base valide d'un itemset $X$ est définie par:

$$vdb(X) = K_{VM} \backslash Des(X, K_{VM}).$$

La base valide d'une règle R : $(X \Rightarrow Y)$ est définie par :

[2]Robust Association Rules.
[3]Missing Values Completion.
[4]Valid database.

$$vdb(X \Rightarrow Y) = K_{VM} \backslash Des(XY, K_{VM}).$$

Exemple 9 Il faut remarquer qu'une base valide n'est pas liée aux valeurs que peuvent prendre les attributs mais aux attributs eux-mêmes, i.e., la base valide de $(X_1 = a)$ est la même que celle de $(X_1 = b)$. Dans la figure 2.3, nous représentons par $vdb_1$, la base valide de $X_1$. Tandis que, $vdb_4$ désigne celle de $X_4$ relativement au contexte incomplet $K_{VM}$ donné par la Figure 2.1 (Gauche) (page 24).

| | $X_1$ | $X_2$ | $X_3$ | $X_4$ |
|---|---|---|---|---|
| 1 | a | c | c | E |
| 2 | a | b | h | C |
| 3 | a | b | h | D |
| 4 | b | b | d | E |
| 5 | b | c | c | C |
| 6 | b | c | d | D |
| 7 | b | c | g | F |
| 8 | Desactivée | | | |
| 9 | a | b | h | ? |
| 10 | Desactivée | | | |
| 11 | b | c | i | ? |
| 12 | a | b | g | ? |

| | $X_1$ | $X_2$ | $X_3$ | $X_4$ |
|---|---|---|---|---|
| 1 | a | c | c | e |
| 2 | a | b | h | c |
| 3 | a | b | h | d |
| 4 | b | b | d | e |
| 5 | b | c | c | c |
| 6 | b | c | d | d |
| 7 | b | c | g | f |
| 8 | ? | b | c | c |
| 9 | Desactivée | | | |
| 10 | ? | c | h | e |
| 11 | Desactivée | | | |
| 12 | Desactivée | | | |

Fig. 2.3 Gauche : $vdb_1$ - Base valide relative à l'attribut $X1$. Droite : $vdb_4$ - Base valide relative à l'attribut $X_4$ .

L'idée principale de RAR pour rendre l'extraction de règles associatives robuste aux valeurs manquantes, est d'évaluer les itemsets dans leurs bases valides [41]. La base valide représente le plus grand sous-ensemble de données où l'évaluation du support d'un itemset $X$ est non affecté par les valeurs manquantes. Ainsi, la décision de présence de l'itemset $X$ sera faite uniquement en fonction de l'information disponible. Pour cela, des modifications des notions de support et de con ance ont été introduites pour les adapter aux bases valides.

Définition 21 Soit $K_{VM}$(X) le sous-ensemble de transactions de $K_{VM}$ contenant l'itemset $X$. Le support d'un itemset $X$ dans sa base valide est défini comme suit :

$$Support(X) = \frac{|K_{VM}(X)|}{|vdb(X)|}$$

La confiance d'une règle R : $(X \Rightarrow Y)$ est définie par :

$$Confiance(R) = \frac{|K_{VM}(XY)|}{|K_{VM}(X)| - |Des(Y, K_{VM}) \cap K_{VM}(X)|}$$

Exemple 10 Les tables 2.3 et 2.4 illustrent le résultat (seulement quelques exemples d'itemsets et de règles) avec les nouvelles définitions du support et de la confiance à partir du contexte $K_{VM}$ donné par la Figure 2.1 (Gauche) (page 24) pour une valeur de minsup égale à 10% et une valeur de minconf égale à 20%.

| Itemsets | transactions désactivées | support |
|---|---|---|
| $(X_1 = a)$ | $\{8, 10\}$ | 5/(12-2) |
| $(X_1 = b)$ | $\{8, 10\}$ | 5/(12-2) |
| $(X_2 = b)$ | $\emptyset$ | 7/12 |
| $(X_4 = c)$ | $\{9, 11, 12\}$ | 3/(12-3) |
| $(X_4 = d)$ | $\{9, 11, 12\}$ | 2/(12-3) |
| $(X_4 = e)$ | $\{9, 11, 12\}$ | 3/(12-3) |
| $(X_4 = f)$ | $\{9, 11, 12\}$ | 1/(12-3) |
| $(X_1 = a \wedge X_4 = c)$ | $\{8, 9, 10, 11, 12\}$ | 1/(12-5) |

Tab. 2.3 Exemples de calcul du support de quelques itemsets selon la redéfinition du support.

Cependant, ces nouvelles définitions peuvent introduire des erreurs dans le cas où la base valide d'un itemset n'est pas représentative de la base initiale. Pour illustrer ce risque, considérons l'exemple suivant :

Exemple 11 Considérons l'exemple du contexte $K_{VM}$ donné par la Table 2.5. Le support de l'itemset $(X_1 = a \wedge X_2 = c)$ est égal à 2/(8-6). Ceci indique que 100% des transactions contiennent cet itemset. Cependant, en réalité les transactions $t_5$, $t_6$, $t_7$ et $t_8$ ne peuvent pas contenir cet itemset. Ceci est dû au fait que la base valide de l'itemset doit être un échantillon représentatif de la base totale.

| Règle ($X \Rightarrow Y$) | transactions désactivées pour $X$ | transactions désactivées pour $Y$ | support | con ance |
|---|---|---|---|---|
| $(X_1 = a) \Rightarrow (X_4 = c)$ | $\{8, 10\}$ | $\{9, 11, 12\}$ | 1/(12-5) | 1/(5-2) |
| $(X_1 = b) \Rightarrow (X_4 = c)$ | $\{8, 10\}$ | $\{9, 11, 12\}$ | 1/(12-5) | 1/(5-1) |
| $(X_1 = a) \Rightarrow (X_4 = d)$ | $\{8, 10\}$ | $\{9, 11, 12\}$ | 1/(12-5) | 1/(5-2) |
| $(X_1 = b) \Rightarrow (X_4 = d)$ | $\{8, 10\}$ | $\{9, 11, 12\}$ | 1/(12-5) | 1/(5-1) |

Tab. 2.4 Exemples de calcul de la confiance de quelques règles selon la redéfinition de la confiance.

| | $X_1$ | $X_2$ |
|---|---|---|
| 1 | $\emptyset$ | c |
| 2 | a | $\emptyset$ |
| 3 | a | c |
| 4 | a | c |
| 5 | $\emptyset$ | d |
| 6 | b | $\emptyset$ |
| 7 | b | $\emptyset$ |
| 8 | b | $\emptyset$ |

Tab. 2.5 Contexte incomplet ($K_{VM}$).

Ce dernier point va permettre d'éviter de travailler sur des échantillons trop petits. Pour cette raison, l'auteur a introduit une nouvelle mesure qui est la représentativité.

**Définition 22 (Représentativité)** Un itemset $X$ doit avoir une représentativité supérieure à une valeur seuil fixée par l'utilisateur. La mesure de représentativité est définie comme suit :

$$Representativite(X) = \frac{|vdb(X)|}{|K_{VM}|}$$

Ainsi, cette nouvelle mesure introduit une contrainte d'utilisation. En e et, une évaluation correcte du support d'un itemset ne pourra se faire que si la base valide est un échantillon représentatif de la base totale. Pour cette même raison, la définition de transaction désactivée semble être illégitime. En effet, si nous considérons l'exemple de la transaction $t_2$ : $(X_1 = a, X_2 =?)$. La transaction $t_2$ sera considérée comme désactivée pour les itemsets $(X_1 = a \wedge X_2 = c)$ et $(X_1 = b \wedge X_2 = c)$. Cependant, le fait de considérer $t_2$ désactivée pour l'itemset $(X_1 = b \wedge X_2 = c)$ semble être inapproprié car quel que soit la valeur cachée derrière la valeur manquante, $t_2$ ne peut pas contenir cet itemset. Puisque, la valeur de $X_1$ est égale à la valeur "$a$" dans $t_2$, alors que l'itemset présente la valeur "$b$" pour le même attribut.

Par conséquent, il nous semble qu'il existe une contradiction dans la définition même d'une transaction désactivée. D'une part, l'auteur a défini une base valide comme étant le sous-ensemble de données où la décision est possible. Cela implique que toutes les transactions désactivées sont considérées comme étant le sous-ensemble de données où la décision est impossible. D'autre part, l'auteur fait intervenir des données en tant que transactions désactivées, où la décision est possible. Cette contradiction a été introduite pour ne pas nuire à la représentativité d'une base. Pour cette même raison, Kryszkiewicz a montré que cette approche n'est pas valide puisqu'elle ne vérifie pas certaines propriétés définies dans [27].

RAR est une méthode d'extraction de connaissances, sous forme de règles associatives, qui prend en considération la présence de valeurs manquantes. Cette méthode se base sur l'évaluation du support des itemsets dans des bases valides où la décision est à priori possible. Nous avons montré que cette méthode introduit une contradiction dans la notion de prise de décision. Cela a été justifié par la contrainte représentativité. Cette même contrainte pose un problème. En effet, il n'est pas facile de fixer le seuil de représentativité. Il faut une certaine expertise de la part des utilisateurs pour pouvoir juger du taux de représentativité nécessaire.
Dans ce qui suit, nous allons détailler la méthode MVC, associée à RAR, pour la complétion des valeurs manquantes.

### La méthode MVC

Une fois les règles associatives ont été extraites à partir d'un contexte incomplet en utilisant la méthode RAR, l'approche MVC procède par l'application d'un processus de complétion pour compléter les valeurs manquantes.
Cependant, il faut noter que la méthode RAR procède par l'extraction des règles associatives à partir des itemsets fréquents. Cette technique d'extraction engendre un nombre très élevé de règles [22]. Pour cela, et a n de réduire les règles extraites, une technique de filtrage a été définie selon deux niveaux [41] :

- Intervenir au niveau de la production de règles, a n de ne produire que les règles pertinentes.
- Proposer des critères d'intérêt par le biais de métriques statistiques telles que, J- mesure [51] et score-VM [41], afin de résoudre le problème de conflit.

1. Réduction de la production de règles

   A ce niveau, l'auteur propose de ne produire que les règles présentant :

   - Une confiance élevée[5]:

   Par exemple, si la complétion d'une donnée va se baser sur une règle ayant une confiance égale à 40%. Cette complétion va introduire une erreur de 60% [41]. Ainsi, l'auteur justifie l'utilisation des règles de confiance élevée pour une complétion able des valeurs manquantes.

   - Une corrélation positive :

   Le critère $RI$[6] a été introduit par Shapiro, a n de mesurer la corrélation induite entre $X$ et $Y$ d'une règle $R : (X \Rightarrow Y)$.

   Définition 23 (Intérêt d'une règle) [37] Le critère $RI$ d'une règle $R : (X \Rightarrow Y)$ est défini par :

$$RI = |XY| - \frac{|X||Y|}{N}$$

   où $N$ désigne le nombre de tuples qui vérifient $X$ et $Y$.

   Lorsque ($RI = 0$), X et Y sont indépendants, et la règle $R$ n'est pas intéressante. Lorsque ($RI > 0$) (resp. ($RI < 0$)), X est positivement, (resp. négativement) corrélé à Y. L'auteur propose dans ce cas d'utiliser les règles dont le critère $RI$ est positif.

2. Résolution de conflit

   Vu le nombre très élevé de règles qui peuvent être extraites et a n de résoudre le problème de conflit, l'approche MVC procède par l'introduction de mesures permetant d'évaluer la pertinence d'une règle par rapport à une autre lors de l'étape de complétion. Ces mesures sont la J-mesure [51] et le Score-VM [41].

---

[5] Une confiance est considérée comme élevée si elle $\in$ à l'intervalle [95%,100%].

[6] L'acronyme RI désigne Rule Interest.

**Définition 24 (J-Mesure)** [51] La J-Mesure relativement à une règle $R : (X \Rightarrow Y)$ est définie comme suit :

$$J(x;y) = p(y)[p(x/y)log(\frac{p(x/y)}{p(x)}) + (1 - p(x/y))log(\frac{1-(p(x/y))}{(1-p(x))})]$$

Le terme entre crochet est connu sous le nom d'entropie croisée. Une valeur J-Mesure élevée indique que $X$ est un bon prédicteur de $Y$.

**Définition 25 (Score-VM)** [41] Le score-VM d'une règle $R$ dans un contexte incomplet $K_{VM}$ représente, en pourcentage, le nombre de valeurs manquantes que $R$ peut compléter dans $K_{VM}$.

L'intérêt de ce score est de se focaliser sur les règles complétant le maximum de valeurs manquantes. Ainsi, la complétion se fera avec un nombre minimum de règles et donc un minimum d'effort de validation de la part de l'utilisateur, puisque l'auteur autorise l'implication de ce dernier lors de l'étape de complétion.

Cependant, bien qu'elle fait partie de la deuxième stratégie, à savoir son adaptation à l'aspect incomplet lors de l'extraction des connaissances et sa constitution d'un processus de complétion apte à déterminer les valeurs manquantes, RAR_MVC souffre des limites suivantes :

1. C'est une approche qui s'avère invalide dans sa première étape, à savoir l'étape de calcul du support [29].
2. Elle exige qu'une base valide soit représentative, sinon l'approche devient inappropriée [41].
3. Génère un nombre prohibitif de règles. Ceci aura comme conséquence d'engendrer un problème de conflit difficilement maîtrisable et affectant la fiabilité du processus MVC lors de la complétion.

Dans ce qui suit, nous présentons une deuxième approche faisant partie de de la deuxième stratégie.

**Approche 2 : La méthode $AR_{MVC}$**

$AR_{MVC}$ [7] est une deuxième approche proposée dans le cadre de la deuxième stratégie

[7] L'acronyme $AR_{MVC}$ désigne Association Rules based approach for Missing Values Completion.

[53]. Motivés par le fait qu'ignorer toute observation manquante conduit à un résultat aberrant, les auteurs de cette approche ne procèdent pas par la suppression de toutes les données incomplètes. L'extraction des règles associatives est faite directement à partir d'une base totale, données manquantes comprises. De même, cette approche présente un processus de complétion des valeurs manquantes. Ce processus est basé essentiellement sur les métriques statistiques suivantes : support, confiance et lift [13].

Dans ce qui suit, nous présentons le principe de fonctionnement de $AR_{MVC}$. Ce principe est constitué de deux étapes, présentées comme suit [53] :

Étape 1. Extraction des règles associatives à partir d'un contexte incomplet : la procédure d'extraction des règles associatives est faite à partir de la base totale sans ignorer les observations incomplètes. Cependant, chaque valeur manquante ne sera pas considérée lors de l'extraction.

Exemple 12 Considérons l'exemple du contexte incomplet $K_{VM}$ présenté par Figure 2.1 (Gauche) (page 24). Ainsi, la table donnée par Table 2.6 représente les données à partir desquelles l'extraction sera faite. Chaque couple (attribut=valeur) est considéré comme un item où les valeurs manquantes ne sont pas considérées.

Étape 2. La procédure de complétion : une règle associative est présentée sous la forme $R : H(R) \Rightarrow d(R, X)$ où $H(R)$ et $d(R, X)$ dénotent respectivement la prémisse et la conclusion de $R$ et où $X$ est un attribut.

Le pseudo-code de la procédure de complétion de $AR_{MVC}$ est donné par l'Algorithme 1. Le fonctionnement de cet algorithme est présenté comme suit :

1. Pour chaque transaction présentant une valeur manquante, déterminer les règles vérifiant la transaction et permettant la complétion de cette valeur manquante.
2. Toutes les règles vérifiant la transaction et permettant la complétion de la valeur manquante sont évaluées selon la métrique Score.
3. La règle présentant la métrique Score la plus élevée est utilisée pour la complétion.

La mesure Score est définie comme suit [53] : $Score(R) = \frac{appl(R)}{w} \times lift(R)^{\frac{w}{n}}$.

où :

$w$ est la taille de la prémisse.

$appl(R)$ est défini par : $appl(R) = I(R) \times \sum_{i=1..|w|} appl(X_i = v_i)$ où

| | |
|---|---|
| 1 | $(X_1 = a),(X_2 = c),(X_3 = c),(X_4 = e)$ |
| 2 | $(X_1 = a),(X_2 = b),(X_3 = h),(X_4 = c)$ |
| 3 | $(X_1 = a),(X_2 = b),(X_3 = h),(X_4 = d)$ |
| 4 | $(X_1 = b),(X_2 = b),(X_3 = d),(X_4 = e)$ |
| 5 | $(X_1 = b),(X_2 = c),(X_3 = c),(X_4 = c)$ |
| 6 | $(X_1 = b),(X_2 = c),(X_3 = d),(X_4 = d)$ |
| 7 | $(X_1 = b),(X_2 = c),(X_3 = g),(X_4 = f)$ |
| 8 | $(X_2 = b),(X_3 = c),(X_4 = c)$ |
| 9 | $(X_1 = a),(X_2 = b),(X_3 = h)$ |
| 10 | $(X_2 = c),(X_3 = h),(X_4 = e)$ |
| 11 | $(X_1 = b),(X_2 = c),(X_3 = i)$ |
| 12 | $(X_1 = a),(X_2 = b),(X_3 = g)$ |

Tab. 2.6 Contexte d'extraction relatif au contexte $K_{VM}$ donné par la Figure 2.1 (Gauche, page 24).

$$I(R) = \begin{cases} 1 & \text{si } H(R) \text{ vérifie } t. \\ 0 & \text{sinon.} \end{cases}$$

et

$$appl(X_i = v_i) = \begin{cases} 0,5 & \text{si } X_i \text{ présente une valeur manquante dans } t. \\ 1 & \text{sinon.} \end{cases}$$

Cependant, bien que $AR_{MVC}$ présente l'avantage de faire partie de la deuxième stratégie, il est néanmoins important de signaler que cette approche souffre des limites suivantes :

1. Aussi bien que RAR_MVC, $AR_{MVC}$ procède par l'extraction des règles associatives à partir des itemsets fréquents. Par conséquent, il y aura génération d'un nombre exorbitant de règles lors de l'étape de complétion. Il a été montré que ce nombre de règles affecte considérablement l'efficacité du processus de complétion [53]. En effet, ceci implique un problème de conflit difficilement maîtrisable.
2. Nous remarquons que même le problème de conflit n'a pas été mentionné par les auteurs. Même si implicitement, la métrique Score est dédiée à la résolution de ce problème. Cependant, nous constatons aussi que $AR_{MVC}$ procède par le calcul de Score sans se soucier de l'existence même de conflits lors de la complétion.
3. De plus, nous constatons une anomalie dans la mesure même. En effet, la mesure

Algorithme : $AR_{MVC}$
Données : - $K_{MV}$ : Contexte incomplet
- $RA$ : L'ensemble des règles associatives
Résultats : $K_{VM}$ complété
début
  pour chaque transaction manquante $t$ de $K_{MV}$ faire
    pour chaque attribut $X_i$ de $t$ présentant une valeur manquante faire
      $max = 0$;
      pour chaque règle $R$ de $RA$ telle que $X_i$ apparaît dans la conclusion faire
        si $R$ vérifie $t$ alors
          $score = Calculer_Score(R)$;
          si score>max alors
            remplacer la valeur de l'attribut $X_i$ par la conclusion de $R$;
            $max = score$;
  retourner ($K_{VM}$ complété);
n

Algorithme 1 : L'algorithme $AR_{MVC}$

Score attribue à toute valeur manquante un poids égal à 0,5. Nous pensons que ceci entraînera l'utilisation de règles peu ables lors de la complétion. En effet, cette attribution permettra l'utilisation de règles dont toute la partie prémisse est manquante dans la transaction. De telles règles seront considérées vérifiant la transaction.

4. $AR_{MVC}$ est supposée être une solution venant pallier les limites des approches classiques par le fait qu'elle ne rejette pas les données incomplètes. Cependant, il est indéniable d'affirmer que c'est une approche qui prend en considération les valeurs manquantes. En effet, il s'agit bien d'extraire des connaissances sans suppression préalable des données incomplètes. Cependant, cette extraction est faite avec rejet de toute valeur manquante.

## 2.3 Discussion

Au cours de notre présentation des différentes approches proposées dans le cadre de complétion des valeurs manquante, utilisant la technique des règles associatives, nous avons regroupé ces différentes approches dans deux grandes stratégies. La Table 2.7 donne les différentes approches présentées et la Table 2.8 résume les principales caractéristiques de ces différentes approches.

| | |
|---|---|
| Approche 1 | Complétion des valeurs manquantes par par l'exploitation des règles associatives [8]. |
| Approche 2 | Extraction des règles associatives pour la prédiction des valeurs manquantes [24]. |
| Approche 3 | RAR_MVC [41]. |
| Approche 4 | $AR_{MVC}$ [53]. |

Tab. 2.7 Les différentes approches

A partir de la Table 2.8, il est facile de voir que la première stratégie se base en premier lieu à ignorer définitivement toute observation incomplète. Ensuite à appliquer un processus de complétion, i.e., définir une valeur de remplacement pour chaque valeur manquante. Ceci est fait par une exploitation des règles extraites à partir d'un sous-ensemble de données, supposé complet. Cependant, en réalité, ce sous-ensemble de données sur lequel s'est basé l'extraction est loin d'être complet. Il représente en effet, seulement une partie des données. L'extraction est alors faite à partir d'un sous-ensemble de la base dit biaisé, i.e., non représentatif [48]. Ceci implique, la détérioration de la qualité des règles extraites [41]. Par conséquent, la fiabilité du processus de complétion des valeurs manquantes est considérablement affectée.
Les approches faisant partie de la deuxième stratégie viennent pallier la faiblesse des méthodes dites classiques. Elles se basent sur le fait qu'il n'est pas raisonnable d'ignorer définitivement des objets au prétexte qu'ils ne sont pas complets [46]. Ces approches proposent donc, d'extraire des connaissances directement à partir de l'ensemble total de données, même celles incomplètes. Ceci constitue l'avantage de ces méthodes. Ainsi, d'une part l'ensemble total de données sera exploité, d'autre part, les connaissances ainsi obtenues seront qualifiées de ables pour la complétion des valeurs manquantes. Cependant, nous avons vu dans la partie consacrée à la présentation des approches faisant partie de cette stratégie, que ces approches souffrent de certaines limites. En effet, nous avons montré que la méthode RAR_MVC (Approche 3) exige qu'une base valide (à partir de laquelle se fera l'évaluation du support des itemsets) soit représentative [41]. De même, nous avons constaté qu'elle introduit une contradiction dans la définition de transaction désactivée. De plus, il a été démontré que cette méthode est invalide lors de l'étape de calcul du support [29]. De même, nous avons constaté que 'approche $AR_{MVC}$ (Approche 4) se propose d'extraire des connaissances sans suppression préalable des données incom-plètes pour ne pas nuire à la représentativité de la base. Cependant, nous remarquons que cette extraction est faite sans considération des valeurs manquantes.

| | | | | |
|---|---|---|---|---|
| | Stratégie 1 | | Stratégie 2 | |
| | Approche 1 | Approche 2 | Approche 3 | Approche 4 |
| Exploitation de l'ensemble total de données | Non | Non | Oui | Oui |
| Résolution de conflit | - | réduction de la partie conclusion | Score-VM [41] J-Mesure [51] | Score[53] |
| Génération des règles associatives à partir de | - | itemsets fréquents | itemsets fréquents | itemsets fréquents |

Tab. 2.8 Caractéristiques des différentes approches

Outre le fait de distinguer les différentes approches présentées selon la stratégie adoptée. Il est important de signaler, néanmoins, les similarités et les différences au sein d'une même stratégie. Les critères présentant ces différents points sont présentés comme suit :

- Résolution de conflit : le problème de conflit se traduit par l'existence de plusieurs règles, où chaque règle propose une valeur de remplacement différente.

- Technique de génération des règles associatives : la technique de génération des règles associatives fût introduite pour la première fois par Agrawal et al. [1], à partir des itemsets fréquents.

Cependant, même si elles ne procèdent pas par une suppression préalable des données incomplètes, il est important de signaler que le problème majeur des approches faisant partie de la "stratégie 2" reste le problème de conflit lors de l'étape de complétion. Ce problème est dû essentiellement à la technique de génération des règles, qui se fait à partir des itemsets fréquents. Cependant, même avec l'utilisation de mesures de résolution de conflits telles que J-Mesure [51], Score-VM [41] et Score [53], ce problème demeure important.

## 2.4 Conclusion

Dans ce chapitre, nous avons présenté un état de l'art des différents travaux proposés dans le cadre de complétion des valeurs manquantes, utilisant la technique des règles associatives. Le constat majeur, c'est que les approches faisant partie de la première stratégie se sont focalisées sur la complétion des valeurs manquantes, sans tenir compte de l'aspect incomplet des données. Il faut signaler de même que même si les approches faisant partie de la deuxième stratégie prennent en considération l'aspect incomplet des données, ces approches souffrent d'un problème de conflit. Le conflit reste le problème majeur de ces différentes approches lors de l'étape de complétion.

Dans le chapitre suivant, nous proposons une nouvelle approche de complétion des valeurs manquantes, basée sur l'exploitation d'une base générique de règles associatives.

# Chapitre 3

# Une nouvelle approche de complétion des valeurs manquantes

## 3.1 Introduction

La présence de valeurs manquantes dans les bases de données est un problème qui s'est toujours posé lors de l'exploitation des données [46]. En effet, dans un domaine tel que la fouille de données, cet aspect incomplet pose un réel dé lors de l'extraction des connaissances. Dans le chapitre 2, nous avons présenté les travaux proposés dans la littérature, où l'essence même de motivation à la complétion des données manquantes, est l'exploitation des relations induites entre ces données. Une étude critique de ces différents travaux a per- mis de dégager le constat suivant : les approches que nous avons regroupées sous le label de "stratégie 1" se sont essentiellement intéressées à compléter les valeurs manquantes sans se soucier de l'aspect incomplet des données. En revanche, le défi majeur face aux valeurs manquantes, ne serait-il pas l'exploitation même des données incomplètes ? Conscients de cet enjeu, des travaux faisant partie de la "stratégie 2", se sont focalisées sur l'extraction des connaissances à partir de la base totale, incluant les données manquantes. Cependant, ces approches pèchent par leur inefficacité dans la résolution du problème de conflit. En effet, ce problème présente un impact capital lors de l'étape de complétion, i.e., pouvant conduire à la détérioration de l'efficacité du processus.

Dans ce chapitre, nous proposons une nouvelle approche de complétion des valeurs manquantes, intitulée $GBAR_{MVC}$ [6]. Notre approche se propose :

- En premier lieu, d'extraire des connaissances à partir de l'ensemble total des données, sans ignorer celles incomplètes. Ainsi, les règles extraites seront qualifiées relativement à la base totale.
- En second lieu, nous proposons un processus de complétion des valeurs manquantes basé sur une base générique de règles associatives, permettant la réduction du nombre de règles. Cette réduction sert à favoriser l'efficacité du processus de complétion. En outre, nous introduisons également une nouvelle mesure intitulée Robustesse permet- tant la sélection de la règle la plus able lors de la complétion, en cas de conflit.

Ce deuxième point constitue l'intérêt de notre travail. En effet, les travaux antérieurs de complétion des valeurs manquantes se sont focalisées sur l'extraction de l'ensemble total des règles associatives. Ces règles sont coûteuses à extraire et leur nombre exorbitant handicape une exploitation judicieuse [45]. Nous pensons qu'il s'agit d'une contribution à la présentation d'un processus de complétion able. Ainsi, tout en se focalisant sur un ensemble de règles de taille beaucoup plus réduit, nous réduisons considérablement les conflits. Ces conflits sont inévitables à toute méthode de complétion. Cependant, il est important de signaler que le processus de complétion de $GBAR_{MVC}$ se démarque de plus par une prise en considération préalable des valeurs manquantes. Ceci est fait grâce à la redéfinition de la notion de presque-fermeture [48].

Dans ce qui suit, nous montrons l'intérêt de l'utilisation d'une base générique de règles associatives pour la complétion des valeurs manquantes.

## 3.2 Intérêt d'une base générique de règles associatives pour la complétion des valeurs manquantes

Dans cette section, nous allons montrer l'intérêt de l'utilisation d'une base générique de règles associatives pour la complétion des valeurs manquantes. Pour cela, nous allons considérer l'exemple suivant. Soit la transaction manquante $t$ donnée par la Table 3.1. Sachant que la valeur manquante de $t$ est soit l'item "A", soit l'item "B". Nous allons essayer de comparer l'utilisation de l'ensemble total de règles associatives, noté $RA$ et l'utilisation d'une base générique de règles associatives qu'on notera par $RG$. Cependant, il est important de signaler que nous ne présentons que les règles concluant sur les items "A" et "B". L'ensemble de règles associatives $RA$ sont données par la Figure 3.1 (Gauche). Tandis que la base générique de règles associatives $RG$ est présentée par la Figure 3.1 (Droite).

| | | | | | |
|---|---|---|---|---|---|
| ? | C | D | E | F | G |

Tab. 3.1 Transaction manquante

| | | | |
|---|---|---|---|
| $R_1$ | D ⇒ A | $R_5$ | E G ⇒ B |
| $R_2$ | D F ⇒ A | $R_6$ | C G ⇒ B |
| $R_3$ | D G ⇒ A | $R_7$ | C E G ⇒ B |
| $R_4$ | D F G ⇒ A | | |

| | |
|---|---|
| $R_1$ | D ⇒ A |
| $R_5$ | E G ⇒ B |
| $R_6$ | C G ⇒ B |

Fig. 3.1 Gauche : Ensemble de règles associatives ($RA$). Droite : Base générique de règles associatives ($RG$).

D'après cet exemple, nous constatons que toutes les règles de $RA$ peuvent être utilisées pour compléter la valeur manquante de $t$. Cependant, nous remarquons que les règles $R_2$, $R_3$ et $R_4$ ne constituent pas de règles intéressantes par rapport à la règle $R_1$ lors de la complétion. En effet, la règle $R_1$ peut être interprétée de la manière suivante : "Il suffit que la transaction présente l'item "D" pour que la règle $R_1$ peut être utilisée pour compléter la valeur manquante". Cependant, les règles $R_2$, $R_3$ et $R_4$ présentent plus de contraintes (matérialisées par les items de la prémisse), à satisfaire afin de pouvoir être utilisée lors de la complétion. De même, si nous considérons que la règle $R_1$ ne peut pas être utilisée pour compléter la transaction $t$, il devient alors inutile de connaître si les règles $R_2$, $R_3$ et $R_4$ peuvent en être utilisées. Puisque toutes ces règles contiennent la même contrainte imposée par $R_1$. Ce constat est de la même façon observé pour la règle $R_7$ qui n'est pas intéressante par rapport aux règles $R_5$ et $R_6$. Cependant, si nous considérons l'ensemble de règles de la base générique $RG$, nous constatons que toutes les règles que nous avons qualifiées de non intéressantes ne figurent pas dans $RG$. Ceci est la caractéristique essentielle des bases génériques. En effet, une base générique est constituée dans sa partie prémisse de générateurs minimaux. Ces générateurs minimaux offrent l'avantage qu'ils présentent le minimum de contraintes à satisfaire lors de la complétion. Ainsi, l'utilisation d'une base générique de règles montre les intérêts suivants quant à la complétion des valeurs manquantes :

- Renvoie un nombre minimal de règles, démunies de redondance.
- Ces règles sont les plus intéressantes à la complétion, i.e., présentent le minimum de contraintes à satisfaire lors de la complétion des valeurs manquantes. Ceci est garanti par la minimalité de la taille des prémisses des règles, constituées de générateurs minimaux.
- Réduit considérablement les conflits entre les règles. Ainsi, une réduction des conflits assurera plus d'efficacité lors de l'étape de complétion.

## 3.3 Extraction de la base générique de règles associatives en présence de valeurs manquantes

Les limites des travaux proposés dans la littérature et dédiés à la complétion des valeurs manquantes nous ont motivées à la proposition d'une nouvelle approche de complétion. Notre objectif porte essentiellement sur deux axes :

- Nous ne souhaitons pas procéder à la complétion sans aucune prise en considération préalable de toute valeur manquante.
- Nous nous démarquons lors de l'étape de complétion par l'utilisation d'une base générique de règles associatives dans le but de réduire le problème de conflit.

Le problème de dérivation de la base générique de règles associatives repose sur les concepts clés d'itemsets fermés [36] et de leurs générateurs minimaux associés [4]. Ainsi pour faire l'extraction de cette base générique, nous proposons d'utiliser la redéfinition de la notion de presque-fermeture [48] que nous avons présentée au chapitre 1 (page 20). Cette redéfinition est nécessaire pour pouvoir aménager l'extraction de la base générique de règles associatives en présence de valeurs manquantes. Nous avons eu recours à cette redéfinition car elle consiste en une méthode de traitement des valeurs manquantes. En effet, dans ce travail l'auteur décrit les dommages causés par les valeurs manquantes sur l'ensemble des itemsets fermés et leurs générateurs minimaux associés, ensuite il montre la consistance des motifs obtenus avec la nouvelle définition tant d'un point théorique qu'expérimental. Le résultat est que l'ensemble des motifs extraits à l'aide de la nouvelle définition sont plus fiables en présence de valeurs manquantes.

### 3.3.1 Pseudo-fermeture dans un contexte incomplet

L'objectif est d'extraire les itemsets pseudo-fermés[1] et leurs générateurs minimaux associés à partir d'un contexte incomplet. Ainsi, nous adoptons la même stratégie optimiste de calcul de la presque-fermeture, adoptée dans [48], lors de l'extraction des motifs $\delta$- libres. Lors du calcul de la pseudo-fermeture, si un item est manquant, alors il est supposé présent. L'adoption de telle stratégie n'est pas surprenant dans le cadre d'extraction des connaissances à partir d'un contexte présentant des valeurs manquantes. En effet, dans [28], Kryszkiewicz propose de considérer deux stratégies lors de l'évaluation du support d'un itemset. Une stratégie dite optimiste, pour laquelle l'itemset est supposé présent et une stratégie dite pessimiste, pour laquelle l'itemset est supposé absent.

Dans ce qui suit, nous introduisons des définitions respectivement de transaction Certaine, transaction Probable et celle relative aux transactions Probablement.

Définition 26 (Transaction $Certaine(X)$) Une transaction $t$ est dite *Certaine* dans un contexte incomplet $K_{VM}$, relativement à un itemset $X$, si $t$ contient l'itemset $X$. L'ensemble de transactions Certaine relatives à un itemset $X$, noté $Certaine(X)$, est défini comme suit :

$$Certaine(X) = \{t \in T \mid \forall i \in X, i \text{ est présent dans } t\}.$$

Définition 27 (Transaction $Probable(i)$) Une transaction $t$ est dite $Probable(i)$ dans un contexte incomplet $K_{VM}$, relativement à un item $i$, si $i$ est manquant dans $t$.

Il est important de signaler que la définition de transaction Probable coïncide avec celle de transaction désactivée proposées dans [28, 41] relativement à un item. Cependant, nous introduisons la définition de transaction Probablement relativement à un itemset $(Xi)$ qui ajoute à la définition de transaction désactivée une contrainte de présence de $X$.

Définition 28 (Transaction $Probablement(X, i)$) Une transaction $t$ est dite *Probablement* dans un contexte incomplet $K_{VM}$, relativement à un itemset $(Xi)$ si $t$ est

[1] Un itemset pseudo-fermé désigne un itemset fermé dans un contexte incomplet.

$Certaine(X)$, tel que $t$ est $Probable(i)$. L'ensemble de transactions $Probablement$, noté $Probablement(X, i)$, est défini comme suit :

$$Probablement(X, i) = \{t \in T \mid t \in Certaine(X) \cap Probable(i)\}.$$

Considérons l'exemple du contexte complet $K$ donné par la Figure 3.2 (Gauche). Chaque tuple est représenté par quatre attributs $X_1$, $X_2$, $X_3$ et $X_4$. À chaque attribut est associé un domaine de valeurs : $dom(X_1) = \{A, B\}$, $dom(X_2) = \{C, D\}$, $dom(X_3) = \{E, F, G\}$ et $dom(X_4) = \{K, L\}$. Le contexte transactionnel associé au contexte $K$ est donné par la Figure 3.2 (Droite) où chaque couple (attribut, valeur) est codifié par la valeur même de l'attribut, constituant un item. La figure 3.3 (Gauche) représente le contexte $K_{VM}$ où des valeurs manquantes ont été aléatoirement introduites. Le contexte transactionnel associé est donné par la figure 3.3 (Droite) où chaque valeur manquante indique la présence d'un seul item parmi ceux manquants.

Exemple 13 La transaction $t_3$ est considérée comme $Certaine(AC)$ relativement au contexte $K_{VM}$ (Figure 3.3 (Droite)), puisque $AC$ est présent dans $t_3$. Cependant $t_3$, est considérée comme $Probable(E)$ puisque $E$ est manquant. La transaction $t_3$ est alors considérée $Probablement(AC, E)$.

| | $X_1$ | $X_2$ | $X_3$ | $X_4$ |
|---|---|---|---|---|
| 1 | A | C | E | K |
| 2 | B | C | E | L |
| 3 | A | C | E | K |
| 4 | A | D | F | L |
| 5 | B | C | F | L |
| 6 | B | C | F | K |
| 7 | A | D | G | L |
| 8 | B | D | G | L |

| | A | B | C | D | E | F | G | K | L |
|---|---|---|---|---|---|---|---|---|---|
| 1 | × | | × | | × | | | × | |
| 2 | | × | × | | × | | | | × |
| 3 | × | | × | | × | | | × | |
| 4 | × | | | × | | × | | | × |
| 5 | | × | × | | | × | | | × |
| 6 | | × | × | | | × | | × | |
| 7 | × | | | × | | | × | | × |
| 8 | | × | | × | | | × | | × |

Fig. 3.2 Gauche : Contexte d'extraction complet $K$. Droite : Contexte d'extraction transactionnel associé.

Dans ce qui suit, nous rappelons la définition de la presque-fermeture présentée dans [11].

| | $X_1$ | $X_2$ | $X_3$ | $X_4$ |
|---|---|---|---|---|
| 1 | A | C | E | K |
| 2 | B | C | E | L |
| 3 | A | C | ? | K |
| 4 | A | D | F | ? |
| 5 | B | C | F | L |
| 6 | ? | C | F | K |
| 7 | A | D | G | L |
| 8 | B | ? | G | L |

| | A | B | C | D | E | F | G | K | L |
|---|---|---|---|---|---|---|---|---|---|
| 1 | × | | × | | × | | | × | |
| 2 | | × | × | | × | | | | × |
| 3 | × | | × | | ? | ? | ? | × | |
| 4 | × | | | × | | × | | ? | ? |
| 5 | | × | × | | | × | | | × |
| 6 | ? | ? | × | | | × | | × | |
| 7 | × | | | × | | | × | | × |
| 8 | | × | ? | ? | | | × | | × |

Fig. 3.3 Gauche : Contexte d'extraction incomplet $K_{VM}$. Droite : Contexte d'extraction transactionnel associé.

**Définition 29 (Presque-fermeture)** Soit $\delta$ un entier positif, la presque-fermeture d'un itemset $X$ dans un contexte complet $K$, notée par $AC(X)$ est définie comme suit : $AC(X) = X \cup \{i | i \in \mathcal{I} \wedge supp(X) - supp(Xi) \leq \delta\}$ où $supp(X)$ désigne le support absolu de $X$.

Cette définition indique que lorsque un item $i \in AC(X)$, cela signifie que $i$ appartient à toutes les transactions contenant $X$ avec un nombre d'exceptions inférieur ou égal à $\delta$.

**Exemple 14** Considérons l'exemple du contexte complet $K$ donné par la Figure 3.2 (Droite). D'après la définition 29, nous avons $AC(AC) = ACEK$ pour $\delta = 0$. En effet, $E$ et $K$ appartiennent à toutes les transactions contenant l'itemset $AC$ (0 exceptions).

Cependant, il est important de signaler que pour ($\delta = 0$), la notion de presque-fermeture correspond au cas particulier de fermeture dans un contexte complet. Le concept de presque-fermeture a été redéfini pour faire l'extraction des motifs $\delta$-libres à partir d'un contexte incomplet [45, 48]. Cependant, nous verrons qu'en présence de valeurs manquantes et lorsque ($\delta = 0$), le concept de presque-fermeture ne correspond plus au concept de fermeture comme cela a été considéré dans [48]. Nous désignerons par pseudo-fermeture, le cas d'un contexte incomplet ($\delta = 0$) ou ($\delta > 0$) afin de faire la distinction avec le concept de presque-fermeture dans un contexte complet. Cependant, dans le cadre de notre travail, nous nous intéresserons plus particulièrement au cas où ($\delta = 0$). La Table 3.2 illustre cette distinction.

| | Contexte complet | Contexte incomplet |
|---|---|---|
| $\delta = 0$ | fermeture | pseudo-fermeture |
| $\delta > 0$ | presque fermeture | pseudo-fermeture |

Tab. 3.2 Distinction de la notion de fermeture

Dans ce qui suit, nous présentons les définitions de la pseudo-fermeture et d'un générateur minimal dans un contexte incomplet.

**Définition 30 (Pseudo-fermeture)** La pseudo-fermeture d'un itemset $X$ dans un contexte incomplet $K_{VM}$, notée $PF(X)$, est définie comme suit :

$$PF(X) = X \cup \{i \,|\, i \in \mathcal{I} \;\wedge\; supp(X) - supp(Xi) = |Probablement(X, i)|\}.$$

L'idée sous-jacente de la pseudo-fermeture, est que lors du calcul, si un item est déclaré manquant alors il est supposé présent. Ceci est fait par une prise en considération des transactions contenant $X$ pour lesquelles $i$ est déclaré manquant. Cette prise en considération est matérialisée par le nombre de transactions $Probablement(X, i)$.

**Exemple 15** Considérons l'exemple du contexte incomplet $K_{VM}$ donné par la Figure 3.3 (Droite). Nous avons $supp\,(AC) - supp\,(ACE) = 1$. Le nombre 1 représente le nombre de transactions $Probablement(AC, E)$. De plus, $supp\,(AC) - supp\,(ACK) = 0$. Comme nous avons $|Probablement(AC, K)| = 0$. Par conséquent, les items $E$ et $K$ appartiennent à $PF(AC)$, alors $PF(AC) = ACEK$.

**Définition 31 (Générateur minimal dans un contexte incomplet)** Un itemset $g$ est dit générateur minimal dans un contexte incomplet s'il n'est pas inclus dans la pseudo-fermeture de l'un de ses sous-ensembles de taille $|g| - 1$.

Toutefois, il est important de signaler que les générateurs minimaux dans un contexte incomplet vérifient une propriété importante, celle de l'anti-monotonie (c.f., Théorème 1). En effet, cette propriété s'avère efficace lors de l'extraction, puisqu'elle permet l'élagage de l'espace de recherche. Ceci est justifié d'une part par le fait que les motifs $\delta$-libres vérifient cette propriété [11] et d'autre part par le fait que la pseudo-fermeture dans un contexte incomplet correspond à une presque-fermeture ($\delta = 0$) dans le cas d'un contexte complet.

| $GM$ | support $GM$ | pseudo-fermé | $GM$ | support $GM$ | pseudo-fermé |
|---|---|---|---|---|---|
| A | 4 | A | AG | 1 | ADGL |
| B | 3 | BCL | AL | 1 | ADGL |
| C | 5 | C | BE | 1 | BCEL |
| D | 2 | ADL | BF | 1 | BCFL |
| E | 2 | CE | BG | 1 | BGL |
| F | 3 | F | CF | 2 | BCF |
| G | 2 | DGL | CL | 2 | BCL |
| K | 3 | ACK | DF | 1 | ADF |
| L | 4 | L | EK | 1 | ACEK |
| AC | 2 | ACEK | EL | 1 | BCEL |
| AE | 1 | ACEK | FK | 1 | CFK |
| AF | 1 | AFD | FL | 1 | BCFL |

Tab. 3.3 L'ensemble des générateurs minimaux $GM$ et leurs pseudo-fermés associés extraits du contexte $K_{VM}$ donné par la Figure 3.3 (Droite) pour une valeur de *minsup* égale à 1.

En effet, les exceptions (matérialisées par le nombre $\delta$ dans un contexte complet) peuvent être assimilées aux valeurs manquantes présentes dans les données (dans un contexte incomplet).

Théorème 1 (Anti-monotonie) [11] Soit $K_{VM}$ un contexte incomplet. $\forall X\ Y \in \mathcal{I}$ tel que $X \subset Y$. Si $Y$ est un générateur minimal alors $X$ est aussi un générateur minimal.

Exemple 16 L'ensemble des générateurs minimaux $MG$ ainsi que leurs pseudo-fermés associés, extraits à partir du contexte d'extraction incomplet $K_{VM}$ représenté par la Figure 3.3 (Droite) et une valeur de *minsup* égale à 1 est donné par la Table 3.3.

Dans ce qui suit, nous montrons que la pseudo-fermeture ne correspond plus au cas particulier de fermeture comme c'est le cas dans un contexte complet.

Proposition 1 La pseudo-fermeture dans un contexte incomplet n'est pas un opérateur de fermeture.

Preuve.
Un opérateur de fermeture implique que le support de tout générateur minimal est égal au support de sa fermeture. Cependant, la pseudo-fermeture adopte une stratégie optimiste. Lors du calcul de la pseudo-fermeture d'un générateur minimal, si un item est déclaré manquant alors il est supposé présent. Ceci implique que le support du générateur en question n'est pas forcément égal au support de sa pseudo-fermeture. Par conséquent, la pseudo-fermeture n'est pas un opérateur de fermeture. Si nous considérons l'exemple du générateur minimal $AC$, nous remarquons que $PF(AC) = ACEK$. Cependant, nous avons *supp* $(AC) = 2$, tandis que *supp* $(ACEK) = 1$.

■

Une fois que les générateurs minimaux et leurs pseudo-fermés ont été extraits, la base générique de règles associatives est directement générée. En effet, chaque règle sera constituée dans sa partie prémisse d'un générateur minimal, où la partie conclusion est le pseudo-fermé associé. Dans ce qui suit, nous présentons la définition de la base générique de règles associatives.

### 3.3.2 Base générique de règles associatives pseudo-exactes en présence de valeurs manquantes

Dans cette section, nous utilisons la définition de la base générique de règles associatives exactes introduite dans [4, 5] que nous adaptons à un contexte incomplet.

Définition 32 (Base générique de règles associatives pseudo-exactes) Soit $PFF$ l'ensemble des itemsets pseudo-fermés fréquents extraits à partir d'un contexte incomplet $K_{VM}$. Pour chaque itemset pseudo-fermé fréquent $c$, notons par $GM_c$ l'ensemble des générateurs minimaux associés à $c$. La base générique de règles associatives pseudo-exactes est définie comme suit :

$$BG = \{R : g \Rightarrow (c - g) \mid c \in PFF,\ g \in GM_c \text{ et } g = c^{(2)}\}.$$

Dans ce qui suit, nous associons à chaque item le couple (attribut, valeur) correspondant relativement au contexte complet $K$ donné par la Figure 3.2 (Gauche).

[2] la condition $g = c$ permet de ne pas retenir les règles de la forme $g \Rightarrow \emptyset$.

Exemple 17 La base générique de règles associatives pseudo-exactes extraite à partir du contexte incomplet $K_{VM}$ présenté par la Figure 3.3 (Droite) est donnée par la Table 3.4.

| Règles associatives pseudo-exactes | Support | Confiance |
|---|---|---|
| $R_1$ :$(X_1, B) \Rightarrow (X_2, C) \wedge (X_4, L)$ | 2 | $\frac{2}{3}$ |
| $R_2$ :$(X_2, D) \Rightarrow (X_1, A) \wedge (X_4, L)$ | 1 | $\frac{1}{2}$ |
| $R_3$ :$(X_3, E) \Rightarrow (X_2, C)$ | 2 | 1 |
| $R_4$ :$(X_3, tt) \Rightarrow (X_2, D) \wedge (X_4, L)$ | 1 | $\frac{1}{2}$ |
| $R_5$ :$(X_4, K) \Rightarrow (X_1, A) \wedge (X_2, C)$ | 2 | $\frac{2}{3}$ |
| $R_6$ :$(X_1, A) \wedge (X_2, C) \Rightarrow (X_3, E) \wedge (X_4, K)$ | 1 | $\frac{1}{2}$ |
| $R_7$ :$(X_1, A) \wedge (X_3, E) \Rightarrow (X_2, C) \wedge (X_4, K)$ | 1 | 1 |
| $R_8$ :$(X_1, A) \wedge (X_3, F) \Rightarrow (X_2, D)$ | 1 | 1 |
| $R_9$ :$(X_1, A) \wedge (X_3, tt) \Rightarrow (X_2, D) \wedge (X_4, L)$ | 1 | 1 |
| $R_{10}$ :$(X_1, A) \wedge (X_4, L) \Rightarrow (X_2, D) \wedge (X_3, tt)$ | 1 | 1 |
| $R_{11}$ :$(X_1, B) \wedge (X_3, E) \Rightarrow (X_2, C) \wedge (X_4, L)$ | 1 | 1 |
| $R_{12}$ :$(X_1, B) \wedge (X_3, F) \Rightarrow (X_2, C) \wedge (X_4, L)$ | 1 | 1 |
| $R_{13}$ : $(X_1, B) \wedge (X_3, tt) \Rightarrow (X_4, L)$ | 1 | 1 |
| $R_{14}$ :$(X_2, C) \wedge (X_3, F) \Rightarrow (X_1, B)$ | 1 | 1 |
| $R_{15}$ :$(X_2, C) \wedge (X_4, L) \Rightarrow (X_1, B)$ | 2 | 1 |
| $R_{16}$ :$(X_2, D) \wedge (X_3, F) \Rightarrow (X_1, A)$ | 1 | 1 |
| $R_{17}$ :$(X_3, E) \wedge (X_4, K) \Rightarrow (X_1, A) \wedge (X_2, C)$ | 1 | 1 |
| $R_{18}$ :$(X_3, E) \wedge (X_4, L) \Rightarrow (X_1, B) \wedge (X_2, C)$ | 1 | 1 |
| $R_{19}$ :$(X_3, F) \wedge (X_4, K) \Rightarrow (X_2, C)$ | 1 | 1 |
| $R_{20}$ :$(X_3, F) \wedge (X_4, L) \Rightarrow (X_1, B) \wedge (X_2, C)$ | 1 | 1 |

Tab. 3.4 La base générique de règles associatives pseudo-exactes extraite à partir du contexte incomplet $K_{VM}$ présenté par la Figure 3.3 (Droite).

Dans la section suivante, nous présentons le processus de complétion des valeurs manquantes $GBAR_{MVC}$ basé sur l'exploitation de la base générique de règles associatives pseudo-exactes, extraite à partir d'un contexte incomplet.

## 3.4 $GBAR_{MVC}$ : processus de complétion des valeurs manquantes

Dans cette section, nous présentons l'approche $GBAR_{MVC}$[3] que nous proposons pour la complétion des valeurs manquantes. Ce processus se base, d'une part sur la base générique de règles associatives pseudo-exactes, déjà extraite. D'autre part, nous proposons une nouvelle mesure intitulée Robustesse. Cette nouvelle mesure évalue la force d'une règle à compléter une valeur manquante. En effet, il s'agit de prendre en considération le degré de corrélation entre la prémisse et la conclusion d'une règle, grâce à la mesure lift [13]. De plus, cette nouvelle mesure tient compte du degré de correspondance d'une règle avec une transaction manquante. Ceci est fait grâce à une nouvelle mesure que nous proposons intitulée Correspondance.

Pour la complétion des valeurs manquantes, nous utiliserons des règles de la forme prémisse $\Rightarrow (X_i, v_i)$ où prémisse est la conjonction d'éléments de la forme $(X_j, v_j)$ avec $i = j$. Dans ce qui suit, nous rappelons la définition de consistance d'une règle $R$ avec une transaction $t$ présentant au moins une valeur manquante [53] et nous introduisons de nouvelles définitions relatives aux mesures Correspondance et Robustesse.

Définition 33 [53] Une règle $R$ : prémisse $\Rightarrow (X_i, v_i)$ est dite consistante avec une transaction $t$ présentant une valeur manquante sur l'attribut $X_i$, s'il n'existe aucun élément $(X_j, v_j)$ de la prémisse de $R$ qui di ère de la valeur de $X_j$ dans $t$.

### La mesure Correspondance

Nous proposons une nouvelle mesure intitulée Correspondance qui permet d'évaluer le degré de correspondance ou d'interprétation d'une règle $R$ par rapport à une transaction $t$ présentant au moins une valeur manquante. En effet, il s'avère efficace lors de la complétion de prendre en considération le degré de correspondance de la règle avec la transaction manquante [24, 53]. L'idée sous-jacente de la mesure Correspondance est d'utiliser lors de la complétion les règles présentant une correspondance maximale.

[3] L'acronyme $GBAR_{MVC}$ désigne Generic Basis of Association Rules based approach for Missing Values Completion.

**Définition 34** La mesure Correspondance d'une règle $R$ : prémisse $\Rightarrow (X_i, v_i)$ avec une transaction $t$ présentant une valeur manquante sur l'attribut $X_i$, est définie comme suit :

$$Correpondance(R, t, X_i) = \begin{cases} 0 & \text{si } R \text{ n'est pas consistante avec } t \\ \sum verifie(X_j, v_j) & \text{sinon.} \end{cases}$$

où

$$verifie(X_j, v_j) = \begin{cases} 0 & \text{si } X_j \text{ présente une valeur manquante dans } t \\ 1 & \text{sinon.} \end{cases}$$

**Exemple 18** Considérons l'exemple de la transaction $t_6 : (X_1, ?)(X_2, C)(X_3, F)(X_4, K)$. La règle $R_{16} : (X_2, D) \wedge (X_3, F) \Rightarrow (X_1, A)$ n'est consistante avec $t_6$, puisque la valeur de l'attribut $X_2$ est $C$ dans $t_6$. Cette valeur est différente de la valeur de l'attribut $X_2$ dans $R_{16}$ qui est égale à $D$. Par conséquent, $Correspondance(R_{16}, t_6, X_1) = 0$. Cependant, si nous considérons l'exemple de la règle $R$: $(X_2, C) \wedge (X_3, F) \Rightarrow (X_1, B)$, nous avons $Correspondance(R, t_6, X_1) = \frac{1}{2}$, puisque l'item $(X_2, C)$, ainsi que $(X_3, F)$ sont présents dans $t_6$.

L'intérêt de la mesure Correspondance s'avère utile à cause justement des valeurs manquantes. En effet, deux règles présentant un conflit peuvent être consistantes pour compléter une valeur manquante, la Correspondance permettra de tenir compte de l'absence d'items. Par conséquent, il s'agira de privilégier la règle présentant une plus grande correspondance avec la transaction manquante.

### La mesure Lift

Une deuxième mesure qui s'avère intéressante lors de la complétion est la mesure lift introduite par Brin [13]. Cette mesure vient contredire le fait que la mesure confiance permet de se focaliser sur les règles intéressantes. Il a été démontré que les règle de confiance élevée ne sont pas forcément les plus intéressantes [23]. En effet, une règle $R : X \Rightarrow Y$ indique que la connaissance de "$X$ implique celle de $Y$" dans une certaine proportion donnée par la confiance. Ainsi, la confiance mesure la probabilité conditionnelle de $Y$ sachant $X$. Cependant, elle ne mesure pas la réelle dépendance entre $X$ et $Y$. En revanche, la mesure lift permet de mesurer la corrélation induite entre la prémisse $X$ et la conclusion $Y$ d'une règle $R : X \Rightarrow Y$. Nous pensons donc qu'il s'agit d'une mesure utile pour notre problématique puisqu'elle va nous permettre d'affirmer que la présence de $X$ favorise la présence de $Y$ selon une valeur retournée par le lift. Le recours à une telle mesure n'est pas surprenant. En effet, cette même mesure a été utilisée dans [53] lors de l'étape de complétion. De plus, une mesure de même interprétation sémantique, intitulée $RI$ introduite par Shapiro, a été utilisée dans l'approche RAR_MVC dans le but d'élaguer les règles qualifiées de non intéressantes.

**Définition 35** [13] Le lift d'une règle $R$ : $(X \Rightarrow Y)$, noté $Lift(R)$, est calculé de la manière suivante :

$$Lift(R) = \frac{supp(XY)}{supp(X) \times Supp(Y)}.$$

Si $Lift(R) = 1$, alors $X$ et $Y$ sont dits indépendants. Si $Lift(R) < 1$, $X$ et $Y$ sont dits négativement corrélés. Par contre, si $Lift(R) > 1$, alors $X$ et $Y$ sont dits positivement corrélés [13].

L'objectif lors de notre complétion des valeurs manquantes est d'utiliser la règle qui maximise à la fois le critère Correspondance et la mesure lift. Pour cela, nous introduisons la mesure Robustesse qui permet d'évaluer la force d'une règle à compléter une valeur manquante. En effet, cette mesure est en fonction de la Correspondance et du lift. Elle prend donc en considération d'une part, le degré de correspondance de la règle avec la transaction manquante et d'autre part, la corrélation induite entre la prémisse et la conclusion de la règle.

**Définition 36** La Robustesse d'une règle $R$ à compléter une transaction $t$ présentant une valeur manquante sur l'attribut $X_i$ est définie comme suit :

$$Robustesse(R, t, X_i) = Correspondance(R, t, X_i) \times lift(R).$$

Dans ce qui suit, nous présentons l'algorithme de complétion des valeurs manquantes de $GBAR_{MVC}$. Le pseudo-code est donné par l'Algorithme 2.

Le principe de fonctionnement de l'algorithme $GBAR_{MVC}$ est décrit dans ce qui suit :

- Pour chaque attribut manquant $X_i$ de $t$, l'algorithme sélectionne les règles à partir de la base générique de règles associatives pseudo-exactes, $BG$, concluant sur $X_i$ et qui sont consistantes avec $t$. Nous désignerons par $R_{probables}(t, X_i)$ l'ensemble de ces règles (lignes 3-7).

- Si l'ensemble $R_{probables}(t, X_i)$ est vide, alors il n'existe aucune règle permettant la complétion de $X_i$ (lignes 8-9).
- Si l'ensemble de règles de $R_{probables}(t, X_i)$ concluent sur une même valeur $v$, alors $v$ est utilisée pour compléter la valeur manquante de $X_i$ (lignes 11-12).
- Sinon, l'ensemble de règles de $R_{probables}(t, X_i)$ présentent un conflit. Dans ce cas, l'algorithme procède au calcul de la mesure Robustesse pour chaque règle appartenant à $R_{probables}(t, X_i)$ (lignes 14-18).
- La règle présentant la valeur Robustesse la plus élevée est utilisée pour compléter la valeur manquante de $X_i$ (ligne 19).

Exemple 19 Le déroulement de l'algorithme 2 sur le contexte incomplet $K_{VM}$ donné par la figure 3.3 (Gauche, page 52) est décrit dans ce qui suit :

- La transaction $t_3$ : $(X_1, A)(X_2, C)(X_3, ?)(X_4, K)$ présente une valeur manquante sur l'attribut $X_3$. Les règles $R_6$ et $R_{10}$ contiennent l'attribut $X_3$ dans la partie conclusion. Cependant, seule la règle $R_6$ :$(X_1, A) \wedge (X_2, C) \Rightarrow (X_3, E) \wedge (X_4, K)$ est consistante avec $t_3$, i.e., $R_{probables}(t_3, X_3) = \{R_6\}$. Par conséquent, $t_3$ devient : $(X_1, A)(X_2, C)$ $(X_3, E)(X_4, K)$.
- La transaction $t_4$ est : $(X_1, A)(X_2, D)(X_3, F)(X_4, ?)$. $t_4$ présente une valeur manquante l'attribut $X_4$. Les règles concluant sur $X_4$ sont données par la Figure 3.4 (Gauche). Seule la règle $R_2$ appartient à $R_{probables}(t_4, X_4)$. Alors, $R_2$ est utilisée pour compléter la valeur manquante de la transaction $t_4$. Par conséquent, $t_4$ devient : $(X_1, A)(X_2, D)(X_3, F)(X_4, L)$.
- La transaction $t_6$ : $(X_1, ?)(X_2, C)(X_3, F)(X_4, K)$ présente une valeur manquante sur l'attribut $X_1$. Les règles concluant sur $X_1$ sont données par la Figure 3.4 (Droite). Seules les règles $R_2$ et $R_3$ sont consistantes avec $t_6$. Ces règles constituent l'ensemble $R_{probables}(t_6, X_1)$. L'ensemble $R_{probables}(t_6, X_1)$ est donnée par la Table 3.6. Comme $R_2$ et $R_3$ concluent sur deux valeurs différentes de l'attribut $X_1$ i.e., présentent un conflit, nous procédons au calcul de la mesure Robustesse respectivement pour $R_2$ et $R_3$.
- La règle $R_3$ : $(X_2, C) \wedge (X_3, F) \Rightarrow (X_1, B)$ présente la valeur Robustesse la plus élevée, $R_3$ est donc utilisée pour compléter la valeur manquante de $X_1$. La transaction $t_6$ devient alors $(X_1, B)(X_2, C)(X_3, F)(X_4, K)$.
- La transaction $t_8$ $(X_1, B)(X_2, ?)(X_3, G)(X_4, L)$ présente une valeur manquante sur

**Algorithme** : $GBAR_{MVC}$

**Données** : - $K_{MV}$ : Contexte incomplet

- $BG$ : Base générique de règles associatives pseudo-exactes

**Résultats** : $K_{VM}$ complété

**début**

  **pour chaque** transaction manquante $t$ de $K_{MV}$ **faire**

    **pour chaque** attribut $X_i$ de $t$ présentant une valeur manquante **faire**

      **pour chaque** règle $R$ de $BG$ telle que $X_i$ apparaît dans la conclusion **faire**

        **si** $R$ est consistante avec $t$ **alors**

          $R_{probables}(t, X_i) = R_{probables}(t, X_i) \cup R$;

      **si** $|R_{probables}(t, X_i)| = 0$ **alors**

        $V_{completion} = \emptyset$;

      **sinon**

        **si** $R_{probables}(t, X_i)$ concluant sur la même valeur $v$ **alors**

          $V_{completion} = v$;

        **sinon**

          $max = 0$;

          **pour chaque** règle $r$ de $R_{probables}(t, X_i)$ **faire**

            Calculer_Robustesse(r);

            **si** r.Robustesse> max **alors**

              $V_{completion}$=*r.conclusion*;

              $max = r.Robustesse$;

      $t.X_i = V_{completion}$;

  retourner ($K_{VM}$ complété);

**n**

Algorithme 2 : L'algorithme $GBAR_{MVC}$

l'attribut $X_2$. Les règles concluant sur $X_2$ sont données par la Table 3.5. L'ensemble $R_{probables}(t_8, X_2)$ est donné par la Table 3.7. La règle $R_3$ est utilisée pour la complétion de $X_2$. Par conséquent $t_8$ devient : $(X_1, B)(X_2, D)(X_3, G)(X_4, L)$.

Finalement, nous remarquons que 100% des valeurs manquantes, figurant dans le contexte $K_{VM}$ donné par la Figure 3.3 (Gauche, page 52), ont été correctement complétées.

| | |
|---|---|
| $R_1$ | $(X_1, B) \Rightarrow (X_4, L)$ |
| $R_2$ | $(X_2, D) \Rightarrow (X_4, L)$ |
| $R_3$ | $(X_3, G) \Rightarrow (X_4, L)$ |
| $R_4$ | $(X_1, A) \wedge (X_2, C) \Rightarrow (X_4, K)$ |
| $R_5$ | $(X_1, A) \wedge (X_3, E) \Rightarrow (X_4, K)$ |
| $R_6$ | $(X_1, A) \wedge (X_3, G) \Rightarrow (X_4, L)$ |
| $R_7$ | $(X_1, B) \wedge (X_3, E) \Rightarrow (X_4, L)$ |
| $R_8$ | $(X_1, B) \wedge (X_3, F) \Rightarrow (X_4, L)$ |
| $R_9$ | $(X_1, B) \wedge (X_3, G) \Rightarrow (X_4, L)$ |
| $R_1$ | $(X_2, D) \Rightarrow (X_1, A)$ |
| $R_2$ | $(X_4, K) \Rightarrow (X_1, A)$ |
| $R_3$ | $(X_2, C) \wedge (X_3, F) \Rightarrow (X_1, B)$ |
| $R_4$ | $(X_2, C) \wedge (X_4, L) \Rightarrow (X_1, B)$ |
| $R_5$ | $(X_2, D) \wedge (X_3, F) \Rightarrow (X_1, A)$ |
| $R_6$ | $(X_3, E) \wedge (X_4, K) \Rightarrow (X_1, A)$ |
| $R_7$ | $(X_3, E) \wedge (X_4, L) \Rightarrow (X_1, B)$ |
| $R_8$ | $(X_3, F) \wedge (X_4, L) \Rightarrow (X_1, B)$ |

Fig. 3.4 Gauche : Règles concluant sur $X_4$. Droite : Règles concluant sur $X_1$.

| | | | |
|---|---|---|---|
| $R_1$ | $(X_1, B) \Rightarrow (X_2, C)$ | $R_8$ | $(X_1, A) \wedge (X_4, L) \Rightarrow (X_2, D)$ |
| $R_2$ | $(X_3, E) \Rightarrow (X_2, C)$ | $R_9$ | $(X_1, B) \wedge (X_3, E) \Rightarrow (X_2, C)$ |
| $R_3$ | $(X_3, G) \Rightarrow (X_2, D)$ | $R_{10}$ | $(X_1, B) \wedge (X_3, F) \Rightarrow (X_2, C)$ |
| $R_4$ | $(X_4, K) \Rightarrow (X_2, C)$ | $R_{11}$ | $(X_3, E) \wedge (X_4, K) \Rightarrow (X_2, C)$ |
| $R_5$ | $(X_1, A) \wedge (X_3, E) \Rightarrow (X_2, C)$ | $R_{12}$ | $(X_3, E) \wedge (X_4, L) \Rightarrow (X_2, C)$ |
| $R_6$ | $(X_1, A) \wedge (X_2, F) \Rightarrow (X_2, D)$ | $R_{13}$ | $(X_3, F) \wedge (X_4, K) \Rightarrow (X_2, C)$ |
| $R_7$ | $(X_1, A) \wedge (X_3, G) \Rightarrow (X_2, D)$ | $R_{14}$ | $(X_3, F) \wedge (X_4, L) \Rightarrow (X_2, C)$ |

Tab. 3.5 Règles concluant sur $X_2$.

Dans la section suivante, nous proposons d'étudier la complexité théorique de l'algorithme $GBAR_{MVC}$ (Algorithme 2, page 61).

| | Correspondance | Lift | Robustesse |
|---|---|---|---|
| $R_2 : (X_4, K) \Rightarrow (X_1, A)$ | 0,25 | 1,33 | 0,33 |
| $R_3 : (X_2, C) \wedge (X_3, F) \Rightarrow (X_1, B)$ | 0,50 | 1,33 | 0,66 |

Tab. 3.6 $R_{probables}(t_6, X_1)$.

| | Correspondance | Lift | Robustesse |
|---|---|---|---|
| $R_1 : (X_1, B) \Rightarrow (X_2, C)$ | 0,25 | 1,06 | 0,265 |
| $R_3 : (X_3, G) \Rightarrow (X_2, D)$ | 0,25 | 2 | 0,5 |

Tab. 3.7 $R_{probables}(t_8, X_2)$.

## 3.5 Etude de la complexité de $GBAR_{MVC}$

A n d'étudier la complexité théorique de l'algorithme $GBAR_{MVC}$, considérons un contexte incomplet $K_{VM}=(O, I, R)$ et notons par $t_{premisse}$, la taille maximale des prémisses des règles de $BG$.

Le calcul de la complexité de l'algorithme $GBAR_{MVC}$ au pire des cas est présenté comme suit :

- Le test de consistance d'une règle $R$ avec une transaction $t$ consiste à vérifier s'il existe au moins un élément $(X_j, v_j)$ de la prémisse de $R$ en conflit avec la valeur de $X_j$ dans $t$. Par conséquent, ce test est de l'ordre de $O(t_{premisse})$ (ligne 6).
- L'ajout d'une règle $R$ à l'ensemble $R_{probables}(t, X_i)$ est de l'ordre $O(1)$ (ligne 7).
- Le corps de la boucle pour (ligne 5) est alors de l'ordre de $O(t_{premisse})$. Cette boucle se répète autant de fois qu'il y a de règles concluant sur l'attribut $X_i$ dans $BG$. Par conséquent, cette boucle (lignes 5-7) est de l'ordre de $O(t_{premisse} \times NR_i)$, où $NR_i$ désigne le nombre de règles de $BG$ concluant sur l'attribut $X_i$.
- Après avoir construit l'ensemble $R_{probables}(t, X_i)$, l'algorithme procède par tester si la cardinalité $R_{probables}(t, X_i)$ est égale 0. Ce test est de l'ordre de $O(1)$ (ligne 8). Si la condition est vérifiée, la partie alors est de l'ordre de $O(1)$ (ligne 9).
- La partie sinon (ligne 10), consiste à vérifier si l'ensemble des règles de $R_{probables}(t, X_i)$ concluent sur une même valeur. Cette opération est de l'ordre de $O(|R_{probables}(t, X_i)|)$. Au pire des cas, le nombre des règles de $R_{probables}(t, X_i)$ est égal au nombre de toutes les règles concluant sur $X_i$.

Par conséquent, cette opération, au pire des cas, est de l'ordre de $O(NR_i)$. La partie alors (ligne 12) se fait en $O(1)$.

- La partie sinon (ligne 13) consiste à parcourir l'ensemble $R_{probables}(t, X_i)$ afin de sélectionner la règle présentant la valeur Robustesse maximale. L'instruction donnée par la ligne 14 est de l'ordre de $O(1)$. Le corps de la boucle pour (lignes 16-17-18) est de l'ordre de $O(1)$. Par conséquent, cette boucle, au pire des cas, est de l'ordre de $O(NR_i)$.
- Finalement, le corps de la boucle pour (linge 5) est de l'ordre de $O(t_{premisse} \times NR_i)$.

Ce traitement est répété autant de fois qu'il y d'attributs $X_i$ manquants dans la transaction $t$. Ainsi, cette boucle est de l'ordre de $\sum_{X_i \in I_m} O(t_{premisse} \times NR_i)$ où $I_m$ désigne l'ensemble des attributs manquants. Au pire des cas, i.e., tous les attributs sont affectés par des valeurs manquantes, ce traitement (ligne 4-19) est de l'ordre de $\sum_{X_i \in I} O(t_{premisse} \times NR_i) \cong O(t_{premisse}) \times \sum_{X_i \in I} O(NR_i) \cong O(t_{premisse} \times |BG|)$

- De même, le traitement sus-décrit (ligne 4-19) se fait pour chaque transaction présen- tant au moins une valeur manquante. Ainsi, la boucle pour (lignes 3-19) est de l'ordre de $O(t_{premisse} \times |BG| \times N_m)$ où $N_m$ désigne le nombre de transactions manquantes. Au pire des cas, i.e., nous considérons que toutes les transactions sont manquantes, ce traitement (lignes 3-19) est de l'ordre de $O(t_{premisse} \times |BG| \times |K_{VM}|)$.

Finalement, la complexité théorique, au pire des cas, de l'algorithme $GBAR_{MVC}$ est de l'ordre de $O(t_{premisse} \times |BG| \times |K_{VM}|)$. Cependant, il est important de signaler que l'algorithme $AR_{MVC}$ (Algorithme 1 page 42) présente une complexité théorique au pire des cas de l'ordre $O(t_{premisse} \times |RA| \times |K_{VM}|)$. Étant donné que $|RA| \geq |BG|$ et $t_{premisse}$ des règles de $RA \geq$ à $t_{premisse}$ des règles de $BG$. Ainsi, $GBAR_{MVC}$ présente une complexité théorique plus faible que $AR_{MVC}$. Ce résultat n'est pas surprenant. En effet, ce résultat théorique est garanti par l'utilisation de la base générique de règles associatives. En effet, ces règles sont en nombre plus réduit par rapport à l'ensemble total des règles associatives. De plus, ces règles présentent une prémisse minimale, constituée de générateurs minimaux. Ainsi réside l'intérêt même de l'utilisation d'une base générique pour la complétion des valeurs manquantes, comme nous l'avons montré dans la section 3.2 (page 47).

## 3.6 Conclusion

Dans ce chapitre, nous avons présenté une nouvelle approche de complétion des valeurs manquantes appelée $GBAR_{MVC}$. Cette approche présente l'avantage qu'elle ne procède pas par une suppression préalable des données manquantes et qu'elle prend en considération la présence des valeurs manquantes grâce à la redéfinition de la presque-fermeture. De plus, la particularité de $GBAR_{MVC}$ est qu'elle se base sur l'utilisation d'une base générique de règles assurant une réduction des conflits lors de l'étape de complétion. Une nouvelle mesure intitulée Robustesse a été de même présentée permettant de mesurer la force d'une règle à compléter une valeur manquante en cas de conflit.

Dans le chapitre suivant, nous allons mener une série d'expérimentations afin d'évaluer $GBAR_{MVC}$ et de confirmer certains résultats.

# Chapitre 4

# Evaluations expérimentales

## 4.1 Introduction

Dans le chapitre précédant, nous avons introduit une nouvelle approche de complétion des valeurs manquantes, appelée $GBAR_{MVC}$. La particularité de $GBAR_{MVC}$ est qu'elle ne procède pas à la complétion des valeurs manquantes d'une manière brutale. En effet, nous avons montré que le processus de complétion de $GBAR_{MVC}$ est préalablement associé à une prise en considération de l'aspect incomplet des données. Ceci nous permet l'obtention d'un ensemble de règles ables utilisées lors de l'étape de complétion. De plus, l'approche $GBAR_{MVC}$ est basée sur un ensemble de règles dites génériques, assurant la réduction des conflits entre les règles lors de la complétion.

Dans ce chapitre, nous allons essayer d'évaluer expérimentalement l'approche $GBAR_{MVC}$. L'objectif de cette évaluation porte essentiellement sur les axes suivants :

- L'approche $GBAR_{MVC}$ offre- t-elle une complétion fiable des valeurs manquantes ?
- L'approche $GBAR_{MVC}$ permet-elle toujours la complétion des valeurs manquantes ?
- L'approche $GBAR_{MVC}$ est- t-elle sensible au nombre de valeurs manquantes introduites ?
- Comment peut-on placer $GBAR_{MVC}$ comparativement aux approches de complétion existantes dans la littérature ?

## 4.2 Evaluation de $GBAR_{MVC}$

Pour répondre à toutes ces questions, nous allons tout d'abord définir d'une part, le protocole de validation qui nous permettra d'évaluer $GBAR_{MVC}$. D'autres parts, nous allons fixer les critères sur lesquels portera l'évaluation de $GBAR_{MVC}$.

### 4.2.1 Protocole de validation

Les approches de complétion des valeurs manquantes sont généralement évaluées par le biais de deux techniques [46]. La première technique est celle de la validation croisée, qui est une technique utilisée dans le domaine de la classification supervisée. La validation est alors analogue aux méthodes de classification, i.e., par la mesure de la précision d'un test de classification. La deuxième technique, consiste en l'introduction artificielle de valeurs manquantes à partir d'une base complète, dite base de référence, et la comparaison du résultat obtenu par rapport à cette référence.
Nous avons choisi de valider notre travail par le biais de la deuxième technique. Ce choix est dicté par le handicap que présentent les méthodes de classification lors de la construction de l'échantillon d'apprentissage. En effet, cet échantillon doit être le plus représentatif possible. Ceci semble constituer une problématique à part entière. Afin de valider $GBAR_{MVC}$, nous utilisons donc une base complète dans laquelle nous introduisons artificiellement des valeurs manquantes suivant ces différents taux : 5%, 10%, 15% et 20% par attribut. Le résultat obtenu par $GBAR_{MVC}$ est alors comparé à la base de référence (sans valeurs manquantes).

Après avoir défini le protocole de validation que nous adoptons pour l'évaluation de $GBAR_{MVC}$. Dans ce qui suit, nous définissons les critères d'évaluation du protocole.

### 4.2.2 Critères d'évaluation

Les critères d'évaluation associés au protocole de validation sont présentés comme suit :

- Pourcentage de valeurs manquantes correctement complétées : ce premier critère d'évaluation est classique dans toute approche de complétion. Dans tout le reste du chapitre, nous désignerons ce critère par la précision. La précision est définie comme suit :
  Précision = (Nombre de valeurs manquantes correctement complé-tées/Nombre de valeurs manquantes complétées)×100.

- **Pourcentage de valeurs manquantes complétées** : ce deuxième critère est moins utilisé mais il constitue un critère important lors de l'évaluation. Ce critère mesure le pourcentage de valeurs manquantes que l'approche a pu compléter, parmi le nombre total des valeurs manquantes. Le pourcentage de complétion est défini comme suit : Pourcentage de complétion = (Nombre de valeurs manquantes complétées/Nombre de valeurs manquantes totales)×100.

A n de comparer $GBAR_{MVC}$ par rapport aux approches de complétion existantes dans la littérature, nous avons choisi l'approche $AR_{MVC}$. Ce choix se justifie par le fait qu'il est plus équitable de comparer $GBAR_{MVC}$ à une approche faisant partie de la deuxième stratégie. $AR_{MVC}$ étant le seule approche existante à part RAR_MVC, qui impose une contrainte, à savoir la représentativité comme nous l'avons présenté au chapitre 2 (page 36).
Dans la section suivante, nous donnons une brève description de l'environnement matériel et logiciel utilisé lors de notre travail.

## 4.3 Environnement d'expérimentation

### 4.3.1 Environnement matériel et logiciel

Toutes les expérimentations ont été menées sur un PC muni du système d'exploitation Linux sous la distribution RedHat, d'un processeur Pentium IV ayant une fréquence d'horloge de 2,4 GHz et 512 Mo de mémoire RAM. Les programmes ont été compilés avec le compilateur gcc 3.3.1.

### 4.3.2 Programmes réalisés

Dans le cadre de notre travail, nous avons implémenté avec le langage C++, ces différents programmes :

- $GBAR_{MVC}$ : il s'agit de l'implantation de l'approche $GBAR_{MVC}$.
- $AR_{MVC}$ : il s'agit de l'implantation de l'approche $AR_{MVC}$.
- addVM: c'est le programme que nous avons réalisé pour introduire artificiellement des valeurs manquantes sur un certain nombre d'attribut et selon un taux donné.

- Validation : c'est le programme permettant de valider notre travail. Nous avons utilisé ce programme aussi bien pour la validation de l'approche $GBAR_{MVC}$, ainsi que celle de $AR_{MVC}$.

Cependant, il est important de signaler que nous avons utilisé le prototype mvminer, mis à notre disposition par Mr. François Rioult[1] afin d'extraire les générateurs minimaux ainsi que leurs pseudo-fermés associés. Nous utilisons ce prototype pour la construction de la base générique de règles associatives pseudo-exactes. De même, nous avons utilisé le programme apriori[2] pour l'extraction des règles associatives utilisées par l'approche $AR_{MVC}$.

### 4.3.3 Bases de test

Nous avons mené une série d'expérimentations sur différentes bases benchmark[3]. Les caractéristiques de ces bases sont données par la Table 4.1.

| Base | Nombre de transactions | Nombre d'items | Nombre d'attributs |
|---|---|---|---|
| Mushroom | 8124 | 128 | 23 |
| Zoo | 101 | 56 | 28 |
| Tic-tac-toe | 958 | 58 | 29 |
| House-votes | 435 | 36 | 18 |
| Monks2 | 432 | 38 | 19 |

Tab. 4.1 Caractéristiques des bases de test.

Dans ce qui suit, nous allons mener cinq séries d'expérimentation pour évaluer $GBAR_{MVC}$. Les trois premières séries consistent à évaluer le pourcentage de complétion et la précision de $GBAR_{MVC}$ en fonction de la variation respective du taux de valeurs manquantes, de la valeur de minsup et du nombre d'attributs manquants. Par la quatrième série, nous souhaitons étudier l'évolution du temps de complétion en fonction de la variation respective du taux de valeurs manquantes et du nombre d'attributs manquants. Une dernière série d'expérimentation est de même présentée, par laquelle nous souhaitons comparer les performances de $GBAR_{MVC}$ vs. $AR_{MVC}$.

[1]Pour lequel nous exprimons toute notre gratitude.

[2]Disponible à l'adresse suivante : http://fuzzy.cs.uni-magdeburg.de/ borgelt/software.html.

[3]Disponibles à l'adresse suivante : http://www.ics.uci.edu/~mlearn/MLRepository.html.

## 4.4 Variation du taux de complétion et de la précision en fonction de la variation du taux de valeurs manquantes

Dans cette section, nous allons étudier la capacité de $GBAR_{MVC}$ à compléter les valeurs manquantes et sa précision en fonction de la variation du taux de valeurs manquantes. La Table 4.2 donne cette variation pour une valeur de *minsup* = 10%.

### 4.4.1 Effet de la variation du taux de valeurs manquantes sur le pourcentage de complétion

À la lecture de la Table 4.2, nous constatons que :

- Pour les bases Tic-tac-toe et Monks2, le pourcentage de complétion est de 100% quel que soit le taux de valeurs manquantes introduites. Nous remarquons donc que le pourcentage de complétion n'est pas affecté par l'augmentation du taux de va- leurs manquantes. Quant aux autres bases, Mushroom, Zoo et House-votes, nous constatons qu'il reste un certain nombre de valeurs manquantes qui n'ont pas pu être complétées. Cependant, le pourcentage de complétion n'a pas été également affecté par l'introduction de valeurs manquantes.
- À remarquer aussi que, le pourcentage de complétion, dans certains cas, est non seule- ment non affecté par l'introduction de valeurs manquantes, mais aussi il s'améliore avec l'augmentation de ces dernières.
- Le nombre de règles utilisées lors de la complétion diminue avec l'introduction de valeurs manquantes. Ce constat n'est pas surprenant. Il a été montré que le nombre de règles diminue rapidement avec l'introduction de valeurs manquantes [41], i.e., l'auteur présente une courbe où le nombre de règles passe de 3000 règles à 500 règles avec l'introduction de seulement 5% de valeurs manquantes.

Par conséquent, il nous semble important de signaler que ce dernier constat n'a pas détérioré la capacité de $GBAR_{MVC}$ à compléter les valeurs manquantes. En effet, il a été mentionné que la complétion est liée à l'existence de plusieurs règles [41]. Cependant, nous constatons que la diminution du nombre de règles n'a pas constitué un handicap en la capacité de $GBAR_{MVC}$ à compléter. Cette capacité de $GBAR_{MVC}$ en la complétion, en dépit de l'augmentation du nombre de valeurs manquantes peut être expliquée par la stratégie adoptée. En effet, nous avons vu au chapitre précédant que $GBAR_{MVC}$ procède par une prise en considération des valeurs manquantes grâce à la définition de la pseudo-fermeture. Quant à l'amélioration du pourcentage de complétion, ceci se justifie par l'apparition de nouveaux items dans la pseudo-fermeture avec l'augmentation de valeurs manquantes [45]. Ceci favorise la complétion de plus de valeurs manquantes.

### 4.4.2 Effet de la variation du taux de valeurs manquantes sur la précision

À la lecture de la Table 4.2, nous constatons que la précision de $GBAR_{MVC}$ a été affectée par l'introduction de valeurs manquantes. Ceci reste valable pour l'ensemble des bases de test considérées. En effet, nous remarquons qu'avec l'augmentation du taux de valeurs manquantes, la précision de $GBAR_{MVC}$ diminue. Ceci ne constitue pas une incapacité de $GBAR_{MVC}$ à compléter correctement les valeurs manquantes. En effet, il semble être légitime d'obtenir ce résultat. Ceci s'explique comme suit :

1. Avec l'augmentation de valeurs manquantes, la qualité des données est de plus en plus détériorée. Cette détérioration entraîne par conséquent l'obtention de règles peu ables utilisées à la complétion. Ceci justifie la diminution de la précision de $GBAR_{MVC}$ en fonction de l'augmentation du taux de valeurs manquantes.
2. Partant du principe qu'une complétion correcte est liée à l'absence de conflits, nous remarquons en effet, qu'avec l'augmentation de valeurs manquantes, nous constatons également une augmentation des conflits. Ces conflits expliquent par conséquent, la détérioration de la précision de $GBAR_{MVC}$, avec l'augmentation des valeurs manquantes.

Cependant, à la lecture de la Table 4.2, nous constatons également que même avec l'absence de conflits, la précision de $GBAR_{MVC}$ n'est pas de 100%. En effet, un certain taux d'erreurs persiste. Ce résultat ne nous semble pas une inefficacité liée au processus de complétion de $GBAR_{MVC}$. Ceci se justifie par la nature même des données. En effet, comme nous l'avons affirmé tout au long de ce mémoire, la complétion par la technique des règles associatives repose sur le principe que les données sont corrélées, i.e., suivent un même comportement. Cependant, dans certain cas, des exceptions peuvent exister au sein des données, affectant ainsi l'efficacité du processus de complétion. Ceci peut être mis à profit par le constat suivant : Les bases Mushroom, Zoo, House-votes et Tic-tac-toe présentent une précision respective de 99%, 97%, 95% et 91% en l'absence de conflits. En revanche, Monks2 présente une précision moins bonne (83%). Ce constat explique le taux d'erreurs, lié à la nature des données.

## 4.5 Variation du taux de complétion et de la précision en fonction de la variation de minsup

Dans cette section, nous allons étudier la capacité de $GBAR_{MVC}$ à compléter les valeurs manquantes et sa précision en fonction de la variation de la valeur de minsup. La Table 4.3 présente cette variation pour un taux de valeurs manquantes=20%.

### 4.5.1 Effet de la variation de minsup sur le pourcentage de complétion

À la lecture de la Table 4.3, nous constatons que le pourcentage de complétion des valeurs manquantes est affecté par la variation de minsup, i.e., ce pourcentage diminue avec l'augmentation de la valeur de minsup. Par exemple, dans le cas de la base Tic-tac-toe, une valeur de minsup égale à 30% n'a permis de compléter que 65% des valeurs manquantes. Cependant, si nous baissons la valeur de minsup jusqu'à 10%, nous remarquons que 100% des valeurs manquantes ont pu être complétées. Ce constat est observé pour l'ensemble des bases. Ceci est justifié par :

1. La diminution du nombre de règles, comme nous pouvons le constater par la même table (Table 4.3). Nous retrouvons dans ce cas le principe déjà mentionné stipulant que la complétion de nombreuses valeurs manquantes est liée à l'existence de plusieurs règles. Cependant, à l'inverse de la première série d'expérimentation, cette diminution en nombre de règles a effectivement affecté le pourcentage de complétion.

| Base | Nombre de valeurs manquantes(%) | Nombre de règles | Nombre de con it | Pourcentage de complétion | Précision |
|---|---|---|---|---|---|
| Mushroom | 05 | 028293 | 00 | 74% | 99% |
| | 10 | 027988 | 00 | 76% | 99% |
| | 15 | 027988 | 00 | 78% | 97% |
| | 20 | 024410 | 23 | 80% | 97% |
| Zoo | 05 | 824650 | 00 | 100% | 97% |
| | 10 | 756741 | 05 | 98% | 89% |
| | 15 | 626390 | 10 | 100% | 88% |
| | 20 | 547075 | 19 | 99% | 88% |
| Tic-tac-toe | 05 | 315094 | 000 | 100% | 91% |
| | 10 | 296222 | 003 | 100% | 89% |
| | 15 | 279915 | 015 | 100% | 87% |
| | 20 | 266022 | 311 | 100% | 60% |
| House-votes | 05 | 125909 | 00 | 91% | 95% |
| | 10 | 102310 | 04 | 93% | 90% |
| | 15 | 094246 | 11 | 92% | 87% |
| | 20 | 081162 | 18 | 92% | 82% |
| Monks2 | 05 | 028325 | 000 | 100% | 83% |
| | 10 | 025402 | 040 | 100% | 65% |
| | 15 | 021790 | 048 | 100% | 71% |
| | 20 | 019741 | 108 | 100% | 63% |

Tab. 4.2 Variation du pourcentage de complétion et de la précision de $GBAR_{MVC}$ en fonction de la variation du taux de valeurs manquantes pour une valeur minsup égale à 10%.

2. L'augmentation de la valeur de minsup entraîne la disparition de certaines valeurs d'attribut (matérialisées par des items) qui deviennent infréquentes. Ceci constitue parfois un handicap à la complétion.

### 4.5.2 Effet de la variation de minsup sur la précision

À la lecture de la Table 4.3, nous constatons qu'à l'inverse du pourcentage de complétion, la précision de $GBAR_{MVC}$ augmente avec l'augmentation de minsup. En effet, si nous reprenons l'exemple de la base Tic-tac-toe, une valeur de minsup égale à 10% permet une complétion correcte de 60% des valeurs manquantes. Par contre, une valeur de minsup égale à 30%, permettra une complétion correcte de 100% des valeurs manquantes. Cette observation peut être expliquée par ce qui suit :

- Les règles présentant une valeur minsup plus élevées décrivent des données plus fréquentes, donc plus ables à la complétion. En revanche, une valeur minsup moins élevée conduit à des données moins fréquentes. Ceci entraîne par conséquent à l'utilisation de règles peu ables à la complétion.
- Une diminution de la valeur minsup entraîne une augmentation des conflits. Ces conflits sont à l'origine de la détérioration de la précision avec la diminution de la valeur de minsup.

## 4.6 Variation du taux de complétion et de la précision en fonction de la variation du nombre d'attributs manquants

Dans cette section, nous allons étudier la capacité de $GBAR_{MVC}$ à compléter les valeurs manquantes et sa précision en fonction de la variation du nombre d'attributs manquants. Dans cette série d'expérimentation, nous faisons varier le nombre d'attributs manquants jusqu'atteindre 50% du nombre d'attributs total de la base. Les Tables 4.4, 4.5 et 4.5 illustrent cette variation pour une valeur de *minsup* = 35% et un taux de valeurs manquantes égal à 5%.

| Base | minsup | Nombre de con it | Pourcentage de complétion | Précision |
|---|---|---|---|---|
| Mushroom | 10% | 23 | 80% | 97% |
| | 15% | 00 | 75% | 98% |
| | 20% | 00 | 71% | 98% |
| | 25% | 00 | 55% | 73% |
| | 30% | 00 | 53% | 57% |
| Zoo | 10% | 19 | 99% | 88% |
| | 15% | 13 | 96% | 78% |
| | 20% | 05 | 92% | 88% |
| | 25% | 00 | 89% | 92% |
| | 30% | 00 | 88% | 93% |
| Tic-tac-toe | 10% | 311 | 100% | 060% |
| | 15% | 207 | 100% | 070% |
| | 20% | 083 | 089% | 085% |
| | 25% | 000 | 086% | 096% |
| | 30% | 000 | 065% | 100% |
| House-votes | 10% | 18 | 92% | 82% |
| | 15% | 02 | 45% | 90% |
| | 20% | 00 | 76% | 79% |
| | 25% | 00 | 70% | 74% |
| | 30% | 00 | 33% | 50% |
| Monks2 | 10% | 108 | 100% | 63% |
| | 15% | 104 | 100% | 63% |
| | 20% | 048 | 100% | 75% |
| | 25% | 000 | 100% | 83% |
| | 30% | 000 | 083% | 92% |

Tab. 4.3 Variation du pourcentage de complétion et de la précision de $GBAR_{MVC}$ en fonction de la variation de minsup pour un taux de valeurs manquantes égal à 20%.

### 4.6.1 Effet de la variation du nombre d'attributs manquants sur le pourcentage de complétion

À la lecture des Tables 4.4, 4.5 et 4.6, nous constatons que le pourcentage de complétion diminue avec l'augmentation du nombre de valeurs manquantes à l'exception de la base Zoo. Cependant, la diminution du nombre de règles n'en constitue pas la cause. En effet, nous remarquons que dans le cas de la base Tic-tac-toe (respectivement Monks), le pourcentage de complétion a passé de 85% (resp. 100%) pour un attribut manquant à 59% (resp. 50%) dans le cas de quatre attributs manquants (resp. cinq attributs), pour une même valeur de nombre de règles. Un autre constat observé dans la base House-votes, est que le pourcentage de complétion reste égal à 100% avec l'introduction d'un attribut manquant, ensuite de deux attributs manquants. Ce pourcentage n'a pas diminué en dépit de la diminution du nombre de règles. Ce même constat est observé aussi dans la base Mushroom. Par conséquent, la diminution du nombre de règles ne constitue pas la cause de la détérioration du pourcentage de complétion. Cette détérioration peut s'expliquer par le fait que plus le nombre d'attributs manquants augmente, plus le risque d'avoir des valeurs d'attributs infréquentes. Ceci va empêcher l'apparition d'items dans les conclusions des règles. Cependant, à l'exception de toutes les autres bases, la base Zoo n'a pas présenté une détérioration du pourcentage de complétion. En effet, nous remarquons que le pourcentage de complétion est égal à 80% pour 14 attributs manquants (ce qui représente les 50% des attributs de la base). En effet, nous remarquons que les données de la base Zoo sont très corrélées. Cette caractéristique de la base augmente l'apparition d'items dans la pseudo-fermeture, matérialisés par les conclusions des règles. Ceci favorise la complétion de plus de valeurs manquantes.

### 4.6.2 Effet de la variation du nombre d'attributs manquants sur la précision

À la lecture des Tables 4.4, 4.5 et 4.6, nous constatons que la précision diminue avec l'augmentation du nombre d'attributs manquants. Ceci se justifie comme dans le cas de l'augmentation du taux de valeurs manquantes, par la détérioration de la qualité des règles avec l'augmentation du nombre d'attributs manquants. Cependant, il est important de signaler que la précision de $GBAR_{MVC}$ donne néanmoins de très bons résultats même avec la présence de 50% d'attributs manquants. Ce constat est observé pour les bases Mushroom et Zoo. En effet, la précision reste égale à 83% dans le cas de la base Mushroom, (resp. 97% dans le cas de la base Zoo). Les autres bases présentent un moins bon résultat. Ceci s'explique par le fait que les données des bases Mushroom et Zoo cachent de très bonnes relations, i.e., présentent un même comportement. Ceci n'est pas étonnant, puisque cette même observation a été soulignée dans [41] concernant la base Mushroom.

| Base | Nombre d'attributs manquants | Nombre de règles | Pourcentage de complétion | Précision |
|---|---|---|---|---|
| Mushroom | 01 | 30674 | 99% | 100% |
| | 02 | 29826 | 99% | 100% |
| | 03 | 29754 | 83% | 100% |
| | 04 | 29666 | 80% | 100% |
| | 05 | 29055 | 80% | 099% |
| | 06 | 28893 | 74% | 099% |
| | 07 | 28893 | 78% | 090% |
| | 08 | 28000 | 80% | 091% |
| | 09 | 27870 | 74% | 091% |
| | 10 | 27718 | 77% | 083% |
| | 11 | 27650 | 72% | 083% |
| Zoo | 01 | 54761 | 66% | 100% |
| | 02 | 53787 | 66% | 100% |
| | 03 | 50719 | 77% | 092% |
| | 04 | 48959 | 83% | 095% |
| | 05 | 48168 | 86% | 096% |
| | 06 | 47292 | 88% | 093% |
| | 07 | 46316 | 88% | 094% |
| | 08 | 45499 | 83% | 097% |
| | 09 | 45314 | 85% | 097% |
| | 10 | 45199 | 86% | 096% |
| | 11 | 44877 | 87% | 096% |
| | 12 | 44678 | 79% | 096% |
| | 13 | 44588 | 79% | 096% |
| | 14 | 44518 | 80% | 097% |

Tab. 4.4 Variation du pourcentage de complétion et de la précision de $GBAR_{MVC}$ en fonction de la variation du nombre d'attributs manquants pour un taux de valeurs manquantes égal à 5% et une valeur minsup égale à 35%.

| Base | Nombre d'attributs manquants | Nombre de règles | Pourcentage de complétion | Précision |
|---|---|---|---|---|
| Tic-tac-toe | 01 | 239 | 85% | 100% |
| | 02 | 239 | 72% | 100% |
| | 03 | 239 | 64% | 100% |
| | 04 | 239 | 59% | 100% |
| | 05 | 236 | 64% | 100% |
| | 06 | 235 | 56% | 100% |
| | 07 | 226 | 62% | 100% |
| | 08 | 220 | 54% | 085% |
| | 09 | 210 | 48% | 084% |
| | 10 | 209 | 52% | 087% |
| | 11 | 207 | 47% | 080% |
| | 12 | 207 | 44% | 078% |
| | 13 | 207 | 41% | 069% |
| | 14 | 206 | 38% | 062% |
| House-votes | 01 | 401 | 100% | 100% |
| | 02 | 336 | 100% | 081% |
| | 03 | 304 | 083% | 081% |
| | 04 | 294 | 062% | 081% |
| | 05 | 286 | 040% | 050% |
| | 06 | 286 | 033% | 050% |
| | 07 | 286 | 028% | 050% |
| | 08 | 286 | 025% | 050% |
| | 09 | 286 | 022% | 050% |

Tab. 4.5 Variation du pourcentage de complétion et de la précision de $GBAR_{MVC}$ en fonction de la variation du nombre d'attributs manquants pour un taux de valeurs manquantes égale à 5% et une valeur minsup égale à 35%.

| Base | Nombre d'attributs manquants | Nombre de règles | Pourcentage de complétion | Précision |
|---|---|---|---|---|
| Monks2 | 01 | 64 | 100% | 100% |
| | 02 | 64 | 088% | 100% |
| | 03 | 64 | 066% | 088% |
| | 04 | 64 | 050% | 088% |
| | 05 | 64 | 040% | 088% |
| | 06 | 60 | 066% | 069% |
| | 07 | 60 | 057% | 069% |
| | 08 | 60 | 050% | 069% |
| | 09 | 60 | 044% | 069% |

Tab. 4.6 Variation du pourcentage de complétion et de la précision de $GBAR_{MVC}$ en fonction de la variation du nombre d'attributs manquants pour un taux de valeurs manquantes égal à 5% et une valeur minsup égale à 35%.

## 4.7 Evolution du temps de complétion en fonction de la variation du nombre d'attributs manquants et du taux de valeurs manquantes

Dans cette section, nous allons étudier l'évolution du temps de complétion de $GBAR_{MVC}$ en fonction de la variation respective du nombre d'attributs manquants et du pourcentage de valeurs manquantes.

### 4.7.1 Effet de la variation du nombre d'attributs manquants sur le temps de complétion

Les Figures 4.1 et 4.2 donnent l'évolution du temps de complétion de $GBAR_{MVC}$ en fonction de la variation du nombre d'attributs manquants, pour une valeur minsup égale à 35% et un taux de valeurs manquantes égal à 5%. Dans cette série d'expérimentation, nous désignons par le nombre de règles, les règles dont la partie conclusion présente un attribut manquant. À la lecture des courbes des Figure 4.1 et 4.2, nous constatons que le temps de complétion de $GBAR_{MVC}$ augmente en fonction de l'augmentation du nombre d'attributs manquants. Cependant, il est important de signaler que cette augmentation en termes de temps est liée à l'augmentation du nombre de règles concluant sur des attributs manquants. Ceci se justifie par le résultat théorique de l'étude de complexité de l'algorithme $GBAR_{MVC}$ (page 63). En effet, lors de la section consacrée à la présentation de l'étude de complexité, nous avons montré que la complexité de $GBAR_{MVC}$ est de l'ordre de $O(t_{premisse} \times |BG| \times |K_{VM}|)$, où $|BG|$ constitue le pire des cas. Ce nombre est égal à $\sum_{X_i \in I_m} O(NR_i)$ où $NR_i$ est le nombre de règles concluant sur un attribut $X_i$ manquant, ce que nous avons désigné dans les Figure 4.1 et 4.2 par le nombre de règles. Étant donné que la taille de la prémisse $t_{premisse}$ est restée constante dans l'ensemble des bases, et le terme $|K_{VM}|$ désigne (le pire des cas), le nombre de transactions manquantes. Ceci se matérialise dans ce cas par le pourcentage de valeurs manquantes, fixé à 5%. Par conséquent, l'augmentation du temps est liée à l'augmentation du nombre de règles concluant sur des attributs manquants. Ce nombre est implicitement lié au nombre même d'attributs manquants.

### 4.7.2 Effet de la variation du taux de valeurs manquantes sur le temps de complétion

Les Figures 4.3 et 4.4 donnent l'évolution du temps de complétion de $GBAR_{MVC}$ en fonction de la variation du taux de valeurs manquantes, pour une valeur minsup égale à 35%. À la lecture des courbes des Figures 4.3 et 4.4, nous constatons que le temps d'exécution augmente avec l'augmentation du taux de valeurs manquantes. Ceci est observé malgré la diminution du nombre de règles concluant sur des attributs manquants et la diminution de la taille des prémisses des règles. Cette observation s'explique par l'aug-mentation du nombre de transactions manquantes, matérialisées par le pourcentage de valeurs manquantes que nous faisons varier de 5% à 20%. Ceci se justifie par le résultat théorique issu de l'étude de complexité de l'algorithme $GBAR_{MVC}$ (page 63). En effet, nous avons montré que la complexité de $GBAR_{MVC}$ est en fonction de la taille de la prémisse, du nombre de règles concluant sur des attributs manquants et du nombre de transactions manquantes. Cependant, d'après ce résultat expérimental, nous remarquons que le temps d'exécution a été affecté beaucoup plus par l'augmentation du taux de valeurs manquantes que par la taille des prémisses des règles ou par le nombre même des règles. En effet, ceci peut être expliqué de la manière suivante :

- Si nous considérons l'exemple de la base Zoo, nous remarquons que 5% de valeurs manquantes implique 5 transactions manquantes. Par conséquent, si nous supposons qu'il existe une seule valeur manquante par transaction, il y aura donc 86900 tests de consistance, puisqu'il existe 17380 règles. Cependant, pour un pourcentage de valeurs manquantes égal à 10% (ce qui implique 10 transactions manquantes), il y aura 131120 tests de consistance pour un nombre de règles réduit à 13112.

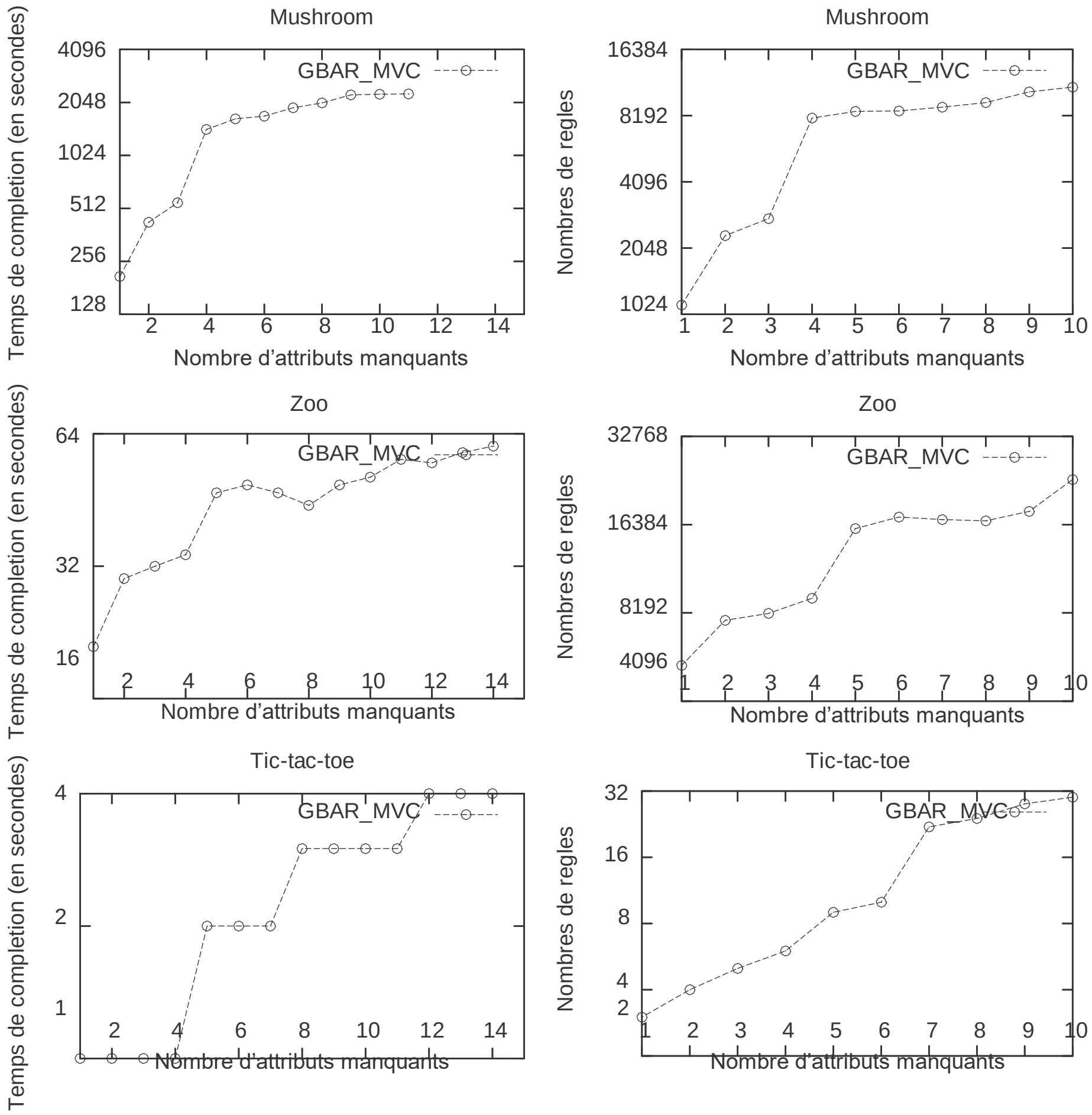

Fig. 4.1 Evolution du temps de complétion et du nombre de règles de $GBAR_{MVC}$ vs. la variation du nombre d'attributs manquants pour un taux de valeurs manquantes égal à 5% et une valeur minsup égale à 35%.

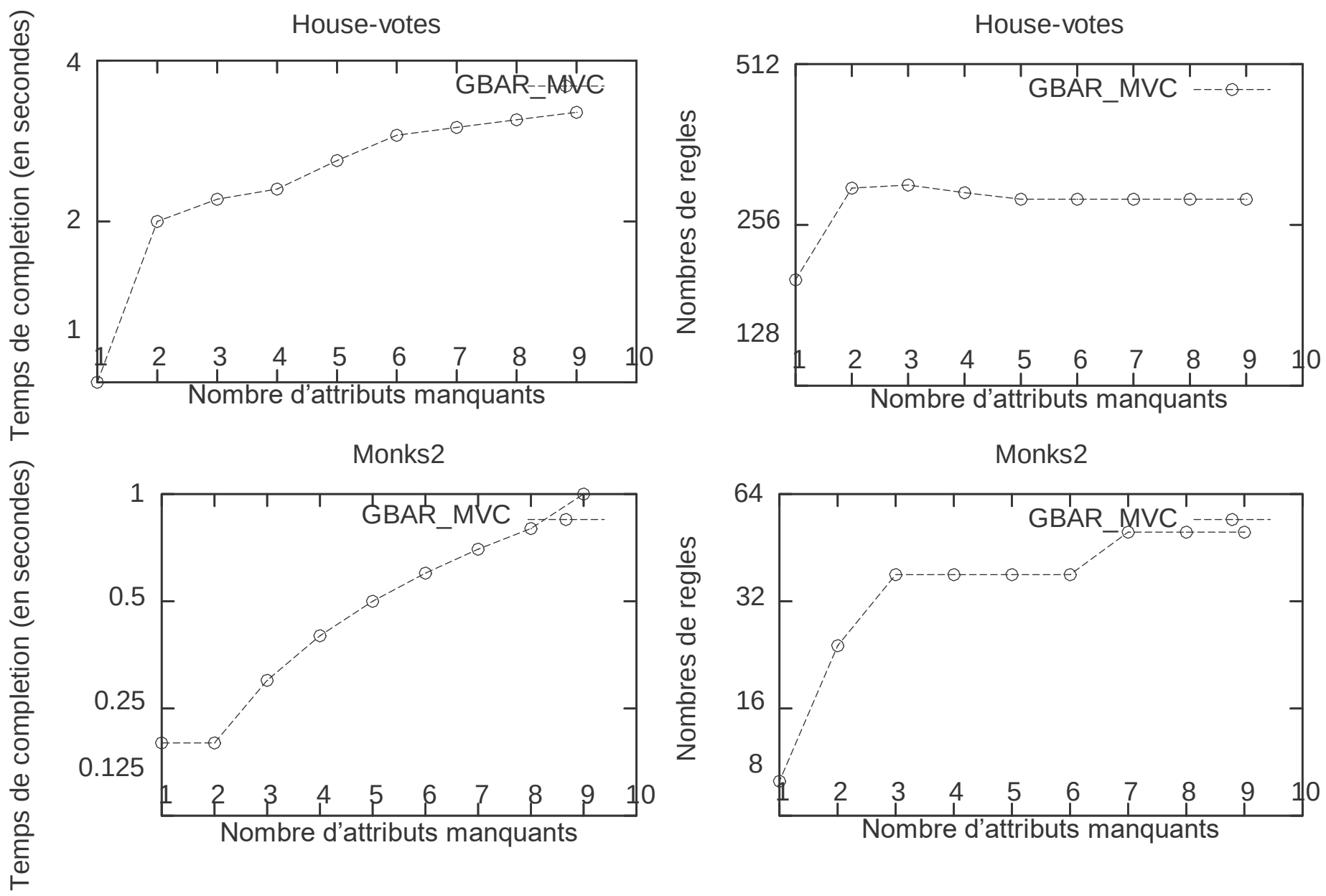

Fig. 4.2 Evolution du temps de complétion et du nombre de règles de $GBAR_{MVC}$ vs. la variation du nombre d'attributs manquants pour un taux de valeurs manquantes égal à 5% et une valeur minsup égale à 35%.

Cette observation a permis de dégager que même si le nombre de règles diminue, le temps d'exécution est en train d'augmenter à cause de l'augmentation du nombre de test de consistance effectuée. Ce nombre est lié à l'augmentation du pourcentage de valeurs manquantes. Toutefois, il est important de signaler que cette augmentation est d'autant plus renforcée, quand il s'agit d'un nombre de règles relativement élevé. Si nous reprenons l'exemple de la base Zoo, pour 5% de valeurs manquantes (impliquent 5 transactions manquantes), il y a un 86900 tests de consistance, puisque le nombre de règles est égal à 17380. En revanche, un même taux de valeurs manquantes sur la base Tic-tac-toe impliquent 47 transactions manquantes et un nombre de tests de consistance égal à 940, puisque il existe seulement 20 règles. Ceci explique le faible temps d'exécution des bases présentant un nombre de règles relativement faibles.

## 4.8 Etude des performances de l'approche $GBAR_{MVC}$ vs. $AR_{MVC}$

Dans cette section, nous allons comparer les résultats de $GBAR_{MVC}$ à ceux obtenus par $AR_{MVC}$. Cette série d'expérimentation nous permet d'une part de comparer les deux critères d'évaluation, à savoir le pourcentage de complétion et la précision des deux approches en fonction de l'augmentation du taux de valeurs manquantes et du nombre d'attributs manquants. D'autre part, cette comparaison porte sur l'évolution en termes de temps de complétion, en fonction des paramètres susmentionnés.

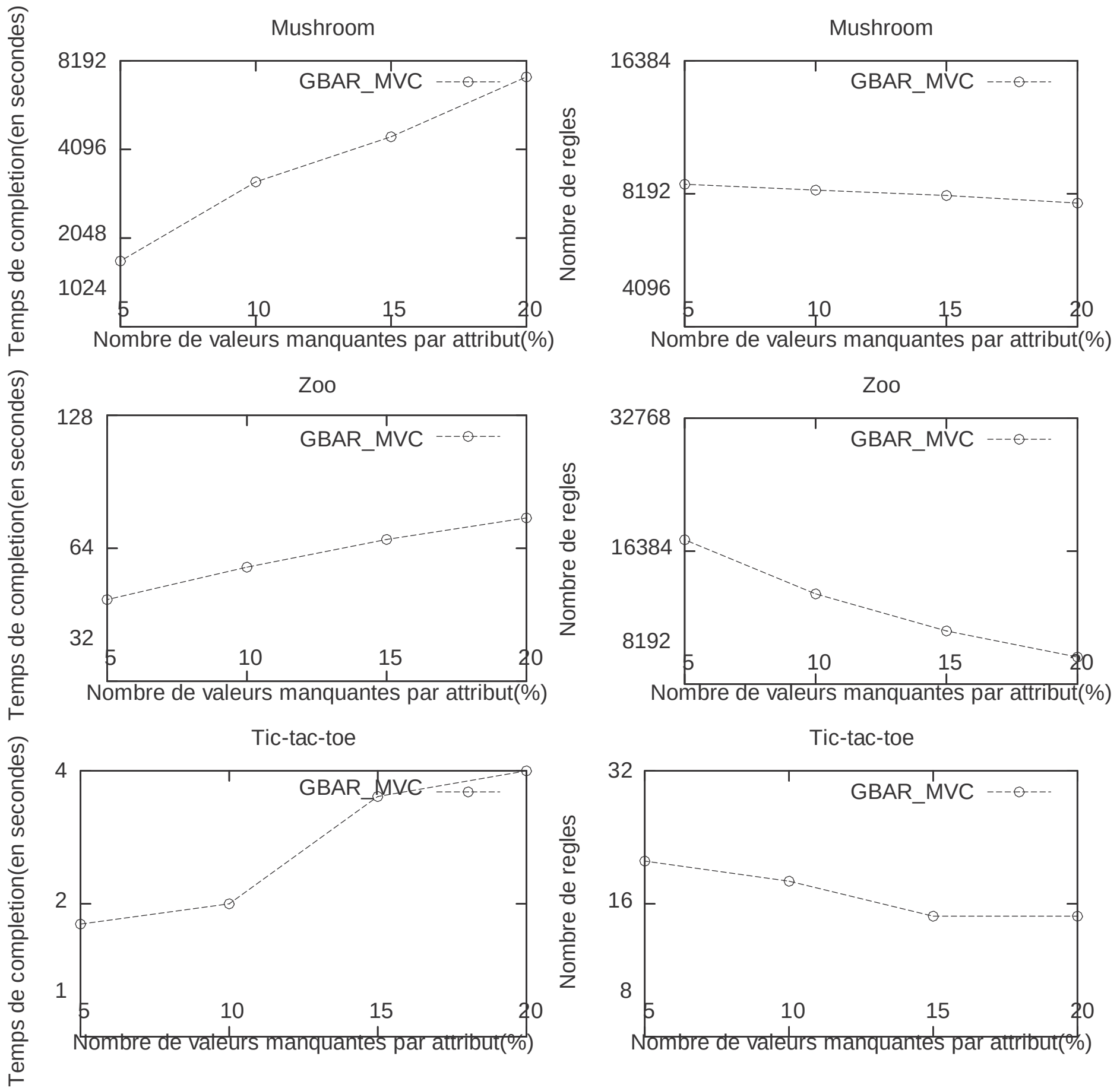

Fig. 4.3 Evolution du temps de complétion et du nombre de règles de $GBAR_{MVC}$ vs. la variation du taux de valeurs manquantes pour une valeur minsup égale à 35%.

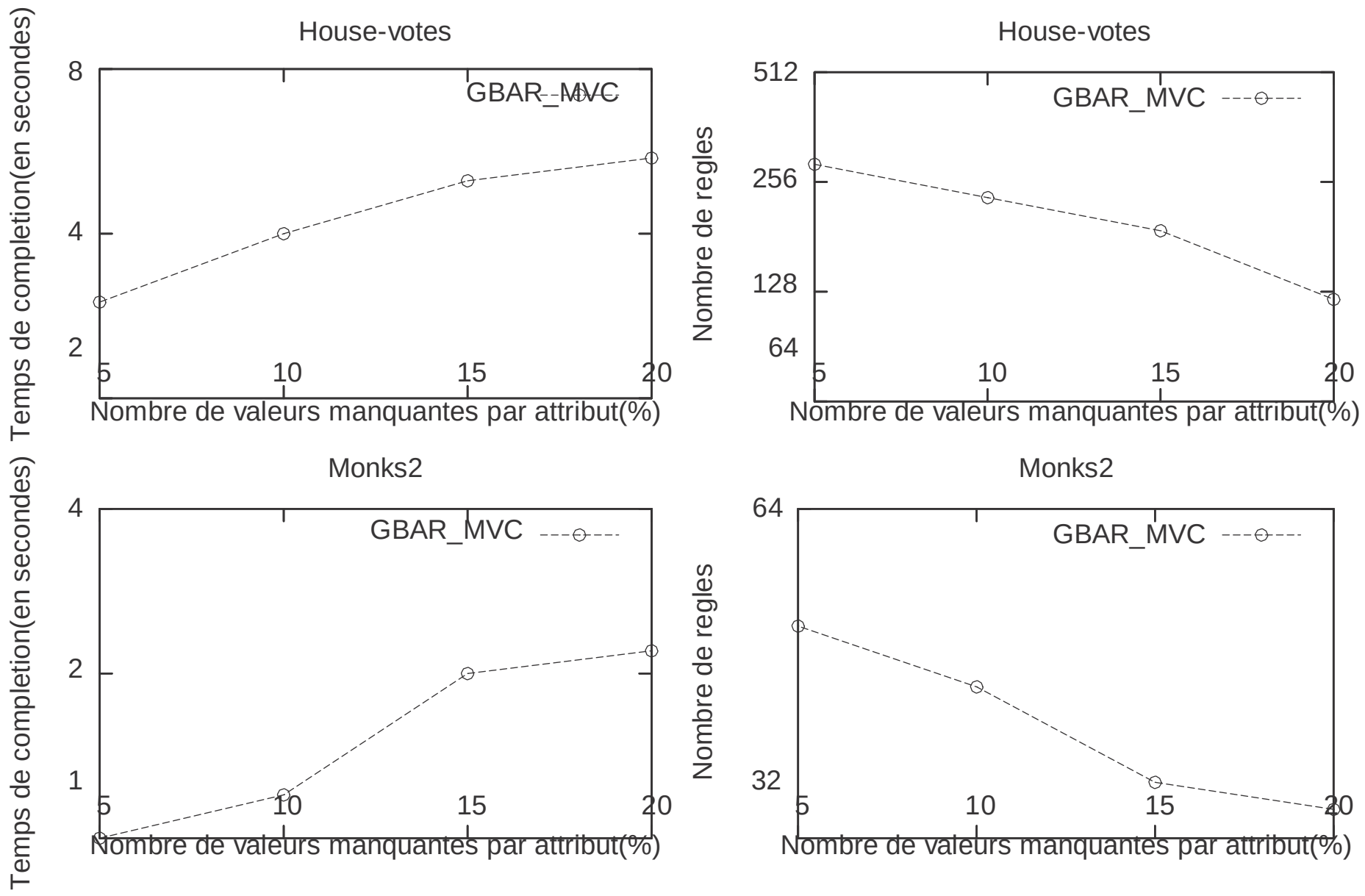

Fig. 4.4 Evolution du temps de complétion et du nombre de règles de $GBAR_{MVC}$ vs. la variation du taux de valeurs manquantes pour une valeur minsup égale à 35%.

### 4.8.1 Comparaison du pourcentage de complétion et de la précision de $AR_{MVC}$ vs. $GBAR_{MVC}$ avec l'augmentation du taux de valeurs manquantes

Les Tables 4.7, 4.8 et 4.9 donnent la variation du pourcentage de complétion et de la précision de $AR_{MVC}$ vs. ceux de $GBAR_{MVC}$ en fonction de la variation du taux de valeurs manquantes pour une valeur de *minsup* = 10%.

À la lecture des Tables 4.7, 4.8 et 4.9, nous remarquons que :

- Dans le cas des bases Mushroom et House-votes, le pourcentage de complétion de $AR_{MVC}$ est meilleur que celui de $GBAR_{MVC}$. En effet, $AR_{MVC}$ a permis de compléter plus de valeurs manquantes que $GBAR_{MVC}$. Cependant, ceci ne s'explique pas par le nombre de règles réduit de $GBAR_{MVC}$ par rapport à ceux de $AR_{MVC}$. En effet, cette observation se justifie par la troisième constatation que nous avons adressée à l'approche $AR_{MVC}$ lors du chapitre 2 (page 43). Ceci est dû à la mesure Score introduite par $AR_{MVC}$. En effet, cette mesure permet l'utilisation de règles dont toute la partie prémisse est manquante dans la transaction. De telles règles ne sont pas utilisées par $GBAR_{MVC}$. Nous pensons qu'il s'agit de règles peu ables à la complétion.
- Cependant, pour les autres bases, Zoo, Tic-tac-toe et Monks2, nous constatons qu'aussi bien que $AR_{MVC}$, $GBAR_{MVC}$ a permis de compléter les 100% des valeurs manquantes. Ceci n'est pas surprenant, même si le nombre de règles de $GBAR_{MVC}$ est relativement faible par rapport à celui de $AR_{MVC}$. Ceci constitue l'avantage de l'approche $GBAR_{MVC}$. En effet, es règles utilisées par $GBAR_{MVC}$ sont démunies de redondance.
- Quant à la précision, nous remarquons que même si $AR_{MVC}$ a permis de compléter plus de valeurs manquantes que $GBAR_{MVC}$ (dans le cas de Mushroom et House- votes), $GBAR_{MVC}$ a présenté une meilleur précision. Ceci est vérifié quel que soit le pourcentage de valeurs manquantes introduites. Ce résultat est de même obtenu pour les bases Zoo, Tic-tac-toe et Monks2. Cette meilleure précision peut être expliquée par ce qui suit :

1. L'extraction des règles utilisées par $GBAR_{MVC}$ est faite avec une prise en considération de l'aspect incomplet des données. Ces règles présentent l'avantage qu'elles sont robustes aux valeurs manquantes, i.e., plus ables en présence de valeurs manquantes. Par conséquent, ces règles offrent une complétion correcte. Ceci se traduit par une meilleure précision que celle offerte par $AR_{MVC}$.
2. Il a été mentionné que la précision de $AR_{MVC}$ dépend fortement du nombre de règles associatives générées [53]. Un nombre de règles assez élevé entraîne par conséquent une inefficacité du processus de complétion. En effet, nous remarquons, par exemple dans le cas de la base House-vote, qu'avec 5% de va- leurs manquantes, la précision de $AR_{MVC}$ est égale à 87%, tandis que celle de $GBAR_{MVC}$ est de 95%. Ceci s'explique d'une part par le nombre de règles relativement élevé de $AR_{MVC}$ par rapport à celui de $GBAR_{MVC}$. D'autre part, par le nombre des conflits présentés par les deux approches. En effet, nous avons constaté lors de la section 4.4.2 de ce chapitre, qu'une complétion correcte des valeurs manquantes est fortement liée à l'absence de conflits. Étant donné que $GBAR_{MVC}$ présente une réduction considérable en nombre de conflits, ceci se matérialise par une meilleure précision que celle offerte par $AR_{MVC}$.
3. Un autre constat justifie l'inefficacité de la complétion de $AR_{MVC}$ par rapport à celle de $GBAR_{MVC}$ est l'utilisation de règles que nous avons jugées peu ables à la complétion. Ces règles présentent une partie prémisse manquante. Nous avons déjà montré que ces règles permettent effectivement la complétion de plus de valeurs manquantes. En contre parties, elles n'offrent pas une complétion able des valeurs manquantes.

| | | Mushroom | | | |
|---|---|---|---|---|---|
| | Nombre de valeurs manquantes (%) | 5 | 10 | 15 | 20 |
| $AR_{MVC}$ | Pourcentage de complétion | 50% | 92% | 99% | 80% |
| | Précision | 42% | 58% | 64% | 66% |
| | Nombre de règles | 79461 | 77161 | 76830 | 68168 |
| | Nombre de conflits | 15 | 26 | 38 | 54 |
| $GBAR_{MVC}$ | Pourcentage de complétion | 74% | 76% | 78% | 80% |
| | Précision | 99 % | 99% | 97% | 97% |
| | Nombre de règles | 28893 | 27988 | 27988 | 24410 |
| | Nombre de conflits | 00 | 00 | 00 | 23 |
| | | Zoo | | | |
| | Nombre de valeurs manquantes (%) | 5 | 10 | 15 | 20 |
| $AR_{MVC}$ | Pourcentage de complétion | 100% | 100% | 100% | 100% |
| | Précision | 55% | 57% | 55% | 66% |
| | Nombre de règles | 3898169 | 3842627 | 3761081 | 3293571 |
| | Nombre de conflits | 18 | 31 | 48 | 36 |
| $GBAR_{MVC}$ | Pourcentage de complétion | 100% | 098% | 100% | 099% |
| | Précision | 97% | 89% | 88% | 88% |
| | Nombre de règles | 0824650 | 0756741 | 0626390 | 0547075 |
| | Nombre de conflits | 00 | 05 | 10 | 19 |

Tab. 4.7 Comparaison du pourcentage de complétion et de la précision de $AR_{MVC}$ vs. ceux de $GBAR_{MVC}$ en fonction de la variation du taux de valeurs manquantes pour une valeur minsup égale à 10%.

| | | | | | |
|---|---|---|---|---|---|
| | Tic-tac-toe | | | | |
| | Nombre de valeurs manquantes (%) | 5 | 10 | 15 | 20 |
| $AR_{MVC}$ | Pourcentage de complétion | 100% | 100% | 100% | 100% |
| | Précision | 86% | 73% | 76% | 71% |
| | Nombre de règles | 632826 | 592115 | 554530 | 528343 |
| | Nombre de conflits | 046 | 112 | 179 | 263 |
| $GBAR_{MVC}$ | Pourcentage de complétion | 100% | 100% | 100% | 100% |
| | Précision | 91% | 89% | 87% | 60% |
| | Nombre de règles | 315094 | 296222 | 279915 | 266022 |
| | Nombre de conflits | 000 | 003 | 015 | 311 |
| | House-votes | | | | |
| | Nombre de valeurs manquantes (%) | 5 | 10 | 15 | 20 |
| $AR_{MVC}$ | Pourcentage de complétion | 95% | 96% | 97% | 98% |
| | Précision | 87% | 77% | 73% | 71% |
| | Nombre de règles | 387342 | 369180 | 335639 | 309617 |
| | Nombre de conflits | 084 | 180 | 298 | 402 |
| $GBAR_{MVC}$ | Pourcentage de complétion | 91% | 93% | 92% | 92% |
| | Précision | 95% | 90% | 87% | 82% |
| | Nombre de règles | 125909 | 102310 | 094246 | 081162 |
| | Nombre de conflits | 000 | 004 | 011 | 018 |

Tab. 4.8 Comparaison du pourcentage de complétion et de la précision de $AR_{MVC}$ vs. ceux de $GBAR_{MVC}$ en fonction de la variation du taux de valeurs manquantes pour une valeur minsup égale à 10%.

| | | Monks | | | |
|---|---|---|---|---|---|
| | Nombre de valeurs manquantes (%) | 5 | 10 | 15 | 20 |
| $AR_{MVC}$ | Pourcentage de complétion | 100% | 100% | 100% | 100% |
| | Précision | 76% | 67% | 65% | 60% |
| | Nombre de règles | 52660 | 45249 | 33815 | 34490 |
| | Nombre de conflits | 044 | 088 | 226 | 348 |
| $GBAR_{MVC}$ | Pourcentage de complétion | 100% | 100% | 100% | 100% |
| | Précision | 83% | 65% | 71% | 63% |
| | Nombre de règles | 28325 | 25402 | 21790 | 19741 |
| | Nombre de conflits | 000 | 040 | 048 | 108 |

Tab. 4.9 Comparaison du pourcentage de complétion et de la précision de $AR_{MVC}$ vs. ceux de $GBAR_{MVC}$ en fonction de la variation du taux de valeurs manquantes pour une valeur minsup égale à 10%.

### 4.8.2 Comparaison du pourcentage de complétion et de la précision de $GBAR_{MVC}$ vs. $AR_{MVC}$ avec l'augmentation du nombre d'attributs manquants

Les Tables 4.10, 4.11 et 4.12 donnent la variation du pourcentage de complétion et de la précision de $GBAR_{MVC}$ vs. ceux de $AR_{MVC}$ en fonction de la variation du nombre d'attributs manquants pour un taux de valeurs manquantes égal à 5% et une valeur de minsup égale à 35%.

À la lecture des Tables 4.10, 4.11 et 4.12, nous remarquons que :

- Dans le cas des bases Mushroom, Zoo, Tic-tac-toe et House-votes et pour un seul attribut manquant, $AR_{MVC}$ n'a pas pu compléter aucune valeur manquante (ceci est indiqué par "-"). Tandis que le pourcentage de complétion de $GBAR_{MVC}$ est égal à 99% (resp. 66%, 85% et 100%) pour la base Mushroom (resp. Zoo, Tic-tac-toe et House-votes). Cette incapacité de $AR_{MVC}$ à compléter les valeurs manquantes s'explique par l'absence de règles concluant sur l'attribut manquant. Ceci n'est pas le cas pour $GBAR_{MVC}$. En effet, $GBAR_{MVC}$ présente un nombre non nul de règles concluant sur l'attribut en question. Ceci constitue l'apport de $GBAR_{MVC}$, qui s'explique par la prise en considération des valeurs manquantes. Cette justification est d'autant plus renforcée par le constat suivant : nous constatons dans la base Zoo (dans le cas de deux attributs manquants), que même si $AR_{MVC}$ présente un nombre beaucoup plus important de règles (60618 règles) par rapport à $GBAR_{MVC}$ (seulement 7727 règles), $GBAR_{MVC}$ a présenté un pourcentage de complétion égal à 66%, tandis que $AR_{MVC}$ n'a complété que les 33% des valeurs manquantes. Cette constatation a été mise à pro t dans la section 3.2 du chapitre 3, où nous avons montré que l'intérêt de l'utilisation d'une base de règles génériques réside sur le fait que nous nous focaliserons sur un ensemble réduit de règles, qui sont les plus intéressantes à la complétion (constituées de générateurs minimaux dans la partie prémisse). De plus, l'avantage de ces règles est qu'elles sont non redondantes. Ceci explique la différence en nombre de règles présentées par $GBAR_{MVC}$ que celles présentées par $AR_{MVC}$. Quant à la différence de précision des deux approches, ceci s'explique comme nous l'avons déjà mentionné, par la présence de règles qui prennent en considération les valeurs manquantes. Ces règles n'en figurent pas dans $AR_{MVC}$.

- Cependant, il est important de signaler qu'au-delà d'un certain nombre d'attribut. $AR_{MVC}$ offre une complétion de plus de valeurs manquantes que $GBAR_{MVC}$. Ce même constat a été observé lors de l'augmentation du pourcentage de valeurs manquantes. Ceci justifie bien le constat que nous déjà explicité. $AR_{MVC}$ dépasse $GBAR_{MVC}$ en la capacité à compléter plus de valeurs manquantes. Ceci s'explique par le fait que $AR_{MVC}$ utilise des règles, jugées non ables par $GBAR_{MVC}$. L'utilisation de telles règles est d'autant plus renforcée par l'augmentation des attributs manquants. Cependant, ces règles ne permettent pas une complétion able des valeurs manquantes. Ceci explique d'autre part la meilleure précision offerte par $GBAR_{MVC}$.

### 4.8.3 Comparaison du temps de complétion de $AR_{MVC}$ vs. celui de $GBAR_{MVC}$ en fonction de l'augmentation du pourcentage de valeurs manquantes et du nombre d'attributs manquants

Les résultats de l'évolution en termes de temps de complétion de $AR_{MVC}$ vs. celui de $GBAR_{MVC}$, en fonction de l'augmentation respective du nombre d'attribut manquant et du taux de valeurs manquantes, sont similaires d'une base à une autre. Cette évolution est donné par les Figures 4.13, 4.14, 4.15 et 4.16. À la lecture des courbes des Figures 4.13 et 4.14, nous constatons que $GBAR_{MVC}$ permet la complétion des valeurs manquantes en un temps plus réduit relativement à $AR_{MVC}$ quel que soit le nombre d'attributs manquants. Ceci s'explique comme nous l'avons déjà montré lors de la section 4.7.1, que le temps de complétion est lié au nombre de règles concluant sur les attributs manquants. Ce résultat a été montré théoriquement lors de l'étude de complexité de l'algorithme $GBAR_{MVC}$ (page 63). Étant donné que le nombre de règles de $GBAR_{MVC}$ est beaucoup plus réduit que celui de $AR_{MVC}$, le temps de complétion de $GBAR_{MVC}$ est par conséquent nettement meilleur. Ceci se justifie par l'augmentation du nombre de test de consistance effectué avec l'augmentation du nombre de règles.

| | | $AR_{MVC}$ | | | $GBAR_{MVC}$ | | |
|---|---|---|---|---|---|---|---|
| | | Nombre de règles | Pourcentage de complétion | Précision | Nombre de règles | Pourcentage de complétion | Précision |
| Mush-room | 01 | 00000 | - | - | 01128 | 99% | 100% |
| | 02 | 05205 | 47% | 100% | 02335 | 99% | 100% |
| | 03 | 07333 | 54% | 100% | 02789 | 83% | 100% |
| | 04 | 09221 | 80% | 098% | 07976 | 80% | 100% |
| | 05 | 09945 | 80% | 091% | 08567 | 80% | 099% |
| | 06 | 10027 | 85% | 091% | 08609 | 74% | 099% |
| | 07 | 10612 | 86% | 089% | 08952 | 78% | 090% |
| | 08 | 11216 | 83% | 088% | 09402 | 80% | 091% |
| | 09 | 13405 | 81% | 083% | 10516 | 74% | 091% |
| | 10 | 15903 | 83% | 077% | 11037 | 77% | 083% |
| | 11 | 16657 | 80% | 078% | 11154 | 72% | 083% |
| Zoo | 01 | 000000 | - | - | 005423 | 66% | 100% |
| | 02 | 060618 | 33% | 100% | 007727 | 66% | 100% |
| | 03 | 059629 | 77% | 092% | 008164 | 77% | 092% |
| | 04 | 073043 | 83% | 095% | 009189 | 83% | 095% |
| | 05 | 133511 | 86% | 096% | 015877 | 86% | 096% |
| | 06 | 197460 | 91% | 090% | 017380 | 88% | 093% |
| | 07 | 211633 | 92% | 089% | 017047 | 88% | 094% |
| | 08 | 245707 | 93% | 091% | 016875 | 83% | 097% |
| | 09 | 257201 | 93% | 092% | 018186 | 85% | 097% |
| | 10 | 300053 | 92% | 091% | 023343 | 86% | 096% |
| | 11 | 303227 | 93% | 090% | 026867 | 87% | 096% |
| | 12 | 327232 | 94% | 089% | 026775 | 79% | 096% |
| | 13 | 339410 | 93% | 090% | 027894 | 79% | 096% |
| | 14 | 367902 | 92% | 091% | 029018 | 80% | 097% |

Tab. 4.10 Comparaison du pourcentage de complétion et de la précision de $AR_{MVC}$ vs. Ceux de $GBAR_{MVC}$ en fonction de la variation du nombre d'attributs manquants pour un taux de valeurs manquantes égal à 5% et une valeur minsup égale à 35%.

| | | $AR_{MVC}$ | | | $GBAR_{MVC}$ | | |
|---|---|---|---|---|---|---|---|
| | | Nombre de règles | Pourcentage de complétion | Précision | Nombre de règles | Pourcentage de complétion | Précision |
| Tic-tac-toe | 01 | 000 | - | - | 003 | 85% | 100% |
| | 02 | 003 | 80% | 100% | 004 | 72% | 100% |
| | 03 | 007 | 64% | 084% | 005 | 64% | 100% |
| | 04 | 024 | 89% | 055% | 006 | 59% | 100% |
| | 05 | 038 | 92% | 064% | 009 | 64% | 100% |
| | 06 | 038 | 76% | 064% | 020 | 56% | 100% |
| | 07 | 041 | 92% | 063% | 022 | 62% | 100% |
| | 08 | 043 | 95% | 065% | 024 | 54% | 085% |
| | 09 | 043 | 84% | 065% | 028 | 48% | 084% |
| | 10 | 046 | 94% | 065% | 030 | 52% | 087% |
| | 11 | 048 | 96% | 066% | 030 | 47% | 080% |
| | 12 | 130 | 88% | 066% | 031 | 44% | 078% |
| | 13 | 212 | 97% | 055% | 032 | 41% | 069% |
| | 14 | 212 | 90% | 055% | 032 | 38% | 062% |
| House-votes | 1 | 000 | - | - | 202 | 100% | 100% |
| | 2 | 242 | 100% | 072% | 300 | 100% | 081% |
| | 3 | 329 | 089% | 057% | 304 | 083% | 081% |
| | 4 | 329 | 067% | 052% | 294 | 062% | 081% |
| | 5 | 329 | 054% | 050% | 286 | 040% | 050% |
| | 6 | 329 | 045% | 050% | 286 | 033% | 050% |
| | 7 | 329 | 038% | 050% | 286 | 028% | 050% |
| | 8 | 329 | 034% | 050% | 286 | 025% | 050% |
| | 9 | 329 | 030% | 050% | 286 | 022% | 050% |

Tab. 4.11 Comparaison du pourcentage de complétion et de la précision de $AR_{MVC}$ vs. ceux de $GBAR_{MVC}$ en fonction de la variation du nombre d'attributs manquants pour un taux de valeurs manquantes égal à 5% et une valeur minsup égale à 35%.

| | | $AR_{MVC}$ | | | $GBAR_{MVC}$ | | |
|---|---|---|---|---|---|---|---|
| | | Nombre de règles | Pourcentage de complétion | Précision | Nombre de règles | Pourcentage de complétion | Précision |
| Monks2 | 1 | 23 | 100% | 100% | 10 | 100% | 100% |
| | 2 | 46 | 100% | 080% | 24 | 088% | 100% |
| | 3 | 54 | 100% | 076% | 38 | 066% | 088% |
| | 4 | 54 | 072% | 073% | 38 | 050% | 088% |
| | 5 | 54 | 070% | 083% | 38 | 040% | 088% |
| | 6 | 54 | 069% | 045% | 50 | 066% | 069% |
| | 7 | 83 | 069% | 042% | 50 | 057% | 069% |
| | 8 | 83 | 054% | 039% | 50 | 050% | 069% |
| | 9 | 83 | 054% | 039% | 50 | 044% | 069% |

Tab. 4.12 Comparaison du pourcentage de complétion et de la précision de $AR_{MVC}$ vs. ceux de $GBAR_{MVC}$ en fonction de la variation du nombre d'attributs manquants pour un taux de valeurs manquantes égal à 5% et une valeur minsup égale à 35%.

Les Figures 4.15 et 4.16 montrent l'évolution du temps de complétion de $AR_{MVC}$ vs. celui de $GBAR_{MVC}$ en fonction de l'augmentation du pourcentage de valeurs manquantes. Nous constatons le même résultat que celui obtenu précédemment, i.e., $GBAR_{MVC}$ offre un temps de complétion plus réduit que $AR_{MVC}$. En effet, nous avons montré lors de la section 4.7.2, que le temps de complétion augmente en fonction du nombre de transactions manquantes, matérialisé par le pourcentage de valeurs manquantes. Nous avons de même montré que le temps de complétion est aussi lié au nombre de règles concluant sur un attribut manquant. Ceci justifie la différence du temps de complétion des deux approches $AR_{MVC}$ et $GBAR_{MVC}$.

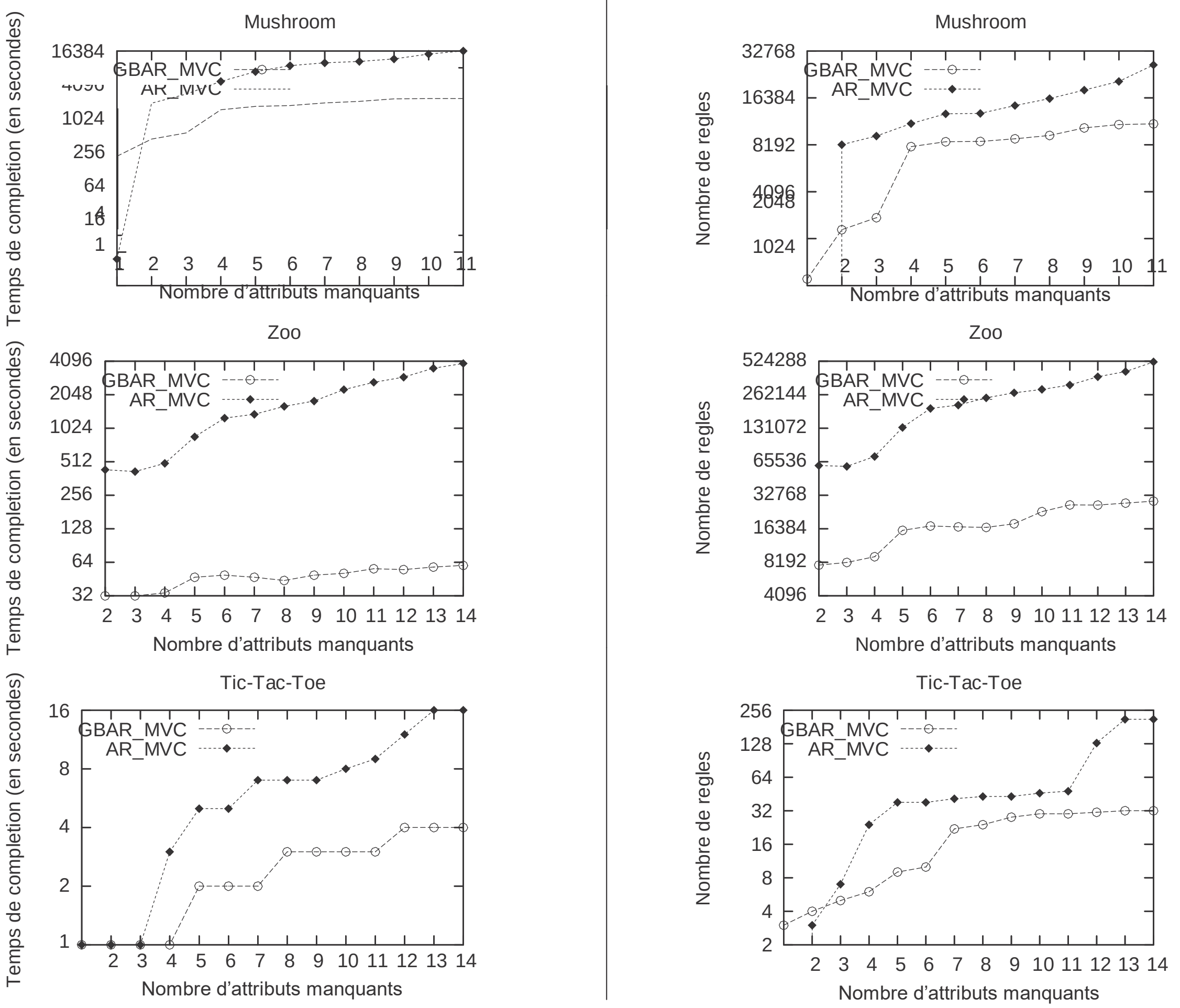

Tab. 4.13 Evolution du temps de complétion et du nombre de règles de $AR_{MVC}$ vs. celui de $GBAR_{MVC}$ en fonction de la variation du nombre d'attributs manquants pour une valeur minsup égale à 35%.

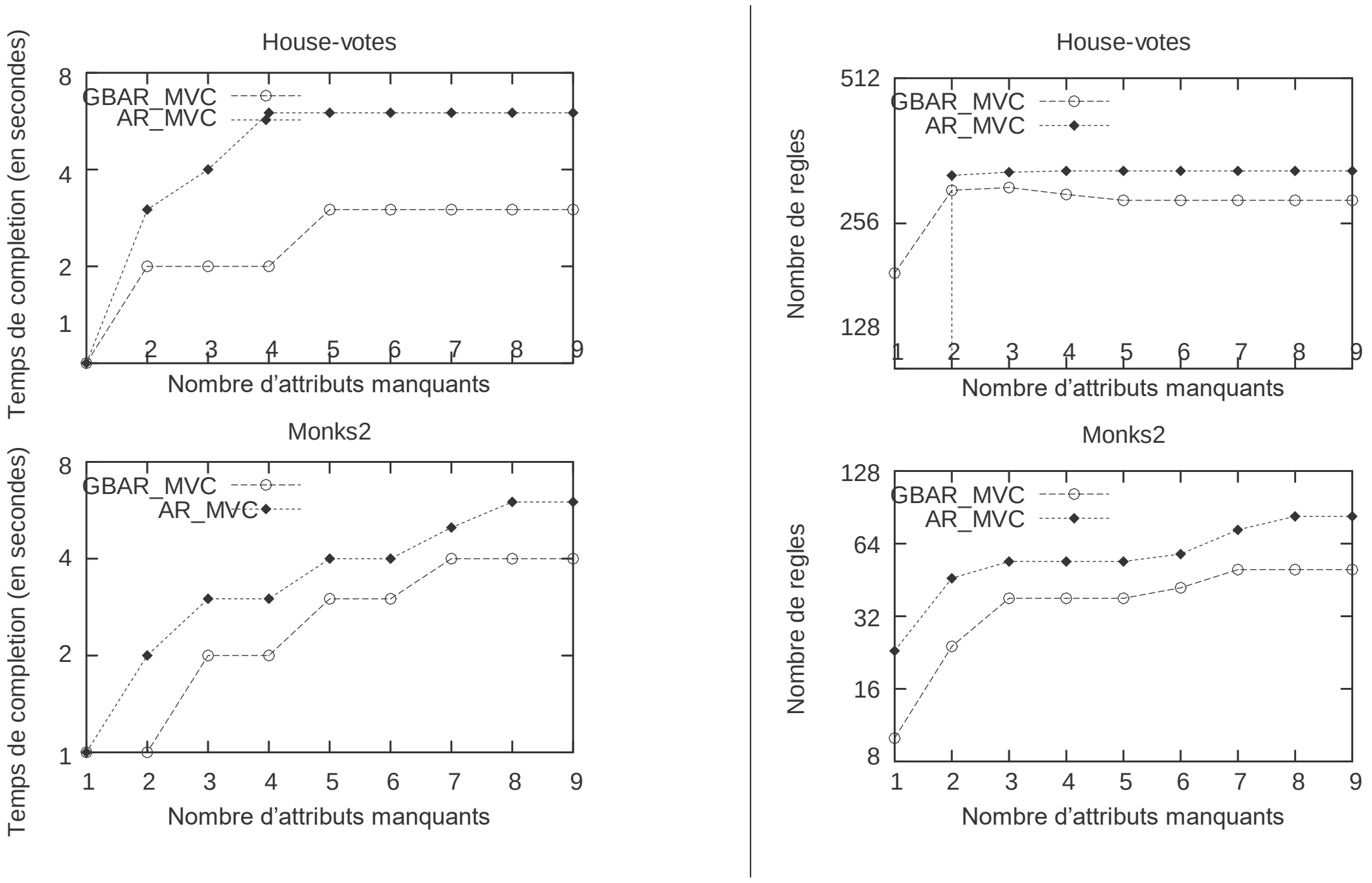

Tab. 4.14 Evolution du temps de complétion et du nombre de règles de $AR_{MVC}$ vs. celui de $GBAR_{MVC}$ en fonction de la variation du nombre d'attributs manquants pour une valeur minsup égale à 35%.

# 4.9 Conclusion

Dans ce chapitre, nous avons présenté une évaluation expérimentale de l'approche $GBAR_{MVC}$. Cette évaluation s'est portée sur plusieurs axes définis sur la base de critères d'évaluation. Ces expérimentations nous ont permis d'une part, d'analyser les caractéristiques de l'approche $GBAR_{MVC}$, et d'autre part de la comparer à une approche existante dans la littérature, à savoir $AR_{MVC}$. Les résultats obtenus montrent que l'approche proposée présente un processus de complétion plus able que celui de $AR_{MVC}$. De plus, ce processus est moins sensible à l'augmentation du nombre de valeurs manquantes.

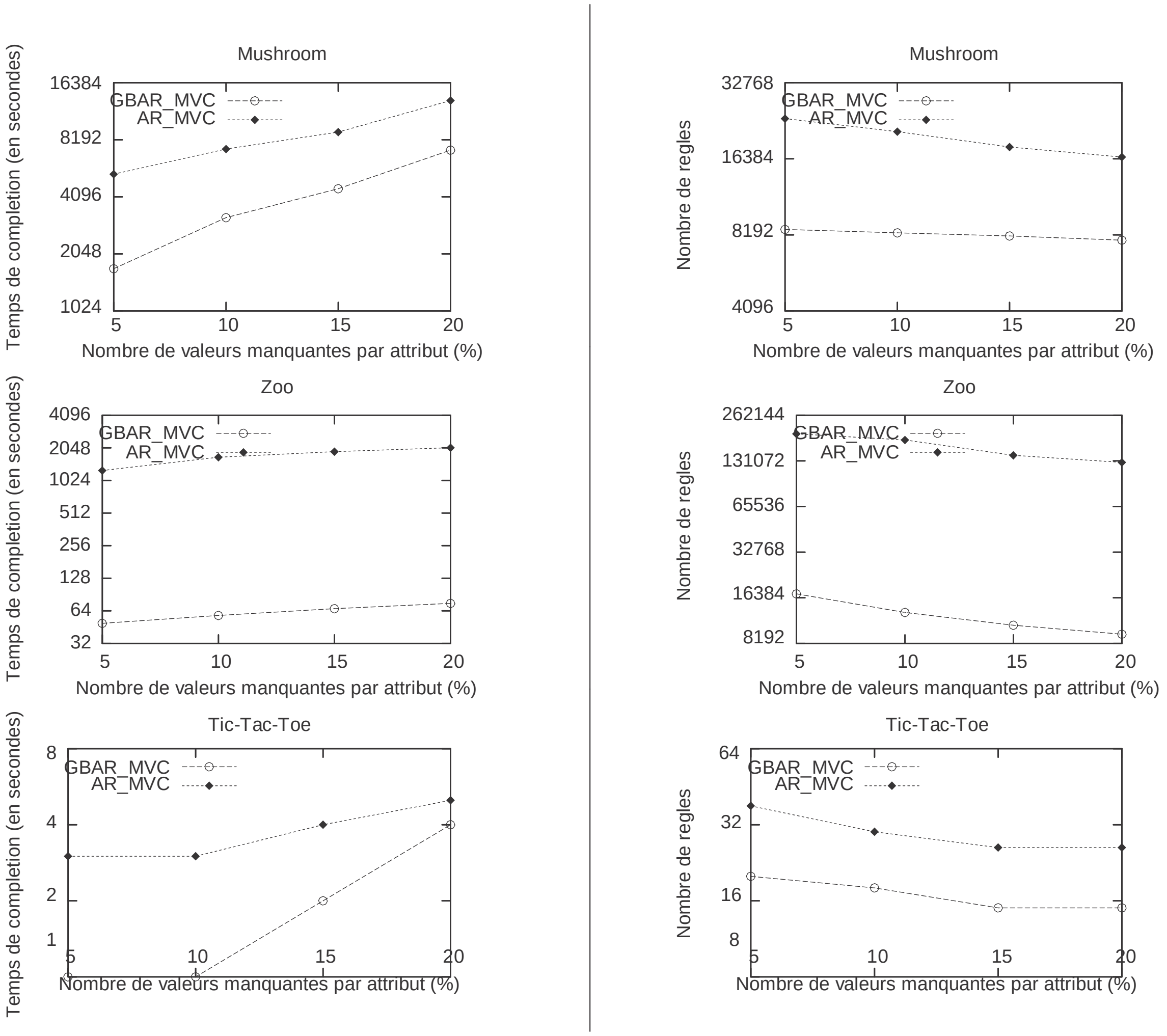

Tab. 4.15 Evolution du temps de complétion et du nombre de règles de $AR_{MVC}$ vs. celui de $GBAR_{MVC}$ en fonction de la variation du taux de valeurs manquantes pour une valeur minsup égale à 35%.

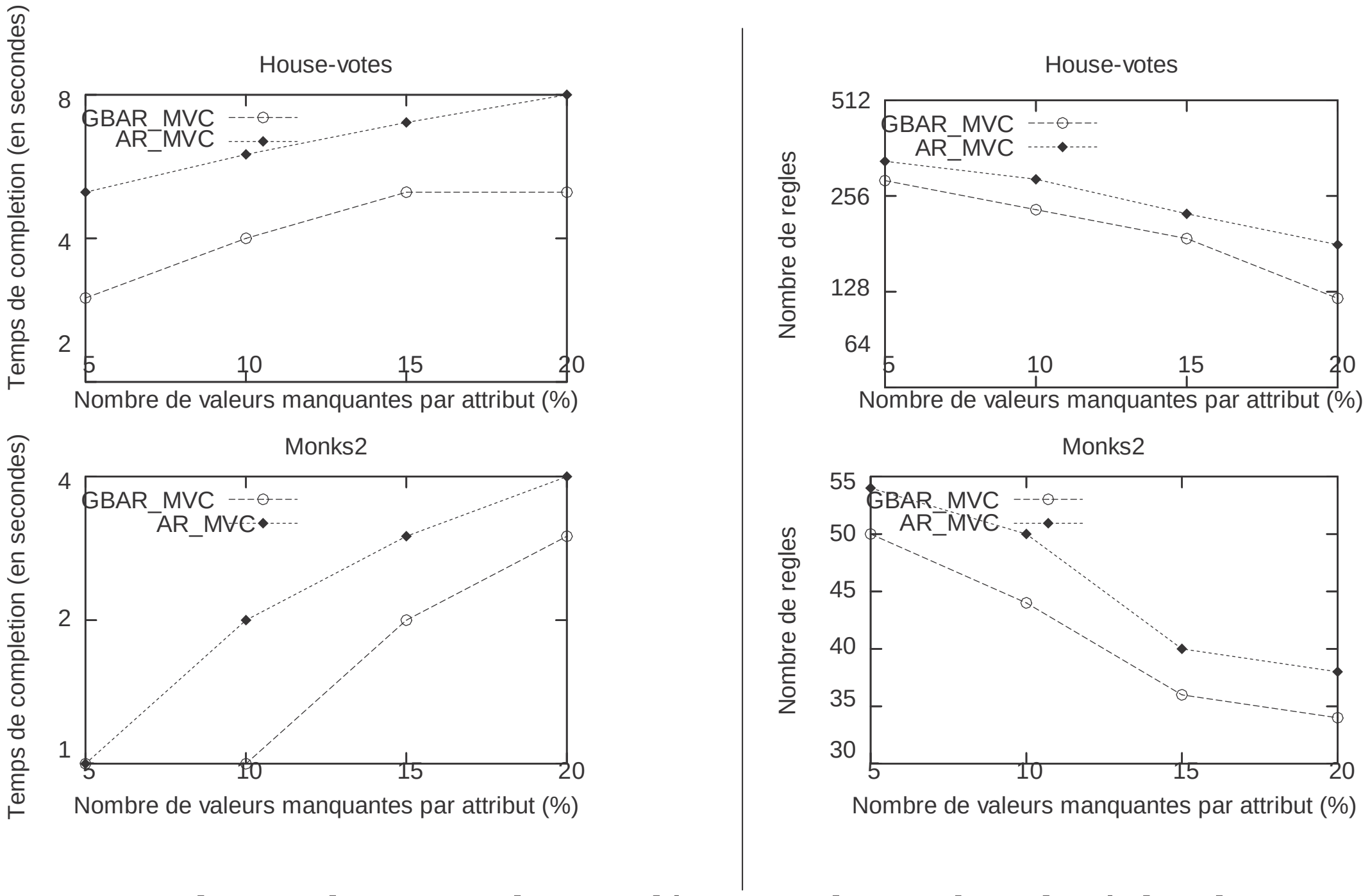

Tab. 4.16 Evolution du temps de complétion et du nombre de règles de $AR_{MVC}$ vs. celui de $GBAR_{MVC}$ en fonction de la variation du taux de valeurs manquantes pour une valeur minsup égale à 35%.

# Conclusion générale

La fouille de données est un domaine de recherche relativement récent. Cependant, ce domaine a connu en un peu de temps un essor important. Ainsi, il n'est pas surprenant d'affirmer, que la fouille de données arrive maintenant à maturité [46]. En effet, le développement des techniques et d'outils d'extraction de connaissances à partir des données se fait maintenant avec de plus en plus d'efficacité. L'attention des chercheurs s'est alors tournée vers ce qu'on appelle des contextes difficiles (présentant des valeurs manquantes) [46], où les techniques classiques de fouille de données n'en restent pas sans faille. C'est dans le cadre de cette problématique que se situent les travaux dédiés à la complétion des valeurs manquantes. Pour ce faire, ces travaux proposent une valeur de remplacement pour chaque valeur manquante. Ceci est fait par le biais d'une exploitation des règles associatives extraites à partir du contexte incomplet. Une étude bibliographique nous a permis de montrer que les approches classiques de complétion ont failli au respect au défi majeur de cette problématique, qui est l'exploitation même des données incomplètes. Cependant, d'autres travaux - plus avertis - se sont focalisés sur l'extraction des règles associatives à partir de la base totale, incluant les données incomplètes. Malheureusement, ces travaux se basent sur l'ensemble total des règles. Ces règles sont coûteuses à extraire et leur nombre présente un obstacle à assurer une complétion able des valeurs manquantes. Cet obstacle se traduit lors de l'étape de complétion par des conflits difficilement maîtrisables.

Afin de pallier les insuffisances présentées par les travaux antérieurs, nous avons proposé une nouvelle approche de complétion des valeurs manquantes, appelé $GBAR_{MVC}$. L'avantage de cette approche réside dans le fait qu'elle soit fondée dans un but à recourir aux inconvénients susmentionnés. En effet, la proposition de $GBAR_{MVC}$ a été motivée sur la base de critères, jugés essentielles à la problématique de complétion, et que nous avons essayé de respecter. Ainsi, $GBAR_{MVC}$ est une approche de complétion non brutale, i.e., le processus de complétion est préalablement associé à une prise en considération de l'aspect incomplet. De plus, cette approche se démarque, des autres approches existantes dans la littérature, par l'utilisation d'une base générique de règles associatives. L'évaluation expérimentale que nous avons menée, en est le résultat de cette motivation. En effet, $GBAR_{MVC}$ a montré un processus de complétion plus fiable que celui de $AR_{MVC}$. Ce processus a également présenté une moindre sensibilité face aux valeurs manquantes. De plus, $GBAR_{MVC}$ dépasse largement $AR_{MVC}$ en termes de temps de complétion.

Les perspectives de recherche ultérieures que nous proposons se résument comme suit :

1. L'intégration d'une approche de calcul de support, qui soit valide, i.e., respecte les propriétés définies dans [29], dans la définition de la pseudo-fermeture. En effet, nous avons vu que cette définition adopte une stratégie optimiste lors du calcul des pseudo-fermés. Cependant, lors du calcul du support des générateurs minimaux, c'est la stratégie pessimiste qui est adoptée. Il serait alors intéressant d'utiliser une valeur approchée du support qui tient compte des valeurs manquantes. Les approches proposées dans ce cadre se basent essentiellement sur l'information disponible [27, 41]. Il serait plutôt judicieux d'intégrer ce qui n'est pas disponible. En effet, des valeurs peuvent ne pas apparaître dans la base incomplète. Une évaluation prenant en considération les valeurs les moins fréquentes constituera une piste à privilégier. Ceci permettra, à notre connaissance de résoudre - ou du moins d'améliorer - le problème de silence, matérialisé par l'incapacité de fournir une valeur de remplacement à une valeur manquante.

2. L'utilisation de la notion de motifs libres-disjonctifs [14] pour la complétion des valeurs manquantes. Ces motifs permettent l'extraction de règles généralisées. Ces règles décrivent non seulement la corrélation entre les données, mais en plus l'absence de corrélations. Un travail intéressant a été proposé récemment dans ce cadre [46]. Ce travail a consisté en l'extraction des motifs $k-$libres [15], qui soient robustes aux valeurs manquantes. Les motifs libres-disjonctifs constituent un cas particulier des motifs $k-$libres, où $k$ est fixé à la valeur 2. Il nous semble alors intéressant d'étudier l'application d'un processus de complétion basé sur de tels motifs.

# Bibliographie

[1] R. Agrawal, T. Imielinski, and A. Swami. Mining association rules between sets of items in large databases. In *Proceedings of the ACM-SIGMOD Intl. Conference on Management of Data, Washington D. C., USA*, pages 207 216, May 1993.

[2] R. Agrawal and R. Srikant. Fast algorithms for mining association rules. In J. B. Bocca, M. Jarke, and C. Zaniolo, editors, *Proceedings of the 20th Intl. Conference on Very Large Databases, Santiago, Chile*, pages 478 499, 1994.

[3] W. Hsu et Y. Ma B. Liu. Integrating Classi cation and association rule mining. In *proceedings of the International Conference in Knowldge Discovery in Databases (KDD'98), New-york, USA*, pages 80 86, 1998.

[4] Y. Bastide. Data mining : algorithmes par niveau, techniques d'implantation et applications. Thèse de doctorat, Ecole Doctorale Sciences pour l'Ingénieur de Clermont-Ferrand, Université Blaise Pascal, France, Décembre 2000.

[5] Y. Bastide, N. Pasquier, R. Taouil, L. Lakhal, and G. Stumme. Mining minimal non-redundant association rules using frequent closed itemsets. In *Proceedings of the International Conference DOOD'2000, LNAI, volume 1861, Springer-Verlag, London, UK*, July 2000.

[6] L. BenOthman and S. BenYahia. Yet another approach for completing missing values. In *Proceedings of the 4th International Conference on Concept Lattices and their Applications (CLA 2006), Hammamet, Tunisia*, 2006.

[7] K. BenSalem. Design and analysis of an iterative algorithm for incomplte data estimation. *International Journal of Computer Mathematics*, 71(1) :71 82, 1999.

[8] S. BenYahia, K. Arour, and A. Jaoua. Completing missing values in databases using discovered association rules. In *Proceedings of the International Conference on Arti- cial and Computational Intelligence for Decision, Control and Automation in Engi-*

*neering and Industrial Applications. Monastir, Tunisia*, pages 138 143, March 22-24 2000.

[9] S. BenYahia and E. Mephu Nguifo. Revisiting generic bases of association rules. In *Proceedings of 6th International Conference on Data Warehousing and Knowledge Discovery (DaWaK 2004), LNCS 3181, Springer-Verlag, Zaragoza, Spain*, 1-3 September 2004.

[10] M. J. A. Berry and G. S. Lino . *Data Mining Techniques : For Marketing, Sales, and Customer Relationship Management, Second Edition.* Wiley Publishing, 2004.

[11] J.-F. Boulicaut, A. Bykowski, and C. Rigotti. Approximation of frequency queries by means of free-sets. In *Proceedings of the International Conference in Principles and Practice of Data Mining and Knowledge Discovery in Databases (PKDD'2000), Lyon, France*, pages 75 85.

[12] J.-F. Boulicaut, A. Bykowski, and C. Rigotti. Free-sets : A condensed representation of boolean data for the approximation of frequency queries. *Data Mining and Knowledge Discovery. Kluwer Academics Publishers*, 7 :5 22, 2003.

[13] S. Brin, R. Motwani, and C. Silverstein. Beyond market baskets : Generalizing association rules to correlations. In Joan Peckham, editor, In *Proceedings of the International Conference on Management of Data (ACM SIGMOD), Tucson, Arizona, USA*, pages 265 276. ACM Press, May 13-15 1997.

[14] A. Bykowski and C. Rigotti. A condensed representation to nd frequent patterns. In *Proceeding of the ACM SIGMOD-SIGACT-SIGART symposium of Principles of Database Systems, Santa Barbara*, USA, pages 267 273, 2001.

[15] T. Calders and B. Goethals. Minimal k-free representation of frequent sets. In *Proceedings of the International Conference on Principles ans Practice of Knowledge Discovery in Databases (PKDD'03), Cavtat-Dubrovnik, Croatia*, pages 71 82, 2003.

[16] B.A Davey and H.A. Priestley. *Introduction to Lattices and Order*. Cambridge University Press, 2002.

[17] A.P. Dempster, N.M. Laird, and D.B. Rubin. Maximum likelihood from incomplete data via the EM algorithm. *Journal of the Royal Statistical Society*, 39(1) :1 38, 1977.

[18] B. Ganter and R. Wille. *Formal Concept Analysis*. Springer-Verlag, 1999.

[19] G. Gasmi. Proposition d'une base générique de règles associatives. Mémoire de mastère, Faculté des Sciences de Tunis, Juillet 2005.

[20] P. Giudici and R. Castelo. Improving Markov Chain Monte Carlo model search for data mining. *Machine Learning*, 50(1 2) :127 158, 2003.

[21] J W. Grzymala-Busse. Three approaches to missing attribute values - a rough set perspective. *Workshop on Foundations of Data Mining, associated with the fourth IEEE International Conference on Data Mining, Brighton, UK, November 2004*.

[22] T. Hamrouni. Extraction des bases génériques informatives de règles sans calcul de fermetures. Mémoire de mastère, Faculté des Sciences de Tunis, Juillet 2005.

[23] J. Han and M. Kamber. *Data Mining : Concepts and Techniques*. Morgan Kaufmann Publishers, 2000.

[24] S. Jami, T. Jen, D. Laurent, G. Loizou, and O. Sy. Extraction de règles d'association pour la prédiction de valeurs manquantes. *ARIMA journal, Numéro spécial CARI'04*, pages 103 124, Novembre 2005.

[25] B. Jeudy and J.-F. Boulicaut. Using condensed representations for interactive association rule mining. *T. Elomaa, H. Mannila, and H. Toivonen, editors, Proceedings of the 6th European Conference on Principles of Data Mining and Knowledge Discovery. Springer, 2002.*, 2431 :225 236.

[26] M. Kang. Statistical analysis of missing data : an overview. *Disponible à l'adresse : http ://www.kines.uiuc.edu/labwebpages/Kinesmetrics/MissingD04_Kang.pdf*.

[27] M. Kryszkiewicz. Probabilistic approach to association rules in incomplete databases. In *Proc. of Web-Age Information Management Conference (WAIM), Shanghai, China, 2000. Lecture Notes in Computer Science, Vol. 1846. Springer*-Verlag (2000).

[28] M. Kryszkiewicz. Association rules in incomplete databases. In *Proceedings of The third Paci c-Asia Conference on Knowldge Discovery and Data mining (PAKDD). Beijing, China, 1999. Lecture Notes in Computer Science, Vol. 1574. Springer*, pages 84 93, 1999.

[29] M. Kryszkiewicz. Concise representation of frequent patterns and association rules. Habilitation dissertation, Institute of Computer Science, Warsaw University of Technology, Poland, August 2002.

[30] M. Kryszkiewicz. Concise representations of association rules. In *D. J. Hand, N.M. Adams, and R.J. Bolton, editors, Proceedings of Exploratory Workshop on Pattern Detection and Discovery in Data Mining (ESF), 2002, LNAI, volume 2447, Springer-Verlag, London, UK*, pages 92 109, 2002.

[31] R. Lefébure and G. Venturi. *Le Data Mining*. Eyrolles, 1999.

[32] R. J.A. Little and D.B. Rubin. *Statistical analysis with missing data*. Wiley, New York, 2002.

[33] W. Z. Liu, A. P. White, S. G. Thompson, and M. A. Bramer. Techniques for dealing with missing values in classi cation. *Second International Symposium on Intelligent Data Analysis, Lecture Notes in Computer Science*, 1280 :527 541, 1997.

[34] M. Magnani. Techniques for dealing with missing values in Knowldge Discovery Tasks. *Disponible à l'adresse : http ://magnanim.web.cs.unibo.it/pub.html*.

[35] N. Pasquier. Datamining : Algorithmes d'extraction et de réduction des règles d'association dans les bases de données. Thèse de doctorat, Ecole Doctorale Sciences pour l'Ingénieur de Clermont Ferrand, Université Clermont Ferrand II, France, Janvier 2000.

[36] N. Pasquier, Y. Bastide, R. Touil, and L. Lakhal. Discovering frequent closed itemsets. In C. Beeri and P. Buneman, editors, *Proceedings of 7th International Conference on Database Theory (ICDT'99), LNCS, volume 1540, Springer-Verlag, Jerusalem, Israel*, pages 398 416, 1999.

[37] G. Piatetsky-Shapiro. Discovery, anlaysis and presentation of strong rules. In Knowledge Discovery in Databases, pages 229 248, 1991.

[38] J. R Quinlan. Induction of decision trees. Machine Learning. *Lecture Notes in Computer Science, 1997*.

[39] J. R Quinlan. Unknown attributes values in induction. In *Proceedings of the International Workshop on Machine Learning, Cornell*, New York, USA, 1987.

[40] J.R Quinlan. C4.5 : Programs for machine learning. Morgan Kaufmann Publishers, 1993.

[41] A. Ragel. *Exploration des bases incomplètes : Application à l'aide au pré-traitement des valeurs manquantes*. Thèse de doctorat, Université de Caen, Basse Normandie, Décembre 1999.

[42] A. Ragel and B. Crémilleux. Treatment of missing values for association rules. In *Proceedings of the International Conference Pacif i c-Asia Conference on Knowlege Discovery and Data Mining (PAKDD'98), Melbourne, Australia, Lecture Notes in Computer Science, Springer-Verlag*, pages 258 270, April 15-17 1998.

[43] A. Ragel and B. Crémilleux. MVC - a preprocessing method to deal with missing values. *Knowledge-Based System Journal*, 12(5-6) :285 291, 1999.

[44] M. Ramoni and P. Sebastiani. Bayesian inference with missing data using bound and collapse. *Journal of Computational and Graphical Statistics*, 9(4) :779 800, 2000.

[45] F. Rioult. Représentation condensée pour les bases de données adéquate aux valeurs manquantes. Mémoire de DEA, Université de Caen, Basse Normandie, Septembre 2002.

[46] F. Rioult. *Extraction de connaissances dans les bases de données comportant des valeurs manquantes ou un grand nombre d'attributs*. Thèse de doctorat, Université de Caen, Basse Normandie, Novembre 2005.

[47] F. Rioult and B. Crémilleux. Extraction de propriétés correctes dans les bases de données incomplètes. *Conférence Francophone sur l'Apprentissage Automatique (CAp'06), 22-24 Mai 2006, Trégastel, France*, pages 347 362.

[48] F. Rioult and B. Crémilleux. Condensed representations in presence of missing values. In *Proceedings of the International symposium on Intelligent Data Analysis, Berlin, Germany*, pages 578 588, 2003.

[49] B.D. Ripley. Pattern recognition and neural networks. Cambridge University Press, 1996.

[50] A. L. Da Silva, G. Saporta, and H. Bacelar-Nicolau. Missing data and imputations methods in partition of variables. In Proceedings of the IXth Conference of the International Federation of Classification Societies, Chicago, pages 631 637, July 2004.

[51] P. Smyth and R.M. Goodman. An information theoretic approach to rule induction from databases. *IEEE Transactions On Knowledge And Data Engineering*, 4(4) :301 316, August 1992.

[52] R. Wille. Restructuring lattices theory : An approach based on hierarchies of concepts. I. Rival, editor, Ordered Sets, Dordrecht-Boston, pages 445 470, 1982.

[53] C. Wu, C. Wun, and H. Chou. Using association rules for completing missing data. In *Proceedings of 4th International Conference on Hybrid Intelligent Systems, (HIS'04), Kitakyushu, Japan*, IEEE Computer Society Press, pages 236 241, 5-8 December 2004.

[54] M. J. Zaki and C. J. Hsiao. Charm : An efficient algorithm for closed itemset mining. In *Proceedings of the 2nd SIAM International Conference on Data Mining, Arlington, Virginia, USA*, pages 34 43, April 2002.

[55] D. Draheim, C. Lutteroth, G. Weber. Generative programming for C. *ACM SIGPLAN Notices*, vol. 40, no. 8, 2005, pp. 29-33

[56] D. Draheim, G. Weber. Modeling submit/response style systems with form charts and dialogue constraints. In: *OTM Confederated International Conferences" On the Move to Meaningful Internet Systems"*, 2003, 267-278

[57] D. Draheim, M. Horn, I. Schulz. The schema evolution and data migration framework of the environmental mass database IMIS. In: *Proceedings of SSDBM 2004 – The 16th International Conference on Scientific and Statistical Database Management*. IEEE Press, 2004, pp. 341 – 344

[58] S. Ben Yahia, and A. Jaoua. Discovering knowledge from fuzzy concept lattice. In *Data mining and computational intelligence,* pp. 167-190. Physica, Heidelberg, 2001.

[59] S. Ben Yahia, T. Hamrouni, and E. Mephu Nguifo. "Frequent closed itemset based algorithms: a thorough structural and analytical survey." *ACM SIGKDD Explorations* Newsletter 8, no. 1 (2006): 93-104.

[60] Gh. Gasmi, S. Ben Yahia, E. Mephu Nguifo, and Y. Slimani. IGB: A New Informative Generic Base of Association Rules. In *Pacific-Asia Conference on Knowledge Discovery and Data Mining*, pp. 81-90. Springer, Berlin, Heidelberg, 2005.

[61] T. Hamrouni, S. Ben Yahia, and Y. Slimani. Prince: An algorithm for generating rule bases without closure computations. In *International Conference on Data Warehousing and Knowledge Discovery*, pp. 346-355. Springer, Berlin, Heidelberg, 2005.

[62] S. Ben Yahia, H. Ounalli, and A. Jaoua. An extension of classical functional dependency: dynamic fuzzy functional dependency. *Information Sciences* 119, no. 3-4 (1999): 219-234.

[63] C. Ch. Latiri, S. Ben Yahia, J. P. Chevallet, and A. Jaoua. Query expansion using fuzzy association rules between terms. Proceedings of *JIM* (2003).

[64] Ali Jaoua, F. Alvi, S. Elloumi, and S. Ben Yahia. Galois Connection in Fuzzy Binary Relations, Applications for Discovering Association Rules and Decision Making. In *RelMiCS*, pp. 141-149. 2000.

[65] I. Bouzouita, S. Elloumi, and S. Ben Yahia. GARC: A new associative classification approach. In *International Conference on Data Warehousing and Knowledge Discovery*, pp. 554-565. Springer, Berlin, Heidelberg, 2006.

[66] S. Ben Yahia, and E. Mephu Nguifo. Revisiting generic bases of association rules. In *International Conference on Data Warehousing and Knowledge Discovery*, pp. 58-67. Springer, Berlin, Heidelberg, 2004.

[67] S. Ben Yahia, and A. Jaoua. A top-down approach for mining fuzzy association rules. In *Proc. 8th Int. Conf. Information Processing Management of Uncertainty Knowledge-Based Systems*, pp. 952-959. 2000.

[68] S. Ben Yahia, and E. Mephu Nguifo. Emulating a cooperative behavior in a generic association rule visualization tool. In *16th IEEE International Conference on Tools with Artificial Intelligence (ICTAI*), pp. 148-155. IEEE, 2004.

[69] T. Hamrouni, S. Ben Yahia, and Y. Slimani. Avoiding the itemset closure computation "pitfall". In *CLA*, vol. 2005, pp. 46-59. 2005.

[70] S. Ben Yahia, Ch. C. Latiri., G. Mineau, and A. Jaoua. Découverte des regles associatives non redondantes: application aux corpus textuels. *Revue d'intelligence artificielle* 17, no. 1-3 (2003): 131-143.

[71] G. Gasmi, S. Ben Yahia, E. Mephu Nguifo, and Y. Slimani. IGB: une nouvelle base générique informative des regles d'association. *Revue I3 (Information-Interaction-Intelligence)*6, no. 1 (2006): 31-67.

[72] Ch. Cherif Latiri, W. Bellegua, S. Ben Yahia, and G. Guesmi. VIE_MGB: A Visual Interactive Exploration of Minimal Generic Basis of Association Rules. In *Proc. of the Intern. Conf. on Concept Lattices and Application (CLA 2005)*, pp. 179-196. 2005.

[73] S. Ben Yahia, K. Arour, A. Slimani, and A. Jaoua. Discovery of compact rules in relational databases. *Information Science Journal* 4, no. 3 (2000): 497-511.

[74] S. Ben Tekaya, S. Ben Yahia, and Y. Slimani. GenAll Algorithm: Decorating Galois lattice with minimal generators. In *CLA*, pp. 166-178. 2005.

[75] S. Ben Tekaya, S. Ben Yahia, and Y. Slimani. Algorithme de construction d'un treillis des concepts formels et de détermination des générateurs minimaux. *Revue Africaine de la Recherche en Informatique et Mathématiques Appliquées* 3 (2005): 171-193.

[76] T. Hamrouni, S. Ben Yahia, and E. Mephu Nguifo. A new exact concise representation based on disjunctive closure. In *Proceedings of the 2nd Jordanian International Conference on Computer Science and Engineering (JICCSE 2006), Al-Balqa, Jordan*, pp. 361-373. 2006.

[77] S. Ben Yahia, and E. Mephu Nguifo. Visualisation des règles associatives: vers une approche métacognitive. In *INFORSID*, pp. 735-750. 2006.

[78] T. Hamrouni, S. Ben Yahia, and Y. Slimani. Prince: Extraction optimisée des bases génériques de règles sans calcul de fermetures. In *23rd French Conference Informatique des Organisations et Systémes d'Information et de Décision (INFORSID'05)*, pp. 353-368. Presse Universitaire de Grenoble, 2005.

[79] G. Gasmi, T. Hamrouni, S. Abdelhak, S. Ben Yahia, and E. Mephu Nguifo. Extracting generic basis of association rules from SAGE data. In *Proceedings of the ECML/PKDD Discovery Challenge Workshop*, pp. 84-89. 2005.

[80] I. Bouzouita, S. Elloumi, and S. Ben Yahia. GARC-M: Generic association rules based classifier multi-parameterizable. In *Proceedings of 4th International Conference of the Concept Lattices and their Applications (CLA 2006), Hammamet, Tunisia*. 2006.

[81] G. Gasmi, S. Ben Yahia, E. Mephu Nguifo, and Y. Slimani. "Discovering "Factual" and "Implicative" generic association rules. *CAP* 2005 (2005): 329-344.